\documentclass{article}
\usepackage[preprint,nonatbib]{neurips_2024}

\usepackage[T1]{fontenc}
\usepackage[utf8]{inputenc}
\usepackage{lmodern}  
\usepackage{microtype}
\usepackage{indentfirst}        
\makeatletter
\@ifundefined{@noticestring}{}{\renewcommand{\@noticestring}{Preprint.}}  
\makeatother

\usepackage{amsmath}
\usepackage{amssymb}
\usepackage{amsthm}
\numberwithin{equation}{section}  

\usepackage{booktabs}
\usepackage{array}
\usepackage{multirow}
\usepackage{graphicx}
\usepackage{rotating}
\graphicspath{{figures/}}
\usepackage{tikz}
\usetikzlibrary{arrows.meta,positioning}
\usepackage{caption}
\usepackage{xcolor}
\newcommand{\fitwidth}[1]{\resizebox{\ifdim\width>\linewidth \linewidth\else\width\fi}{!}{#1}}

\usepackage[section]{placeins}

\usepackage{enumitem}
\definecolor{linkblue}{RGB}{0,68,170}
\usepackage[colorlinks=true,citecolor=linkblue,urlcolor=linkblue,linkcolor=linkblue,
            breaklinks=true]{hyperref}

\usepackage{csquotes}
\usepackage[backend=biber,style=apa,natbib=true,doi=true,url=true]{biblatex}
\usepackage{algorithm}
\usepackage{algpseudocode}
\algrenewcommand\algorithmicrequire{\textbf{Input:}}
\algrenewcommand\algorithmicensure{\textbf{Output:}}
\floatname{algorithm}{Procedure}

\theoremstyle{definition}
\newtheorem{definition}{Definition}

\newcommand{\armOn}{\textsf{On}}  
\newcommand{\armSh}{\textsf{Sh}}  
\newcommand{\armDe}{\textsf{De}}  
\newcommand{\armFr}{\textsf{Fr}}  
\newcommand{\armCt}{\textsf{Ct}}  
\newcommand{\armSf}{\textsf{Sf}}  
\newcommand{\vdMechNull}{\emph{mechanism-null}}
\newcommand{\vdTrainNull}{\emph{train-null}}
\newcommand{\vdReplayNull}{\emph{replay-null}}
\newcommand{\vdStratumNull}{\emph{stratum-null}}
\newcommand{\vdPass}{\emph{pass}}
\newcommand{\vdKill}{\emph{kill}}
\newcommand{\vdConfirmed}{\emph{confirmed}}
\newcommand{\vdDeferred}{\emph{deferred}}

\newcommand{\vdNs}{\emph{n.s.}}
\newcommand{\vdPTwoNull}{\emph{p2-null-offline}}
\newcommand{\vdGOneKill}{\emph{g1-kill}}
\newcommand{\vdDoa}{\emph{doa}}
\newcommand{\vdUnrealizable}{\emph{unrealizable}}
\newcommand{\vdUnderpowered}{\emph{underpowered}}
\newcommand{\genk}[1]{\ensuremath{\mathcal{G}_{#1}}}

\title{\textbf{Form, Not Content? A Preregistered, Placebo-Controlled Evaluation of Learned Error-Conditioned Self-Repair Through Prompts and Weights in Frozen Small Code Models}}

\author{Mehmet \.{I}\c{s}can\thanks{Corresponding author. PythaLab, Y{\i}ld{\i}z Technical
University, Istanbul, Turkey. E-mail: \href{mailto:miscan@yildiz.edu.tr}{\texttt{miscan@yildiz.edu.tr}}.}\\
\small PythaLab, Y{\i}ld{\i}z Technical University, Istanbul, Turkey}
\date{}

\begin{document}
\maketitle

\begin{abstract}
\noindent
Today, frozen small code large language models (LLMs) are increasingly deployed locally, 
yet the information by which a retry is guided after a failed attempt is still measured without 
placebo controls in the self-repair literature. In this work, a failed program is treated as a conjecture 
and an execution counterexample as an oracle-relative refutation, and PoPE (Popperian Placebo-controlled Evaluation) is introduced as 
a methodology for measuring whether evidence that falsifies code generated by an LLM can be used operationally by that same model. 
Within PoPE, error content is paired with channel-specific placebos that preserve the predeclared scaffold components while either ablating task-relevant content or deranging the task--error assignment, and frozen small code models 
(0.5--1.5B) are evaluated under preregistered rules through two channels: a prompt channel and a weight channel based on small-data 
adapter training. In both channels, the generation budget is matched at four generations per arm--unit pair. In the prompt channel, 
public-tier screening unlocks were recorded for 12 units under the content-ablated form placebo and for 10 units under the live 
error-pattern arm on the 40-unit resistant band, and the result was recorded as \emph{mechanism-null}. In the weight channel, 
an 8--8 tie was observed between the error-content adapter and the intervention-free baseline ($p = 1.0$), while the SHA-deranged
placebo adapter remained numerically ahead with 10 public-tier screening unlocks, and content-attributable superiority was not confirmed. 
These results do not constitute evidence of equivalence or non-inferiority. Equivalence was not tested separately. Within the self-audit chain, 
the $+6.0$ percentage-point (pp) delta from a superseded record was retained together with the $+0.001$ adaptive--random control from the same 
record, whereas the effect was recorded as null in the valid draw-budget-matched record. The only confirmed interaction in the program was the 
unit-by-portfolio interaction observed in public-tier dense progress with $p = 0.0001$, and it was not interpreted as evidence of controller 
superiority. In a two-point comparison read from the program cycle logs, an increase in the any-portfolio public-tier unlock rate from $27.5\%^{\dagger}$ to $60\%^{\dagger}$ was observed
when moving from 0.5B to 1.5B, and this observation was retained as a descriptive directional result within the program. The findings are 
restricted to the public-tier screening endpoint, and hidden-tier confirmation was deferred by design. This pattern is interpreted not as 
indicating that compiled criticism disappears as information, but as indicating that its external epistemic role in independently testing a 
new conjecture is lost. When a representation learned from the oracle is written back into the generation state, testing is replaced by conditioning, 
and criticism remains criticism only when each new conjecture is reconnected to an external test. No working JEPA-RL controller is claimed. Instead, PoPE is presented as a placebo-controlled and retestable measurement standard through which this interpretation is operationalized.
\end{abstract}

\section{Introduction}\label{sec:intro}

Code-generating large language models (LLMs) are increasingly being run at smaller scales, on local hardware, and in frozen form, with their weights left unchanged, because of latency, cost, and data-privacy considerations \citep{hui2024,guo2024}. In this deployment regime, retraining is not a practical option. When a generated program fails a public test, the standard mechanism invoked is a retry loop. At that point, the actual decision faced by the practitioner is whether the failed code should be shown to the model again together with execution evidence or whether the same output-generation budget should be allocated to blind resampling. The central question is therefore not so much whether a retry should be performed, but by what information and through which channel the limited draws should be directed.

Within the positive line of work on this retry decision, preference has been given to a learned and error-conditioned controller. In the self-repair literature, improved correction performance has been reported when a failed program is presented to the model again together with execution evidence \citep{madaan2023,shinn2023,chen2024,zhong2024,ding2024}. On the training side of this line of work, learning from execution feedback through reinforcement learning (RL) has been pursued \citep{le2022,shojaee2023,gehring2024,jiang2024}. At its origin, an early formulation was introduced in which a separate corrector module was trained using feedback on defective outputs produced by the base model \citep{welleck2023}. The same idea has also been instantiated through self-training loops constructed from filtered model-generated outputs \citep{zelikman2022,gulcehre2023,singh2024}. Progressive refinement has been presented as a recent positive example of this family \citep{du2025}. In the weight channel, small-data adapter training has been made accessible by parameter-efficient fine-tuning (PEFT), particularly low-rank adaptation (LoRA) and its quantized derivative, QLoRA \citep{hu2021,dettmers2023}. It has also been reported that the behavior of small models can be changed through training \citep{cho2025}. Across all of these lines of work, error content learned from refutations is therefore assumed to constitute a transferable signal in its own right.

This shared assumption has not been validated at every model scale or in every task regime. In controlled re-evaluations, a substantial proportion of reported self-repair gains has been attributed to model capacity and additional sampling budget \citep{olausson2024,huang2024,stechly2025}. A critical survey has further concluded that self-correction obtained through prompted-LLM feedback has not exhibited a general record of success outside exceptionally well-suited tasks \citep{kamoi2024}. On the training side of the same line of evaluation, offline supervised fine-tuning over self-generated correction traces has been reported to be insufficient on its own to instill self-correction because of distribution mismatch and mode collapse \citep{kumar2025}. In small models, the generation distribution can be shifted adversely by the failing code itself \citep{dinh2023}, and a capability gap can remain between generation and self-verification \citep{song2025}. Blind best-of-N sampling, in which no error information is used, is also regarded as a strong baseline in the test-time compute literature \citep{brown2024,snell2025}. For small frozen generators, the critical question is therefore whether the purported gain from a learned controller is attributable to learned error content or instead to the form of the retry and the scale of the generator.

A Popperian framework is adopted for the epistemological decomposition of the retry loop \citep{popper1959,popper1963,lakatos1968}. Under this reading, a generated program can be treated as a conjecture, while a public-test violation can be read as an oracle-relative refutation defined with respect to a computational procedure external to the model. The concept of severity can likewise be approximately operationalized through the execution oracle \citep{mayo2006}. Through this lens, the measurement problem is decomposed into three components: the error content learned from refutations, the form of the retry prompt or adapter, and the scale of the generator.

In the existing learned-controller literature, this decomposition has not been established at four points, and each problem conditions the next. The first problem is form confounding. Under learned selection or learned conditioning, the mechanism claimed to have been learned is presented in the same package as the prompt form through which that mechanism is rendered. It is known that prompt form, including the number of bullet points, ordering, and surface structure, can alter model behavior even when semantic content is weak \citep{sclar2024,brucks2025,khojah2025}. This sensitivity is quantitative rather than merely existential. Performance can be reduced to the level of random guessing through template selection alone, and the effect has been found to be most pronounced in smaller models \citep{voronov2024}. In code tasks, marked performance shifts have likewise been produced through reformatting alone while content is held fixed \citep{he2024}. Among the studies reviewed here, very few positive reports have been evaluated against a form-matched placebo. The second problem is matched-budget and order-statistic confounding. Within-unit adaptation is reported without being separated from the number of additional draws. Because best-of-B is an order statistic, the contribution of an adaptive allocator cannot be distinguished from additional sampling unless a comparison with uniform mixing is performed under the same output-generation budget. Independent arms are assumed in classical allocation theory. Correlated samples drawn from a single frozen generator fall outside that regime \citep{lattimore2020,audibert2010,karnin2013}. Findings that gains from iterative debugging decay rapidly across rounds can also be read as a symptom of the same confound \citep{adnan2025}. The third problem following this budget issue is channel confounding. Error conditioning is tested either in the prompt channel or in the weight channel, while the two channels are never measured jointly on the same units against the same family of placebo controls. A null in the prompt channel therefore remains open to the objection that ``the content would have transferred if it had been written into the weights.'' A positive result in the weight channel remains open to objections based on form and memorization. It has been reported that LoRA trained on small datasets learns less and forgets less \citep{biderman2024}, and that narrowly targeted fine-tuning can impair general correctness \citep{ibrahim2026}. However, a weight-space form control such as a SHA-deranged placebo LoRA has not been included among the studies reviewed here. The fourth problem is scale confounding. A result obtained at a single small scale is interpreted as a result ``about the method,'' even though scale dependence is known to exist \citep{kaplan2020,hoffmann2022}. Unless generator scale is measured as a separate arm, method effects and scale effects remain conflated. When these four problems are considered together, learned error-conditioned self-repair claims cannot be attributed in their current form.

A structural reason prevents this attribution gap from being closed using the existing literature. The cited positive anchors have not been evaluated under the complete combination of conditions studied here. The joint-embedding predictive architecture (JEPA) line has been reported on large-scale image data \citep{lecun2022,bardes2022,assran2023,grill2020}, execution-feedback RL has been reported for models with at least 8B parameters \citep{gehring2024}, self-training has been reported for large model families \citep{singh2024}, and instruction-diversity results have been reported under full fine-tuning \citep{zhang2024a}. The regime comprising frozen 0.5--1.5B code models, a zero-pass-in-pool stratum, and a data budget of a few hundred examples lies in a region that has not been characterized among the studies reviewed here. The selected literature provides few directly comparable records in which the same regime is evaluated through an explicit failure boundary or a preregistered null. The negative characterization provided here therefore does not contradict the existing positive findings. Instead, a regime left outside their scope is measured.

The discipline required to measure this scope gap has not been constructed from scratch. This paper constitutes the fourth stage of a program through which the same measurement discipline has been established in three preceding stages. In the first stage, the measured gain of a Popperian prompt skill was shown to be attributable not to Popperian vocabulary but to scaffold structure \citep{iscan2026a}. In the second stage, it was reported that none of 26 post-hoc falsification operators applied by frozen small code models to their own samples outperformed best-of-N under a matched output-generation budget. This negative result was characterized mechanistically as a coverage wall, capability scissors, and a near-empty consensus trap, while LoRA or soft-prompt adaptation and a stronger generator were identified explicitly as future work \citep{iscan2026b}. In the third stage, the self-repair feedback packet was decomposed under placebo control, the form-not-content wall was defined, and JEPA-RL controllers that had not been preregistered were presented as the starting point for the subsequent evaluation \citep[\S4.1]{iscan2026c}.

For all of these reasons, PoPE is presented in this work as an original evaluation methodology through which learned error-conditioned self-repair can be tested. Through this formulation, the controller line left by the preceding stages as a starting point is subjected to a standalone and preregistered treatment, while the training-based weight channel, the stronger generator, and the third architectural family identified by \citet{iscan2026b} are addressed as the remaining open programmatic branches. Previous controllers are not reported again as new results. Instead, they are carried in the program ledger as inherited stages by reference to \citet{iscan2026c}. Within PoPE, preregistration and placebo control are combined across two deployment channels on frozen small code models. A two-gate decision rule is adopted. A content arm is first required to outperform the intervention-free baseline or blind baseline at the signal gate and is then required to outperform its form-matched placebo at the content gate. An arm that passes the signal gate but fails at the content gate is attributed to form rather than to error content. The measurement instruments are instantiated as the error detector (HEF), diagnostic pipeline (FASTR), diversity schedulers (DS), learnability-gated design (LG), budgeted best-of-B allocator (AEG-BANDIT), JEPA-based error-set architecture (ERA), and the two decisive instruments, namely the learned error-lattice prompt controller (ELF) and the error-lattice QLoRA weight adapter (ELW). Each content arm is mirrored by a form-matched placebo within its own deployment channel. In the prompt channel, a content-ablated placebo (\armSh{}) and a mismatched decoy (\armDe{}) are used. In the weight channel, a SHA-deranged placebo adapter (\armSf{}) and an intervention-free baseline (\armFr{}) are used. All comparisons are performed under a matched output-generation budget ($R = 4$, with four outputs per unit for each arm). By design, negative results are positioned not as unregistered by-products, but as controlled and reusable boundary evidence of the kind that the field has been urged to publish \citep{karl2024}. The AEG result was withdrawn through a matched-budget self-audit and was preserved with a supersession record. Under the naive analysis, an apparently promising allocation gain was withdrawn after the matched draw-budget self-audit, and the correction was documented through a persistent supersession record \citep{leech2024}. No working JEPA-RL controller is claimed in this work. The contribution is instead the measurement apparatus through which each negative result is rendered falsifiable.

The contributions of this measurement apparatus are summarized under five headings:

\begin{itemize}[leftmargin=*]
\item \textbf{C1 --- Measurement of the content-attribution failure boundary across two deployment channels.} In public-tier ($U_p$) screening, superiority attributable to error content was not confirmed in either the prompt channel or the small-data weight channel. This result is reported as a failure boundary restricted to the 1.5B band. No claim of form superiority, equivalence, or general non-transferability is advanced.
\item \textbf{C2 --- A channel-agnostic placebo hierarchy as a methodological contribution.} The \armSh{} placebo is extended with \armDe{} and a SHA-deranged placebo LoRA. The placebo principle is instantiated in weight space as it is in prompt space, and both channels are adjudicated on a common endpoint, under a matched output-generation budget, and according to the same discovery-versus-confirmation and statistical-verdict-versus-audit-verdict rules. The reflexive measurement instrument is thereby extended to PEFT.
\item \textbf{C3 --- Mechanistic attribution through preregistered kill gates.} \vdStratumNull{} (sample depletion), \vdPTwoNull{} (a low-cost offline kill), \vdGOneKill{}$^{\dagger}$ (the gate-blocked build itself being treated as a result), and the realizability gates (\emph{doa}, \emph{unrealizable}, and \emph{underpowered}) are used to distinguish the component to which each null is attributable. A learnability or realizability gate must be passed by a controller before live spend is opened.
\item \textbf{C4 --- Matched-budget allocator self-audit and provenance-tracked supersession.} The naive allocation positive produced by AEG-BANDIT was withdrawn after the pool-cap confound had been identified, and the correction was recorded through a persistent supersession pointer. Adaptive-allocation superiority was not confirmed in the matched-budget replay. Placement before the first draw is retained as a design hypothesis compatible with this \vdReplayNull{} and is not presented as a theorem or a confirmed regularity.
\item \textbf{C5 --- A provenance-tracked audit record and endpoint-separated instrument diagnostics.} Measurements of taxonomy saturation, the exact Hedge bound, outcome-calibrated severity, the C-matrix, and public--hidden calibration are presented in a form that is cleanly separated from the transfer nulls. Explicit defects are preserved within the same record, allowing the frozen results and the audit lineage to be re-examined.
\end{itemize}

To make these contributions traceable, the remainder of the paper is organized section by section. In Section~2, the population and endpoint definitions of the PoPE methodology, the placebo-mirroring scheme, the two-gate decision rule, and the audit invariants are described. In Section~3, the experimental design is presented, and the findings from the two channels, together with the self-audit and standing-diagnostics results, are reported at the level of observation. In Section~4, the findings are discussed in terms of the identified regularities, the operational boundary of learned criticism, comparisons with prior work, and the conditions governing validity. In the final section, the defensible conclusion, the methodological contribution, and future work concerning task diversity and a larger generator are summarized.

\section{Methods}\label{sec:methods}

A single question is addressed by the proposed evaluation method,
Popperian Placebo-controlled Evaluation (PoPE): whether the limited draws are
directed by learned error content or by the form of the retry and the scale of
the generator. PoPE is used to measure whether evidence that falsifies an
LLM's own output can be used operationally by that same LLM. Each learned
component, including the error lattice, adapter, and
allocator, is treated as a test item for this question. The logical backbone
of the method is formed by three distinctions: the distinction between content
and form, under which each content arm is paired with a placebo that carries
the same structure but has its task-relevant content removed; the distinction
between the statistical verdict and the audit verdict, under which the
statistical result of a contrast and the assessment of pipeline integrity are
reported in separate fields and neither is allowed to substitute for the
other; and the distinction between discovery and confirmation, under which no
effect observed during discovery is assigned claim status unless it is
remeasured using fresh seeds and fresh generations \citep{iscan2026c}.

\begin{figure}[t]
\centering
\includegraphics[width=\linewidth]{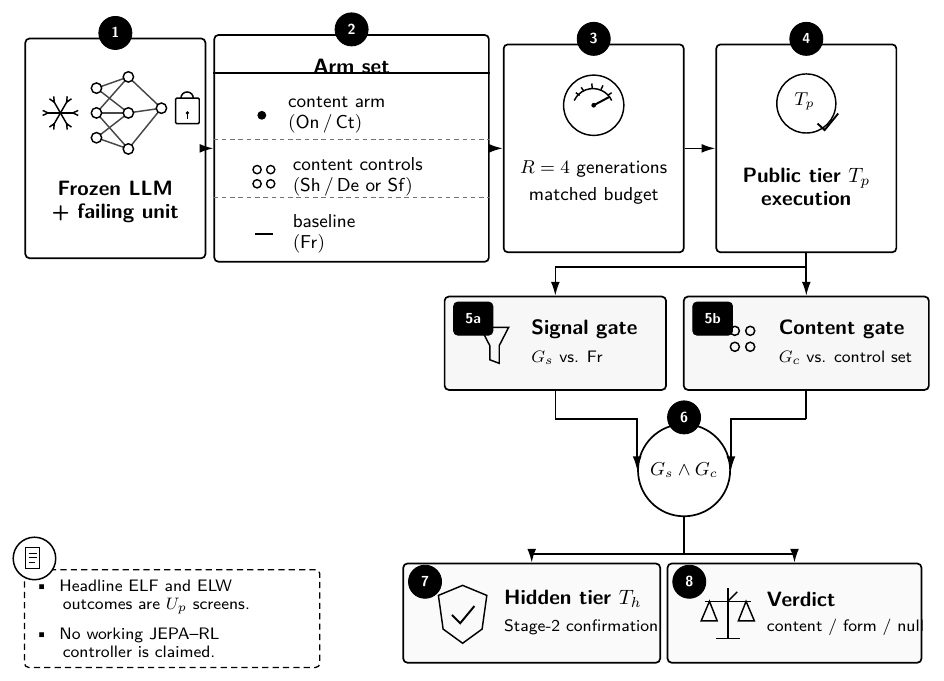}
\caption{The PoPE attribution framework. Each content arm is compared with its
channel-specific placebo twin and \armFr{} under a matched output-generation
budget ($R=4$). Stage-2 hidden confirmation is opened only when
$G_s\wedge G_c$ is satisfied. The headline ELF and ELW results are $U_p$
screens. No working JEPA--RL controller is claimed.}
\label{fig:framework}
\end{figure}

The instrument is summarized in Figure~\ref{fig:framework}. The population and
endpoint hierarchy, arm definitions and mirror chains, matched
output-generation budget, two preregistered statistical families, generation
protocol and audit invariants, discovery-to-confirmation rule, and mechanism
ladder of the preceding instruments are described in this section. Only the
prespecified methodological contract is presented. The design choices were
not justified on the basis of the results, and all numerical fields were
matched to provenance-tracked frozen records.
Procedures~\ref{alg:pope}--\ref{alg:elw} are not repair algorithms. They are
executable evaluation procedures for the measurement instrument.

Figure~\ref{fig:framework} shows that content attribution is made contingent
not on a single difference from a baseline, but on the joint passage of the
signal and content gates.

\subsection{Measurement instrument}\label{sec:instrument}

\subsubsection{Population, endpoints, and unit of analysis}\label{sec:population}

The unit of analysis is the unit. The unit set is denoted by $\mathcal{U}$, and
an individual unit is represented as a task cell immutably keyed by the tuple
$u = \langle b, m, t\rangle$ \citep{iscan2026c}. Here, $b$ denotes the
benchmark, $m$ denotes the model, and $t$ denotes the task identifier within
the benchmark. The notation below is defined once at this point and is held
fixed throughout the section.

Each candidate is evaluated by an execution oracle $\mathcal{O}$ on two test
tiers, the public tier $T_p$ and the prompt-hidden tier $T_h$:
\begin{equation}\label{eq:oracle}
\mathcal{O}(c,\tau)\in\{0,1\},\qquad \tau\in\{T_p,\,T_h\},
\end{equation}
where $\mathcal{O}(c,\tau)=1$ if and only if candidate $c$ passes the executable
test suite of tier $\tau$. Under this notation, each candidate is read as a
conjecture, and each public-test violation is read as an oracle-relative
refutation \citep{popper1963}. Falsification is thereby formalized not as a
logical absolute, but as an executable predicate defined relative to
$\mathcal{O}$. Conditions under comparison are termed arms. The arm set
$\mathcal{A}$ is instantiated as
$\mathcal{A}_{p} = \{\armOn,\armSh,\armDe,\armFr\}$ in the prompt channel and
as $\mathcal{A}_{w} = \{\armCt,\armSf,\armFr\}$ in the weight channel. A
matched output-generation budget of candidates is generated by each arm for
each unit. This budget is denoted by $R$ ($R = 4$,
\S\ref{sec:budget}). For a unit $u$ and arm $a$, the public-tier
mechanism-screen unlock predicate is defined as
\begin{equation}\label{eq:pub4-unlock}
U_{p}(u,a)=\mathbf{1}\big[\exists\, r \le R:\ \mathcal{O}(c_{u,a,r},T_p)=1\big],
\end{equation}
whereas the true-unlock predicate, which constitutes the program endpoint, is
defined as
\begin{equation}\label{eq:true-unlock}
U(u,a)=\mathbf{1}\big[\exists\, r \le R:\ \mathcal{O}(c_{u,a,r},T_p)=1 \wedge \mathcal{O}(c_{u,a,r},T_h)=1\big].
\end{equation}
Here, $c_{u,a,r}$ is the $r$th candidate generated by arm $a$ on unit $u$.
The objects $U_p$ and $U$ are distinct. Only the public tier is examined by
$U_p$, and the hidden tier is never accessed. By contrast, the same candidate
is required by $U$ to pass both tiers. For two paired arms $a,a'$, the
discordant counts $b_{01}$, observed only in favor of $a$, and $b_{10}$,
observed only in favor of $a'$, the net effect
$\mathrm{net}(a,a')=b_{01}-b_{10}$, and the exact one-sided p-value
$p_{+}(a,a')$ are defined by Eq.~\eqref{eq:mcnemar} under the m3 convention.
The two-gate promotion rule is defined using a channel-specific content-control
set,
\[
\mathcal{P}(\armOn)=\{\armSh,\armDe\},
\qquad
\mathcal{P}(\armCt)=\{\armSf\}.
\]
The signal gate compares the content arm with \armFr{}, whereas the content
gate requires the content arm to meet every preregistered threshold in its
channel-specific placebo hierarchy. The signal and content thresholds
$\tau_s(a)$ and $\tau_c(a,p)$ are family-specific:
\begin{equation}\label{eq:gate-signal}
G_s(a)=\mathbf{1}\big[\mathrm{net}(a,\armFr)\ge \tau_s(a)\big],
\end{equation}
\begin{equation}\label{eq:gate-content}
G_c(a)=
\mathbf{1}\!\left[
\bigwedge_{p\in\mathcal{P}(a)}
\mathrm{net}(a,p)\ge\tau_c(a,p)
\right].
\end{equation}
The exact one-sided McNemar tests and within-family Holm results are reported
as inferential diagnostics alongside the frozen promotion thresholds. The
threshold gates are not redefined by these tests. A genuine effect is
attributed to a content arm only when $G_s(a)\wedge G_c(a)$ is satisfied. The
state $G_s(a)\wedge\neg G_c(a)$ is read as form rather than content. This rule
is the operational counterpart of severe-testing logic. Because a claim is
considered severely tested only when it has passed a procedure by which it
would probably have been detected had it been false, credit for content is
made contingent on the gate at which the most plausible alternative
explanation, the form of the retry, is controlled \citep{mayo2006}. The
family-specific threshold values are given in the corresponding family
definitions (\S\ref{sec:elf-family}, \S\ref{sec:elw-family}).

\begin{definition}[dead unit]\label{def:dead}
A unit $u\in\mathcal{U}$ is considered \emph{dead} when none of the candidates
in the cached best-of-$N$ candidate pool ($K_{\mathrm{POOL}} = 8$) passes the
public tier:
\begin{equation}\label{eq:dead-unit}
D(u)=\mathbf{1}\Big[\max_{1\le k\le K_{\mathrm{POOL}}}\mathcal{O}(c^{\mathrm{pool}}_{u,k},T_p)=0\Big].
\end{equation}
This definition denotes a zero-pass state in the finite pool. It does not
indicate that no correct program exists in the model distribution.
\end{definition}

\begin{definition}[public-tier best-of-4 unlock --- mechanism-screen endpoint]\label{def:pub4}
A unit $u$ is considered \emph{public-tier best-of-4 unlocked} for an arm $a$
when $U_p(u,a)=1$, that is, when at least one of the $R$ candidates generated
by $a$ passes the public oracle \eqref{eq:pub4-unlock}. This endpoint is a
mechanism screen. The hidden tier is not accessed.
\end{definition}

The preregistered mechanism screen based on this definition is referred to
throughout the text as the $U_p$ screen.

\begin{definition}[true unlock --- program endpoint]\label{def:trueunlock}
A unit $u$ is considered to have achieved a \emph{true unlock} for an arm $a$ when $U(u,a)=1$,
that is, when at least one regenerated candidate passes both the public tier
and the prompt-hidden tier \eqref{eq:true-unlock}.
\end{definition}

Equation~\eqref{eq:true-unlock} defines the program endpoint, whereas
Eq.~\eqref{eq:pub4-unlock} defines the mechanism screen under which the hidden
tier is not accessed.

The population was not treated as a single homogeneous sample across all
generations. The search-bound, hard-dead, 1.5B-marginal resistant, fresh-unit,
non-evaluation training, and candidate-corpus layers were tracked in separate
manifests. Inclusion and exclusion rules were applied at the manifest level
before generation was initiated. The headline band was selected from the
eligible inventory by deterministic SHA ordering, and parse-only,
entrypoint-only, or timeout-only evidence was not considered sufficient on its
own for semantic-marginal eligibility. Unit, training-pair, and candidate
denominators were thereby retained within their respective layers without
being converted into one another. The shared test population of ELF and ELW
is the \textbf{40-unit 1.5B-marginal resistant band}. The union of 37
P0-marginal units and 9 S2-marginal units was defined to contain 46 units, and
the realized band contained 40 units. The complete flow of the layers and the
supporting populations, including the corresponding population-audit
diagnostic, are provided in the population and corpus accounting table in the
Results section (Table~\ref{tab:consort}).

A single endpoint hierarchy is maintained across these layers. The program
endpoint is \textbf{true unlock} (Definition~\ref{def:trueunlock}). True unlock
was realized in the inherited m3 primary. In LG P0, the unit-by-portfolio
interaction in public-tier dense progress was measured, whereas hidden-partial
calibration was retained as a separate descriptive record. By contrast, the
two headline instruments, the learned error-lattice prompt controller (ELF)
and the error-lattice QLoRA weight adapter (ELW), use a two-stage design.
Stage-1 is the \textbf{$U_p$ screen \eqref{eq:pub4-unlock}}. Stage-2
hidden-tier ($T_h$) true-unlock confirmation was preregistered to be spent only
if the screen was positive, so that under a negative screen it is not spent by
design. No unlock count from ELF or ELW is therefore labeled ``true unlock'' in
this paper. The
correct label is the ``$U_p$ screen'' \eqref{eq:pub4-unlock}. The termination
of hidden spend by a negative screen is the prespecified stopping rule of the
program and is reported as a gate-as-result outcome.

No single pooled omnibus test is run above the endpoint hierarchy. Because
benchmarks, models, strata, and endpoints differ across generations, such a
test would constitute a category error. Instead, a single estimand template is
defined and instantiated separately for each instrument.
\begin{itemize}[leftmargin=*,nosep]
\item \textbf{E-UNLOCK} (the default: net discordant unlock advantage over
paired units, exact one-sided McNemar; LEGACY, live-ollama, FASTR, ELF, and
ELW).
\item \textbf{E-INTERACTION} (LG P0 only: a permutation test for the unit- and
portfolio-centered interaction variance $T_{\mathrm{int}}$ in public-tier
dense fraction. This estimand is neither an unlock-advantage estimand nor a
public--hidden coupling estimand).
\item \textbf{E-ALLOCATION} (AEG only: the adaptive-versus-random net contrast
under an equal draw budget, evaluated through zero-cost offline replay).
\item \textbf{E-OFFLINE-GATE} (DS P2 and FDP G1, for which the firing of the
gate is itself the finding).
\item \textbf{E-MECHANISM-DESCRIPTIVE} (ERA standing diagnostics, the ELF
C-matrix and decoy separation, and LG calibration $r$ at the discovery tier,
always labeled ``descriptive / not confirmatory'').
\end{itemize}

\phantomsection\label{sec:budget}%
All E-UNLOCK comparisons were conducted under a
\textbf{matched output-generation budget ($R = 4$)}. Four output samples were
allocated to each arm for each unit. The global budget caveat is stated once
at this point and is referenced by all E-UNLOCK families. Only the number of
output generations is matched. Input length, tokenization cost, training cost,
wall-clock time, and the number of floating-point operations are not matched,
because code-bearing and error-augmented prompts are structurally longer.
Accordingly, the term ``matched output-generation budget'' is used throughout
the text, and expressions implying a stronger form of equality are avoided.
The arm sets matched by this budget in the two channels are defined in
Table~\ref{tab:arms}.

\subsubsection{Arms and mirror chains}\label{sec:arms}

In both headline channels, the intervention arm is paired with placebos under
which the predeclared scaffold components are held fixed and only the correctness of content is changed. The
placebo twin is not treated as a decorative control. It constitutes the
explicitly named alternative hypothesis. Because the sensitivity of LLMs to
surface form has been documented independently \citep{min2022,webson2022}, any
credit assigned to content without controlling for form would be assigned
against the strongest untested rival explanation. The four paired arms in the
prompt channel (ELF) and the three arms in the weight channel (ELW) are defined
in Table~\ref{tab:arms}. The adapter mechanism in the weight channel is
low-rank adaptation \citep[LoRA;][]{hu2021}, implemented using the quantized
LoRA \citep[QLoRA;][]{dettmers2023} prescription over a 4-bit quantized frozen
base.

\begin{table}[t]
\caption{Arms and mirror chains in the two channels. The
\armOn{}--\armSh{} contrast is the prompt-channel content contrast under which
the predeclared scaffold components are preserved while the task$\to$error
assignment is changed. The \armCt{}--\armSf{} contrast is the weight-channel
content contrast under which the training-surface specification and family marginals are preserved
while the task$\to$error assignment is removed through a SHA-seeded
derangement. \armFr{} is the intervention-free baseline in both channels.}
\label{tab:arms}
\centering\footnotesize\setlength{\tabcolsep}{4pt}
\fitwidth{\begin{tabular}{@{}llp{0.66\linewidth}@{}}
\toprule
Arm & Channel & Conditioning packet / training data \\
\midrule
\armOn{} & prompt & The live learned error-lattice controller. An
error-conditioned repair prompt is generated and updated within the run \\
\addlinespace
\armFr{} & prompt & The pretrained state of the same controller. No online
update is applied \\
\addlinespace
\armSh{} & prompt & A structure-matched content-ablated placebo. The genuine
task prompt, entrypoint, layout, and scaffold are preserved, while only the
public-facts block is replaced by a content-free block with the same layout \\
\addlinespace
\armDe{} & prompt & Donor error facts. Plausible-looking but mismatched error
content taken from another evaluation unit is carried. The task$\to$error
mapping is disrupted while representational form is preserved \\
\midrule
\armCt{} & weight & A QLoRA adapter trained on 1{,}964 genuine
task--failing-code--error-atom pairs \\
\addlinespace
\armSf{} & weight & A SHA-deranged placebo LoRA. Training is performed on a
SHA-seeded complete derangement of the same error blocks. The
task$\to$content mapping is removed while the number of examples and surface
form remain identical to those of \armCt{} \\
\addlinespace
\armFr{} & weight & The base model without an adapter. Byte-for-byte
restoration of the base output is verified in the adapter-effect probe \\
\bottomrule
\end{tabular}}
\end{table}

The error-lattice controller that produces the \armOn{} arm is composed of
five components:
\begin{enumerate}[label=(\roman*),leftmargin=*,nosep]
\item a two-level family/atom error ontology \citep{tambon2024,wang2024};
\item an outcome-calibrated severity value $v$, learned through
corrected-sign projected stochastic gradient descent. The gradient has the
form $(y-p)e$, where $y$ denotes the observed outcome, $p$ denotes the model
prediction, and $e$ denotes the error feature;
\item a Hedge reservoir over a growing error vocabulary, with the same form as
the C1 machinery in \S\ref{sec:era};
\item a LinUCB policy by which error-conditioned routing among six repair
actions is performed within a unit \citep{li2010}; and
\item an interaction matrix $C$ carrying the family co-occurrence structure.
\end{enumerate}
The $C$ matrix is admitted only if it passes a predeclared offline gate whose
threshold is at least a 1\% relative improvement in log loss over the diagonal,
and it was held frozen from its offline fit throughout Stage-1. Online updates to $C$ were
deferred. The sign of the severity update is a SymPy-verified correction to
the reversed sign in the design document and is recorded in the amendment
trail.

Of these five components, component (iv) and the parameterization of the
weight-channel adapter are formally specified below. Repair routing by the
\armOn{} controller is performed through disjoint-linear LinUCB over the six
repair actions $\mathcal{K}=\{1,\dots,6\}$ \citep{li2010}. Initially,
$A_a=I_d$ and $b_a=0$ are maintained for each action. Given the
error-conditioned context $x_{t,a}\in\mathbb{R}^{d}$ in round $t$, the ridge
estimate is calculated as $\hat{\theta}_a=A_a^{-1}b_a$, and the upper
confidence bound is calculated as
\begin{equation}\label{eq:linucb-select}
a_t=\arg\max_{a\in\mathcal{K}}\Big(\hat{\theta}_a^{\top}x_{t,a}
+\beta\sqrt{x_{t,a}^{\top}A_a^{-1}x_{t,a}}\Big).
\end{equation}
Given the selected action $a_t$ and observed reward $\rho_t$, the statistics
are updated as a between-unit batch after completion of the unit:
\begin{equation}\label{eq:linucb-update}
A_{a_t}\leftarrow A_{a_t}+x_{t,a_t}x_{t,a_t}^{\top},\qquad
b_{a_t}\leftarrow b_{a_t}+\rho_t\,x_{t,a_t}.
\end{equation}
Tie-breaks are deterministic. The exploration coefficient $\beta$ and context
dimension $d$ are retained in the frozen policy configuration.

The intervention in the weight channel is quantized low-rank adaptation, under
which a low-rank update is added to the 4-bit quantized frozen base weight
$W$ \citep[QLoRA;][]{dettmers2023}. For rank $r$ and scale $\alpha$, the
effective weight is
\begin{equation}\label{eq:qlora}
W' = W + \frac{\alpha}{r}\,BA,\qquad
B\in\mathbb{R}^{d\times r},\ A\in\mathbb{R}^{r\times k},\ r\ll\min(d,k).
\end{equation}
The weight $W$ is kept frozen and NF4-quantized, and only $A$ and $B$ are
trained. The adapter is learned over the error corpus through token-level
causal cross-entropy:
\begin{equation}\label{eq:qlora-ce}
\mathcal{L}(A,B)=-\frac{1}{\lvert\mathcal{D}\rvert}\sum_{x\in\mathcal{D}}
\sum_{i=1}^{\lvert x\rvert}\log p_{W'}\!\big(x_i \mid x_{<i}\big),
\end{equation}
where $\mathcal{D}$ denotes the error corpus, consisting of genuine content
pairs for \armCt{} and their SHA-deranged placebos for \armSf{}, and $x_i$
denotes the $i$th token. The causal language-modeling loss was calculated only
over target tokens, while prompt and input tokens were masked using the ignore
index ($-100$). The \armCt{} and \armSf{} variants differ only in the
task$\to$content mapping of the examples. The number of examples, surface
form, family marginals, and hyperparameters are identical. The concrete
prescription, including NF4, rank $r$, scale $\alpha$, and the remaining
optimization settings, is provided from a single source in
Table~\ref{tab:experiment-implementation}. Successful training does not
provide evidence of content transfer, since the \armCt{} and \armSf{} variants
differ only in the task$\to$content mapping while the objective, data budget,
and optimization settings are held identical. The realized training-loss
convergence is reported as a diagnostic in the Results
(\S\ref{sec:sensitivity}).

The definition of \armSh{} includes a registered amendment. In the original
pilot, the task itself was also removed from \armSh{}, through the use of a
generic prompt and a fictitious entrypoint. Every subsequent
\armOn{}--\armSh{} comparison would thereby have been turned into a
strawman. The defect was detected during the Cycle-5 liveness pilot and was
corrected through an amendment. The amended rule is the definition given in
Table~\ref{tab:arms}. The original rule, the defect, and the correction are
recorded in sequence in the audit trail in Appendix~\ref{app:a}.

A unit-level split was applied in the weight mirror. Evaluation and reserve
units were excluded from the training corpus through unit identifier, prompt,
and solution-provenance fields. The complete provenance of the training corpus
and the content/shuffled pair counts is provided in
Table~\ref{tab:provenance}.

Contrast integrity is maintained through a minimal-pair discipline. The
\armOn{}--\armSh{} contrast is the prompt-channel content contrast under which
the task prompt, entrypoint, scaffold, and block layout are held fixed and only
error truth is changed. Exact token length and lexical distribution are not
matched byte-for-byte in either channel, and this internal-validity limitation
is considered separately in the Discussion. Under the
\armOn{}--\armDe{} contrast, representational form is preserved and donor
matching is changed. The \armCt{}--\armSf{} contrast is the weight-channel
content contrast under which the training-surface form and family marginals
are preserved and the task$\to$error assignment is changed through
derangement. Because adapter presence and learned content are changed jointly
under the \armCt{}--\armFr{} contrast, that contrast is not described as a
single-component minimal pair. Content attribution is therefore made
contingent on the passage of both gates by \armCt{}
(\S\ref{sec:elw-family}).

Placebo integrity is maintained through executable audits. Byte-level SHA
verification and the forbidden-token audit inherited from m3 were applied to
all prompts rendered in the prompt channel. Function names, assertion
literals, task identifiers, hidden content, and code-smuggling patterns were
screened. Shape-leak and audit-failure invariants were predeclared for
Stage-1. Fixed-point, donor-mismatch, and family-distribution invariants were
predeclared in the weight channel. Under the leakage invariant, solutions from
excluded and evaluation units are not loaded within the policy scope.

\begin{table}[t]
\caption{Learned components of the PoPE instrument. The training signal,
corpus role, and generated output are shown. All components have
descriptive/machinery status. None constitutes a validated deployment reward
(ERA validity: \vdDeferred{}).}
\label{tab:learned-components}
\centering\footnotesize\setlength{\tabcolsep}{4pt}
\fitwidth{\begin{tabular}{@{}p{2.6cm}p{3.4cm}p{2.9cm}p{3.4cm}@{}}
\toprule
Component & Training signal & Corpus & Output \\
\midrule
Error lattice (severity $v$, Hedge, $C$) &
per-candidate execution outcome; corrected-sign projected SGD gradient
$(y-p)e$ &
1{,}920 re-executed cached candidates ($K=93$ categories) &
outcome-calibrated severity, Hedge category weights, admitted interaction
matrix $C$ \\
\addlinespace
QLoRA adapter (\armCt{}, \armSf{}) &
token-level causal cross-entropy \eqref{eq:qlora-ce} &
content pairs (+ SHA-deranged placebo), non-evaluation units;
Table~\ref{tab:provenance} &
rank-$r$ LoRA update $\tfrac{\alpha}{r}BA$ \eqref{eq:qlora} \\
\addlinespace
LinUCB controller &
per-round unlock/dense reward over six repair actions &
within-unit rounds, budget $R$ (online) &
error-conditioned repair-action policy \eqref{eq:linucb-select} \\
\addlinespace
JEPA head (VICReg) &
VICReg composite loss \eqref{eq:vicreg-loss} &
1{,}920-candidate error-set smoke corpus &
$d_z=128$ error-set embedding \\
\addlinespace
ZIB-EI allocator &
ZIB posterior over dense $q$; closed-form EI \eqref{eq:zib-ei} &
S1 cache, 8 i.i.d.\ draws/(unit, arm) &
EI-greedy arm allocation with correlated-evidence correction \\
\bottomrule
\end{tabular}}
\end{table}

\begin{table}[t]
\caption{Training/evaluation data provenance. The training corpus was
separated from the evaluation band through a unit-level split (leakage 0).
The number of evaluation rows was verified against the frozen manifest.}
\label{tab:provenance}
\centering\footnotesize\setlength{\tabcolsep}{4pt}
\fitwidth{\begin{tabular}{@{}p{4.6cm}p{7.4cm}@{}}
\toprule
Item & Count / composition \\
\midrule
Training source &
236 non-evaluation units (46 evaluation/reserve units excluded through
unit-identifier, prompt, and solution provenance) \\
\addlinespace
Training pairs &
1{,}964 content $+$ 1{,}964 SHA-deranged shuffled pairs, $\leq30$/unit,
unit split, leakage 0 \\
\addlinespace
Evaluation band &
40-unit 1.5B-marginal resistant band
($\lvert\mathcal{U}\rvert=40$) \\
\addlinespace
ERA smoke corpus &
1{,}920 re-executed cached candidates from the hard-dead stratum \\
\bottomrule
\end{tabular}}
\end{table}

The content-attribution decision is made using the two-gate promotion rule
\eqref{eq:gate-signal}--\eqref{eq:gate-content}. Content attribution is made
only when both the signal gate ($a>\armFr{}$/blind) and the content gate
($a$ outperforms every control $p\in\mathcal{P}(a)$) are passed. An arm that passes only the
signal gate is read as form. The \armSh{}/\armDe{}/\armFr{} triad is the
channel-agnostic generalization of the single GR-\armSh{} placebo from m3.
The SHA-deranged placebo LoRA extends the same reflexive instrument to
parameter-efficient fine-tuning (PEFT). The statistical machinery for the two
gates is defined in \S\ref{sec:elf-family}, the general execution loop is given
in Procedure~\ref{alg:pope}, and the channel-specific instantiations are given
in Procedures~\ref{alg:elf} and~\ref{alg:elw}.

\begin{algorithm}[t]
\caption{The PoPE two-gate evaluation procedure. The procedure is not a repair
algorithm. It is a channel-agnostic evaluation instrument through which any
learned error-conditioned controller is made falsifiable against its own
form-matched placebo. The ordering signal gate $\to$ content gate $\to$
placebo-hierarchy verdict is applied, and the hidden tier is opened only under
joint promotion.}
\label{alg:pope}
\begin{algorithmic}[1]
\Require Unit set $\mathcal{U}$ (frozen manifest); arm set $\mathcal{A}$
(content arm $a^{\star}$, channel-specific content-control set
$\mathcal{P}(a^{\star})$, baseline \armFr{}); matched output-generation
budget $R$; execution oracle $\mathcal{O}$; public tier $T_p$ and hidden tier
$T_h$; family-level $\alpha$, a preregistered signal threshold
$\tau_s(a^{\star})$, and content thresholds
$\{\tau_c(a^{\star},p):p\in\mathcal{P}(a^{\star})\}$
\Ensure Public unlock $U_p(u,a)$ for each arm and family verdict
$v\in\{\vdPass{},\ \text{form},\ \vdMechNull{}/\vdTrainNull{}\}$
\ForAll{$u\in\mathcal{U}$}
  \ForAll{$a\in\mathcal{A}$}
    \State Render the frozen packet $\pi_{u,a}$ and run the
    placebo-integrity audit (shape leak, derangement, hidden access)
    \For{$r=1$ to $R$}
      \State Derive fresh seed $s_{u,a,r}$ from the frozen namespace;
      $c_{u,a,r}\gets\mathrm{decode}(\pi_{u,a},s_{u,a,r})$
      \State Record the outcome $\mathcal{O}(c_{u,a,r},T_p)$
    \EndFor
    \State Set
    $U_p(u,a)\gets\mathbf{1}[\exists\,r\le R:\ \mathcal{O}(c_{u,a,r},T_p)=1]$
    \eqref{eq:pub4-unlock}
  \EndFor
\EndFor
\State For each content--control pair, calculate the discordant counts
$b_{01},b_{10}$ and $p_{+}$ using Eq.~\eqref{eq:mcnemar}; retain the audit
verdict in a field separate from the statistical verdict
\State Evaluate the signal gate $G_s(a^{\star})$ using
Eq.~\eqref{eq:gate-signal} and the content gate $G_c(a^{\star})$ using
Eq.~\eqref{eq:gate-content}
\State Apply the Holm step-down correction within the family
\eqref{eq:holm}
\If{$G_s(a^{\star})=0$}
  \State Record the verdict as \vdMechNull{} or \vdTrainNull{}
\ElsIf{$G_s(a^{\star})=1$ and $G_c(a^{\star})=0$}
  \State Record the result as consistent with form, without content
  attribution
\Else
  \State Record the screen as \vdPass{}
  \State Access the hidden tier $T_h$ and open Stage-2 true-unlock
  confirmation using $U(u,a^{\star})$ \eqref{eq:true-unlock}
\EndIf
\State Under a negative screen, hidden spend is not incurred by design
(gate-as-result, \S\ref{sec:discovery})
\end{algorithmic}
\end{algorithm}

\begin{algorithm}[t]
\caption{Evaluation procedure for the prompt channel of error-lattice
fine-tuning (ELF). The procedure is not a repair algorithm. It is an evaluation
instrument by which the arm-level public-tier best-of-$R$ unlock outcome is
measured for each dead unit under a matched output-generation budget. It is
the prompt-channel instantiation of Procedure~\ref{alg:pope}.}
\label{alg:elf}
\begin{algorithmic}[1]
\Require The 40-unit 1.5B-marginal resistant band $\mathcal{U}$ (frozen
manifest); arm set
$\mathcal{A}_{p}=\{\armOn{},\armFr{},\armSh{},\armDe{}\}$; budget $R$;
public oracle $(\mathcal{O},T_p)$; frozen decoding parameters
\Ensure $U_p(u,a)$ and the family verdict for each unit and arm
\State Read band $\mathcal{U}$ from the frozen manifest and verify it through
the population audit
\State Fix the four paired arms in deterministic order; apply byte-level SHA
and forbidden-token audits while the prompts are rendered
\State Generate $R$ fresh whole-function candidates for every arm on every
unit; produce a total of
$\lvert\mathcal{A}_{p}\rvert\cdot\lvert\mathcal{U}\rvert\cdot R$ live
generations
\State Test only the public tier; record the per-unit outcome as $U_p(u,a)$
\eqref{eq:pub4-unlock}
\State Evaluate the three preregistered McNemar gates
(\S\ref{sec:elf-family}; Eqs.~\eqref{eq:gate-signal}
and~\eqref{eq:gate-content}) and assign the verdict using the frozen promotion
rule
\State Do not access the hidden tier $T_h$; leave Stage-2 confirmation
contingent on the screen outcome
\end{algorithmic}
\end{algorithm}

\begin{algorithm}[t]
\caption{Training and evaluation procedure for error-lattice weights (ELW).
The procedure is not a repair algorithm. It is an evaluation instrument by
which the same screen endpoint is measured in the weight channel. It is the
weight-channel instantiation of Procedure~\ref{alg:pope}.}
\label{alg:elw}
\begin{algorithmic}[1]
\Require Training source $\mathcal{D}$ consisting of non-evaluation units
(evaluation/reserve units excluded; Table~\ref{tab:provenance}); the 40-unit
ELF band $\mathcal{U}$; arm set
$\mathcal{A}_{w}=\{\armFr{},\armCt{},\armSf{}\}$; frozen QLoRA prescription;
budget $R$; public oracle $(\mathcal{O},T_p)$
\Ensure $U_p(u,a)$ and the family verdict for each unit and arm
\State Derive content pairs from $\mathcal{D}$ under a per-unit cap; generate
shuffled pairs through a SHA-seeded derangement and verify the derangement
audit
\State Train the two QLoRA variants (\armCt{}, \armSf{}) using a symmetric
prescription \eqref{eq:qlora}--\eqref{eq:qlora-ce} for the same number of
optimizer steps
\State Evaluate the three arms in $\mathcal{A}_{w}$ on $\mathcal{U}$ within a
single 4-bit Transformers stack; produce a total of
$\lvert\mathcal{A}_{w}\rvert\cdot\lvert\mathcal{U}\rvert\cdot R$ generations
\State Use the following round distribution: blind rounds (rounds 1--2 and
the early-unlock rows of unit--arm pairs unlocked early) $+$ augmented rounds
(rounds 3--4, using the arm's own preceding failing attempt and genuine error
atoms); the complete accounting is provided in Table~\ref{tab:provenance}
\State Record the per-unit outcome as $U_p(u,a)$
\eqref{eq:pub4-unlock}; evaluate the two preregistered gates
(\S\ref{sec:elw-family}) and the forgetting guard
\State Do not access the hidden tier $T_h$
\end{algorithmic}
\end{algorithm}

\subsection{Statistical analysis, generation protocol, and disclosure}\label{sec:analysis}

\subsubsection{Preregistered statistical families for the prompt and weight channels}\label{sec:elf-family}

The planned ELF family consists of three one-sided contrasts:
\{\armOn{}$-$\armSh{}, \armOn{}$-$\armDe{}, \armOn{}$-$\armFr{}\}. Each
contrast was evaluated using an exact one-sided McNemar test over the pooled
discordant pairs of the paired units \citep{mcnemar1947}. The exact conditional
McNemar procedure was preregistered, and \citet{fagerland2013} is cited to
contextualize its conservatism relative to mid-$p$ and asymptotic alternatives.
The quantity $b_{01}$ denotes the number of
treatment-favored discordant units, for which only \armOn{} unlocks. The
quantity $b_{10}$ denotes the number of control-favored discordant units. The
total number of discordances is denoted by $d=b_{01}+b_{10}$, and the net
effect is denoted by $\mathrm{net}=b_{01}-b_{10}$. The m3 convention is used.
The corresponding fields in the frozen records are labeled in the opposite
direction and were remapped before reporting. The gate-aligned exact p-value
has the following piecewise form:
\begin{equation}\label{eq:mcnemar}
p_{+} = 1 \ \ \text{if } b_{01} \le b_{10}; \qquad
\text{otherwise}\quad
p_{+} = \sum_{k=b_{01}}^{d} \binom{d}{k} \left(\tfrac{1}{2}\right)^{d}.
\end{equation}

Here, $k$ is the number of treatment-favored discordances under the null.
Concordant units do not contribute to the statistic. This choice is consistent
with an error-statistical perspective. Under small discordant counts, use of
the exact binomial reference distribution allows the error probabilities of
the test to be audited without reliance on an asymptotic approximation and
keeps explicit which effect sizes are actually constrained by a low-power
non-rejection \citep{mayo2006}. When the direction is reversed or a tie is
observed, $p_{+}=1$ is assigned so that movement outside the prespecified
superiority direction cannot produce promotion. In the special case
$b_{10}=0$, the expression reduces to $p_{+}=2^{-b_{01}}$. Within-family
multiplicity was controlled using the Holm step-down rule \citep{holm1979},
with $m=3$ and family-level $\alpha=0.05$:
\begin{equation}\label{eq:holm}
p_{(j)} \le \dfrac{\alpha}{m-j+1}.
\end{equation}

Here, $p_{(1)}\le\dots\le p_{(m)}$ are the ordered raw p-values, and $j$ is the
ordered hypothesis index. A hypothesis passes only when its own threshold and
all preceding rejection conditions are jointly satisfied. The gate thresholds
were preregistered. Thresholds of net $\geq+5$ units were defined for
\armOn{}$-$\armSh{} and \armOn{}$-$\armDe{}, and a threshold of net
$\geq+3$ units was defined for \armOn{}$-$\armFr{}. Three states are defined
by the frozen promotion rule. A mechanism-\vdPass{} verdict is assigned if all
primary gates pass. A \vdMechNull{} verdict is assigned if \armOn{} is equal to
or lower than any control. Otherwise, a \emph{mechanism-partial} verdict is
assigned. The Haldane--Anscombe-smoothed matched-pair odds ratio was
predeclared only as a descriptive summary of discordant composition:
\begin{equation}\label{eq:or-ha}
\mathrm{OR}_{\mathrm{HA}} = \dfrac{b_{01}+1/2}{b_{10}+1/2}.
\end{equation}

This quantity is a pooled descriptive summary. It is neither a causal odds
ratio nor a measure of equivalence.

The power calculation and interpretation commitment were preregistered with
the family. At $n=40$, the design was powered for the preregistered
$+5$-unit effects. In practice, the minimum detectable effect coincides with
the gate thresholds. An interpretation was committed in advance for smaller
true effects. A non-significant but direction-preserving result is read as
``failed to confirm'' and is never read as ``refuted.'' An observed tie is not
interpreted as evidence of equivalence or non-inferiority, and no TOST margin
is claimed. The condition under which the mechanism would be falsified was
also named in the preregistration. The falsifier is a non-positive \armSh{}
delta. The audit invariant of the family requires zero shape leakage in the
rendered prompts. The realized shape-leak count is reported in a field
separate from the statistical decision in the Results.

\phantomsection\label{sec:elw-family}%
The weight channel is presented as a \textbf{separate preregistered family}
from the prompt channel. The two families were never merged into a single
joint preregistration or a program-wide omnibus family. Although the narrative
relates the two channels as ``two channels of the same wall,'' two distinct
families are retained statistically. The planned inferential family comprises
the two contrasts \armCt{}$-$\armFr{} and \armCt{}$-$\armSf{} ($m=2$). The
forgetting guard is retained separately as a do-no-harm condition and is not
part of the inferential family. The test machinery is the same as that defined above: the
gate-aligned exact one-sided McNemar test in Eq.~\eqref{eq:mcnemar}, with the
m3 $b_{01}/b_{10}$ convention. Under the promotion rule, net thresholds of at
least $+4$ units are required to be met by \armCt{} in both the
\armCt{}$-$\armFr{} and \armCt{}$-$\armSf{} contrasts. The exact one-sided
McNemar results are reported as inferential diagnostics alongside these
thresholds. Failure of a single gate directs the family verdict to
\vdTrainNull{}, and hidden confirmation is opened only under a joint
\vdPass{}. At $n=40$, the design was powered for the preregistered
$+4$-unit effects. No point-null acceptance rule was defined for an observed
tie or a smaller difference, and the ``failed to confirm'' interpretation
commitment stated above applies without modification.

The guard component of the family is a do-no-harm condition. Over 12 held-in
control units, it was preregistered that the number of blind unlocks under
\armCt{} could be at most two units below that under \armFr{} (allowance 2).
The guard was designed to distinguish a transfer null from a regression caused
by catastrophic forgetting. It is not a statistical superiority claim and
does not rescue failure of a primary gate.

The load-bearing falsification check is the adapter-effect probe. Three
executable probes were predeclared to verify that the null verdict was not an
artifact of the adapter never having been loaded. Under the first probe, the
base output and content-adapter output must differ under the same seed, and
disabling the adapter must restore the base output byte-for-byte. Restoration
was verified by SHA. Under the second probe, sampled augmented rows must
reproduce the stored prompt hash exactly when reconstructed using the
procedure's own builder. The realized reconstruction count is reported in the Results. Under the third
probe, the round/kind distribution must match the protocol exactly: 280 blind
rows, comprising 240 round-1/2 rows and 40 early-unlock rows corresponding to
20 unit--arm pairs, together with 200 augmented rows
(Table~\ref{tab:provenance}). These probes are records of failed refutation,
not proofs.

A symmetric sanity record was retained for training. The two variants were
trained under the same prescription and for the same number of optimizer
steps, and the realized training-loss convergence and NaN-step count are
reported as diagnostics in the Results (\S\ref{sec:sensitivity}). That the SHA-deranged placebo is trained under the same objective, data
budget, and optimization prescription is an informative property of the
design. Dense best-of-4 $q$ contrasts were predeclared as a secondary
analysis using a unit-wise sign-flip permutation and were not included as
hypotheses in the binary unlock family.
The secondary analysis cannot rescue the primary binary gates. The permutation
reference was selected under the same error-statistical discipline. By
constructing the null distribution from the data through unit-wise sign flips,
the false-positive probability is controlled at the design level without an
additional distributional assumption \citep{mayo2006}. All generations read
by both families are outputs of the single frozen generation protocol
described below.

\subsubsection{Generation protocol and audit invariants}\label{sec:protocol}

Sampling parameters were frozen before each run. Ollama service parameters in
the prompt channel were fixed in the pre-run manifest. The temperature used
for blind sampling was the value that had yielded the highest unlock rate
among the scanned temperatures in this regime during the diversity-scheduling
scan (cycle log$^{\dagger}$). In ELF, decoding
parameters were frozen at the action-specification level. Generations in the
weight channel were produced under a fixed decoding configuration within a
single 4-bit Transformers stack. Because all three arms were run using the
same stack and the same decoding configuration, the Ollama/Transformers
confound present in earlier stages was removed in this channel. All decoding
values are provided in Table~\ref{tab:experiment-implementation}.
Arm-specific generation settings were not permitted. Apart from adapter state,
prompt payload, and seed, a common evaluation path was used. Seeds were
derived deterministically:
\begin{equation}\label{eq:seed}
s(u,a,r) = \mathrm{sha256}\big(ns \,\|\, b \,\|\, m \,\|\, t \,\|\, a \,\|\, r\big) \bmod 2^{32}.
\end{equation}

Here, $ns$ is the run-specific frozen namespace string, $\|$ denotes byte
concatenation, and $(b,m,t,a,r)$ is the unit--arm--round key. The scheme was
inherited from Eq.~2.6 of \citet{iscan2026c}, and the complete namespace values
are retained in the frozen configurations. The base seed value for the weight
channel is provided in Table~\ref{tab:experiment-implementation}. The
manifest-resume mechanism prevented the same unit--arm--round key from being
generated twice. The stored seed, prompt hash, output hash, and execution
verdict were recorded jointly.

\phantomsection\label{sec:audit}%
The audit layer is implemented not as an implementation detail, but as the
falsification layer of the method. Two verdicts are reported separately for
each generation. The statistical verdict is the result of the preregistered
tests. The audit verdict is the executable result of the prompt-integrity,
leakage, seed-uniqueness, stack-parity, derangement, shape-leak, hidden-access,
and adapter-effect probes. An audit failure is not removed even when the
statistical thresholds are passed, and no valid promotion can be produced
under such a failure. As-frozen statuses are never overwritten, and the
per-layer dispositions are provided in the audit-status table in the Results
(Table~\ref{tab:audit}). The statistical and audit verdict cells in the
figures were generated from separate frozen fields. These two forms of
evidence were not merged.

The recording discipline also covers missing and incomplete states. Listwise
deletion is not applied. When the expected generation count does not match the
number of manifest rows, the family is not opened to unblinded analysis, and
the resumable runner generates only the missing unit--arm--round keys.
Executor timeout, parse failure, and public-test failure are retained as
outcomes and are not converted into missing values. Hidden-tier values not
generated after a negative public screen are marked ``not spent'' and are not
added to the failure count.

Falsification is applied at two levels under the same discipline. At the
object level, model outputs are tested through execution. Each candidate is
run against the executable tests of the public tier, and a test violation is
recorded as an oracle-relative counterexample. At the meta level, the
researcher's own claims are tested. The verdict for every generation is
subjected to a falsification battery consisting of executable counterexample
probes applied before and after the run: 5 categories and 18 probes for ELF,
adapter-effect, prompt-SHA, and protocol-distribution probes for ELW, and 7
pre-use probes for AEG-BANDIT. A probe result of \emph{holds} constitutes a
record of failed refutation, not proof. A \vdConfirmed{} falsification is
closed through correction and rerunning, and is entered into the amendment
trail in Appendix~\ref{app:a}.

For transparent disclosure, one known open defect remains in the record. The
HEF decoy-synthetic test has failed deterministically since the migration to
Torch 2.12. It is tracked rather than suppressed and is disclosed in
Table~\ref{tab:audit} with the status \emph{as-frozen failing (tracked)}. The
defect is retained as a record that weakens the reproducibility of the HEF
learning signal. The Stage-1 ELF and ELW stacks are not affected.

The final component of the disclosure is the environment and provenance
record. All numbers are retained in frozen result records together with
run-level provenance and execution-environment records. The hardware and software fields reported in the submission are
summarized in Table~\ref{tab:experiment-implementation}. The underlying
environment and provenance records are available as stated under
Reproducibility and Data Availability. Frozen records are not overwritten. Corrections are
linked from the original artifact to a successor artifact through a
supersession pointer. Numbers derived from cycle logs, marked by
$^{\dagger}$, are not contained in the frozen results archive and are reported
by reference to the corresponding program-cycle record.

\subsubsection{Discovery, single promotion, and gate as result}\label{sec:discovery}

An effect observed during discovery is promoted at most once into a
preregistered family, using a frozen hypothesis, fresh seeds, fresh
generations, and a separate output manifest. This rule is the operational form
of the use-novelty principle under which a hypothesis cannot be severely
tested on the same data by which it was generated
\citep{mayo2006,nosek2018}. Discovery outputs, offline fits, smoke statistics,
and pilot amendments are excluded from the confirmation family.
Within-family multiplicity is controlled through Holm
(\S\ref{sec:elf-family}). The two channel families are never pooled. Because
multiple families are present across the program, a descriptive program-wide
sensitivity calculation was also reported so that the single \vdConfirmed{}
positive would not be read as a fishing artifact. This calculation is reported
with the P0 result in \S\ref{sec:standing}.

The complementary rule is the gate-as-result principle. The firing of a
preregistered gate is not treated as an abort. It is treated and reported as a
separate prespecified result category. Examples include the offline
conditioning gate (DS \vdPTwoNull{}), the first-draw learnability gate
(FDP \vdGOneKill{}$^{\dagger}$ for both feature sets), the stratum-sufficiency
rule under which falling below the preregistered minimum stratum size in LG
produces \vdStratumNull{} and terminates downstream spend, the realizability
gates (DCH \vdUnrealizable{}, RIFT \vdDoa{}, and ECK \vdUnderpowered{};
\citealp{iscan2026c}), and the two public-tier screens themselves. Each null
can thereby be attributed mechanistically. A learnability or realizability
gate must first be passed by a controller. A preregistered stop and an
unfinished run are not grouped under the same status. The mechanisms of the
stages to which these gates belong are defined in \S\ref{sec:ladder}, and
their verdicts are reported in the program ledger in the Results
(Table~\ref{tab:ledger}).

\subsection{Preceding instruments and the mechanism ladder}\label{sec:ladder}

The mechanism ladder was constructed not as a chronological experimental log,
but as a sequence of instruments under which the narrowest question left open
by the preceding preregistered verdict is tested at each stage. This structure
is the measurement-level counterpart of the view that a research programme is
evaluated through successive problem shifts rather than isolated experiments
\citep{lakatos1968}. At each step, the content of the preceding refutation is
converted into a boundary condition for the next question. Convergence is
toward the objective of placing a learned error prior somewhere in the
system, and the two decisive stages are the prompt and weight channels.
Between these stages, the error-set architecture (ERA), budgeted best-of-B
allocator (AEG-BANDIT), and joint-embedding predictive architecture (JEPA)
components are included as candidate-controller machinery.
Table~\ref{tab:ladder-short} provides the design view of the ladder and
corresponds to the mechanism/channel rows of Table~\ref{tab:ledger}, the
program ledger in \S\ref{sec:ledger}. No results are reported in this view.

The two headline channels, ELF and ELW, constitute the instrument reached by
the ladder and are defined in
\S\ref{sec:population}--\S\ref{sec:elf-family}. The deployed mechanisms of the
preceding stages are summarized here for methodological completeness. Each
stage has its own preregistered gate, and none is included in either headline
confirmation family. The LEGACY stage, consisting of the
FJR/RIFT/DCH/ECK/EOT/SR prompt-only JEPA-RL controllers, was inherited from
the preceding stage \citep[\S4.1]{iscan2026c}. Because their mechanisms and
verdicts were reported there, they are not reported again here as new results.
The first seven stages of the ladder progressively restrict the measurement
conditions. ERA establishes the error taxonomy, error-weighted reward, and
JEPA abstraction as formal machinery. ELF and ELW write the same error content
into prompts and weights, respectively, so that both channels are subjected
to the same placebo discipline.

\begin{table}[t]
\caption{Design view of the controller mechanism ladder. At each stage, the
narrowest question left open by the preregistered verdict of the preceding
stage is taken up. No results are reported in this view. The verdict and
numerical columns are provided in the program ledger in the Results
(Table~\ref{tab:ledger}). Here, $\mathcal{G}_k$ denotes the $k$th stage of the
program.}
\label{tab:ladder-short}
\centering\small
\fitwidth{\begin{tabular}{@{}llll@{}}
\toprule
Stage & Codename & Channel & Prespecified question of the stage \\
\midrule
\genk{0} & LEGACY (FJR/RIFT/DCH/ECK/EOT/SR) & prompt &
Does learned selection outperform the \armSh{} placebo? \\
\genk{1} & HEF & offline &
Can an error-family signal be learned in principle? \\
\genk{1.5} & live-ollama & prompt &
Does the offline signal transfer to live generation? \\
\genk{2} & FASTR & prompt &
Does the pipeline produce a signal in the realized stratum? \\
\genk{3} & DS & offline &
Can conditioning be screened out offline before live spend? \\
\genk{4} & LG & offline$\to$live &
Is a public-tier dense unit-by-portfolio interaction observed? \\
\genk{4b} & AEG-BANDIT & allocation &
Does within-unit adaptation outperform mixing? \\
\genk{5} & FDP & offline &
Can a prior be learned before the first draw? \\
\genk{6} & ERA & architecture &
Can the error-set machinery be constructed and pass its smoke tests? \\
\genk{7} & \textbf{ELF} & \textbf{prompt} &
Does error content outperform form in the prompt channel? \\
\genk{8} & \textbf{ELW} & \textbf{weight} &
Does error content outperform form in the weight channel? \\
\bottomrule
\end{tabular}}
\end{table}

\paragraph{HEF.}
The hypothesis-error-family detector (HEF) was used to isolate whether an
error-family signal could be learned in principle. The detector was run in a
fully offline synthetic environment against an arm-blind fake oracle. The
existence of learnability could thereby be tested without any live spend. The
known open defect that weakens the reproducibility of this stage is disclosed
in \S\ref{sec:protocol}. The verdict is reported in Table~\ref{tab:ledger}.

\paragraph{live-ollama.}
The same detector was run on a genuine frozen coder to test whether the
offline signal transferred to live generation. The live-generation path of
the Ollama-served frozen models (\S\ref{sec:population}) was established at
this stage in the form inherited by the remainder of the program. All
subsequent prompt-channel stages were run on this path.

\paragraph{FASTR.}
The end-to-end diagnostic pipeline was constructed by the FASTR instrument on
genuine hard-dead units, and whether the pipeline produced a signal in the
realized stratum was evaluated. The corresponding thin stratum is defined in
the layer inventory in \S\ref{sec:population}. The verdict and stratum
accounting of the stage are reported in the Results.

\paragraph{DS.}
Under the diversity-scheduling (DS) instrument, z-conditioning was tied to a
preregistered offline gate. The question was whether conditioning could be
screened out offline before live spend. This stage is the first instantiation
of the E-OFFLINE-GATE estimand template (\S\ref{sec:population}), and the
firing of the gate is reported under the gate-as-result status defined in
\S\ref{sec:discovery}.

\paragraph{LG.}
Under the learnability-gated (LG) instrument, the design was transferred to
fresh units, and the unit-by-portfolio interaction in public-tier dense
progress was tested. Its estimand is E-INTERACTION. Neither unlock advantage
nor public--hidden coupling is estimated. A permutation test is instead
performed on the unit- and portfolio-centered interaction variance
$T_{\mathrm{int}}$ (\S\ref{sec:population}). The stratum-sufficiency rule for
this stage, under which falling below the preregistered minimum stratum size
produces \vdStratumNull{} and terminates downstream spend, is an instance of
the gate-as-result principle in \S\ref{sec:discovery}.

\paragraph{AEG-BANDIT.}
\phantomsection\label{sec:aeg}%
The budgeted best-of-B allocator (AEG-BANDIT) is the measurement instrument by
which the question of within-unit exploitation is isolated. The objective is
to maximize the genuine best-of-B objective over three fixed decoding
portfolios,
\{$P_{\mathrm{conc}}$, $P_{\mathrm{mid}}$, $P_{\mathrm{spread}}$\},
under a budget of $B$ sequential calls per unit:
$\max\, \mathbb{E}[\max_{i \le B} q_i]$. Here, $q\in[0,1]$ is the dense score,
defined as the public-test pass fraction of a candidate. This is a
pure-exploration allocation problem rather than a cumulative-regret bandit.
The fixed-budget best-arm-identification setting is based on
\citet{audibert2010} and \citet{karnin2013}, and the acquisition-function view
is based on \citet{frazier2018}. Independent arms are assumed in classical
bandit theory \citep{lattimore2020}. Because the candidates in this setting
are drawn from a single frozen generator, a separate correction for
correlation is introduced.

Dense outcomes lie in $\{0\}\cup(0,1]$. Under the zero-inflated Beta (ZIB)
posterior, a Bernoulli($\pi$) liveness gate and a
Beta($\alpha,\beta$) dense score over $(0,1]$ are maintained for each
(unit, arm) pair. Per-test passes within a single candidate are strongly
correlated. An observation comprising $m$ public tests is therefore reduced
to $\kappa_{\mathrm{eff}}$ effective Bernoulli observations:
\begin{equation}\label{eq:kappa-eff}
\kappa_{\mathrm{eff}} = \dfrac{m}{1 + (m-1)\,\hat{\rho}}.
\end{equation}

Here, $m$ is the number of public tests per candidate, and $\hat{\rho}$ is the
intraclass correlation. The intraclass correlation was estimated from the S1 pass vectors and
substituted into Eq.~\eqref{eq:kappa-eff}. The realized values are reported
in the Results. Without this correction, the posterior
is approximately $2.5\times$ overconfident.

The arm is selected through the greedy argmax of closed-form
incomplete-beta Expected Improvement (EI). For the best dense score observed
up to that point, $q_\star$, the closed-form ZIB-EI acquisition is
(the derivation is provided in Appendix~\ref{app:e}):
\begin{equation}\label{eq:zib-ei}
\mathrm{EI}_a(q_\star) = \hat{\pi}_a \big[\, \mu_a \big(1 - I_{q_\star}(\alpha_a{+}1, \beta_a)\big) - q_\star \big(1 - I_{q_\star}(\alpha_a, \beta_a)\big) \big].
\end{equation}

Here, $a$ denotes the allocation arm, $\hat{\pi}_a$ denotes the posterior mean
of liveness, $\mu_a=\alpha_a/(\alpha_a+\beta_a)$ denotes the Beta posterior
mean of the arm over $(0,1]$, and $I_x(\cdot,\cdot)$ denotes the regularized
incomplete beta function. Tie-breaks are SHA-256-salted and deterministic.
Before the run, the closed form was verified against numerical integration,
with a maximum absolute error of approximately
$1.0\times10^{-10}$.
The implemented posterior-learning EI-greedy policy does not inherit the
$(1-1/e)$ guarantee from the known-distribution idealization and was evaluated
empirically only as a guarantee-free heuristic.

The decision layer of the instrument is formed by an exact zero-cost offline
replay gate and a matched draw-budget ($B=8$) self-audit. The S1 cache contains
8 i.i.d.\ generations for each (unit, arm) pair. Because i.i.d.\ sampling
implies exchangeability, any adaptive policy for which the number of per-arm
draws does not exceed 8 can be evaluated as an \textbf{exact policy replay}
through draws without replacement from a preshuffled order. It is not
off-policy estimation. The condition of per-arm draws $\leq8$ is the
predeclared validity precondition of the gate. Pool-cap or unequal-draw states
are recorded as budget-audit events that invalidate the primary claim, and
the superseding replay is linked immutably to a new artifact through a
supersession pointer. The decision rule was defined as follows. The paired
contrast in
$\mathbb{E}[\text{best-of-}B\ q]$ per unit, averaged over 40 pool orders,
between AEG and the best-FIXED arm was evaluated using a one-sided sign-flip
permutation test with 4,000 permutations and $\alpha=0.05$. The RANDOM-ADAPT
form control, under which the same machinery is used with random arm
selection, was predeclared to separate gains from mixing from gains from
learning. An identical-arms null-world type-I calibration was also
predeclared. Under the single-promotion rule, no live AEG generation was run
before the offline gate had passed. This instrument constitutes the machinery
of the matched draw-budget ($B=8$) self-audit chain reported in the Results,
\S\ref{sec:selfaudit}.

\paragraph{FDP.}
The First-Draw-Prior (FDP) instrument was used to test, on the public-facing
portion of BigCodeBench (Python), whether a learned prior could be learned
before the first draw \citep{zhuo2024}. The stage was protected by a
preregistered G1 pregate. The learnability of a first-draw prior is tested
offline before live spend is opened, and firing of the gate is reported as a
separate prespecified result category under the gate-as-result status defined
in \S\ref{sec:discovery}.

\paragraph{ERA.}
\phantomsection\label{sec:era}%
The error-set architecture (ERA) is formalized through three coupled
components. The definitions below are stated in the present tense. ERA is
included in this paper only as \textbf{constructed and smoke-tested machinery
together with descriptive standing diagnostics}. It is nowhere presented as
a ``validated JEPA-RL reward.'' The validity test was preregistered but was
left \textbf{\vdDeferred{}}. The deferral rule is given below.

The first component, C1, is the growing error taxonomy. A raw execution record
is converted into a canonical event through a deterministic canonicalizer:
$e=\kappa(\mathrm{raw})=
(\mathrm{phase},\mathrm{exc\_class},\mathrm{tmpl\_hash64})$.
New signatures are admitted to the vocabulary according to their order of
arrival. A countable prior mass is assigned to the category born in position
$m$:
\begin{equation}\label{eq:arrival-prior}
\pi_m = \dfrac{1}{m(m+1)}, \qquad
\sum_{m=1}^{\infty}\pi_m = 1, \qquad
r_K = \dfrac{1}{K+1}.
\end{equation}

Here, $m\in\{1,2,\dots\}$ is the birth order of a category, $K$ is the number
of admitted categories, and $r_K$ is the closed-form total reservoir mass of
categories not yet born. Category weights are maintained through a
\textbf{Hedge} update that remains valid under vocabulary growth:
\begin{equation}\label{eq:hedge-update}
w_t(e) = \dfrac{\pi_e\, e^{\eta_t G_{t-1,e}}}{Z_t}, \qquad
Z_t = \sum_{e' \in \mathcal{E}_t} \pi_{e'}\, e^{\eta_t G_{t-1,e'}} + r_{K_t}.
\end{equation}

Here, $\mathcal{E}_t$ is the error vocabulary at round $t$,
$G_{t-1,e}$ is the cumulative gain of category $e$, $\eta_t$ is the learning
rate, $K_t=|\mathcal{E}_t|$ is the vocabulary size, and $Z_t$ is the
normalizer including the reservoir. For fixed $\eta$, finite-horizon regret is
bounded in closed form relative to every category $e$ that is born:
\begin{equation}\label{eq:hedge-regret}
\mathcal{R}_T(e) \le \dfrac{\ln(1/\pi_e)}{\eta} + \dfrac{\eta T L^2}{8}.
\end{equation}

Here, $T$ is the update horizon, $\mathcal{R}_T(e)$ is the empirical regret
relative to comparator $e$, and $L$ is the bounded-gain cap ($L=1$). The tuned
rate and complete THM-6 derivation are provided in Appendix~\ref{app:f}. It was
predeclared as a smoke gate that the empirical regret for every category born
must remain below this closed-form bound. Taxonomy saturation is monitored
using the incidence-form \textbf{Good--Turing missing-mass} estimator, which
belongs to the nonparametric coverage/lower-bound estimator family for species
richness \citep{good1953,chao1984,gale1995}; convergence guarantees are
discussed by \citet{mcallester2000}:
\begin{equation}\label{eq:good-turing}
\hat{M}_n = \dfrac{N_1}{n}.
\end{equation}

Here, $n$ is the number of candidates, $N_1$ is the number of categories
observed in exactly one candidate, and $\hat{M}_n$ estimates the unseen
incidence mass expected from the next candidate. The candidate was selected
as the sampling unit, and dependent events within the same trace are not
counted as separate i.i.d.\ tokens. A stability gate of $\hat{M}\leq0.05$ was
predeclared. Under the THM-1c growth-slope cross-check, the slope of the growth
curve and $\hat{M}$ are compared as two estimators of the same identity. The
smoke corpus consists of 1,920 cached candidates re-executed from the hard-dead
stratum. The taxonomy counts belong to this 1,920-candidate denominator.

The second component, C2, is the error-weighted RL reward. The reward is
constructed from error-set values through online-adaptive scalarization. The
predeclared smoke invariants require the bounds, cap saturation, and
error/$q$ monotonicity to hold at a rate of 1.0. Detailed derivations are
summarized in Appendix~\ref{app:f}, for which two independently produced
derivation records are retained.

The third component, C3, is the VICReg JEPA head. The JEPA head through which
error sets are abstracted \citep{lecun2022} is trained using
variance--invariance--covariance regularization
\citep[VICReg;][]{bardes2022}. The input is encoded through signed hashing
with $d=4096$, the embedding dimension is $d_z=128$, the target encoder is
updated through an exponential moving average
($\theta^- \leftarrow 0.996\,\theta^- + 0.004\,\theta$), and the masking
probability is 0.5. The composite objective is
\begin{equation}\label{eq:vicreg-loss}
\mathcal{L}_{\mathrm{JEPA}} =
25\,\mathcal{L}_{\mathrm{inv}}
+ 25\,\mathcal{L}_{\mathrm{var}}
+ \mathcal{L}_{\mathrm{cov}}
+ \mathcal{L}_{\mathrm{dec}}
+ \mathcal{L}_{q}.
\end{equation}

Here, $\mathcal{L}_{\mathrm{inv}}$ is the invariance loss between the online
prediction and the target latent, $\mathcal{L}_{\mathrm{var}}$ is the
anti-collapse variance hinge, $\mathcal{L}_{\mathrm{cov}}$ is the
off-diagonal covariance penalty, $\mathcal{L}_{\mathrm{dec}}$ is the
error-profile decoder loss, and $\mathcal{L}_{q}$ is the dense public-score
head loss. The VICReg guarantee concerns collapse avoidance rather than
downstream predictive utility. A non-collapsed embedding is not evidence by
itself. Compression of error sets into a $d_z=128$-dimensional latent does not
establish a tested claim about new outputs. It provides only a representation
on which such a claim could be constructed.

The deferral rule was preregistered. The three-label out-of-fold first-draw
ARM-REGRET validity statistic (EQ-10/EQ-16) was preregistered. The derivation
states, however, that once the input-independence signature is triggered, S-4
and S-5 JEPA liveness cease to constitute evidence in favor of a JEPA claim.
Because this condition was triggered, the validity gate was not spent in this
generation and was redirected to a population carrying cross-unit signal. A
JEPA downstream no-signal flag was observed in ERA's own smoke evaluation,
and the corresponding machinery measurements are reported in the Results and
Appendix~\ref{app:f}. All ERA numbers in this paper are
descriptive machinery measurements, and their status is frozen as
\textbf{\vdDeferred{}(validity)}.

Each of these stages was tied to its own form-matched placebo discipline and
preregistered gate. The narrowest question left open by the ladder was taken
up by the two headline channels. Stage-level verdicts and key numbers are
reported in the program ledger in the Results
(Table~\ref{tab:ledger}, \S\ref{sec:ledger}).


\section{Results}\label{sec:results}

The PoPE experiments were conducted on a class of frozen small code large
language models (LLMs). The headline evaluation design is reported in
Table~\ref{tab:experiment-config}, and the model, decoding, stack, training,
and accounting details are reported in
Table~\ref{tab:experiment-implementation}. The unit of analysis is the unit
($u=\langle b,m,t\rangle\in\mathcal{U}$, task cell
$\langle$benchmark$|$model$|$task$\rangle$). The two headline instruments,
the learned error-lattice prompt controller (ELF) and the error-lattice QLoRA
weight adapter (ELW), were evaluated on the same 40-unit 1.5B-marginal
resistant band under a matched output-generation budget ($R=4$) and the
preregistered $U_p$ screen (Definition~\ref{def:pub4}). The realized
prompt-channel accounting was recorded as 640 generations
($4$ arms $\times$ $40$ units $\times$ $R=4$). The realized ELW accounting was
recorded as 480 generations $=$ 280 blind generations
(240 round-1/2 rows $+$ 40 early-unlock rows $=$ 20 unit--arm pairs)
$+$ 200 augmented generations (rounds 3--4). Generation accounting, decoding
settings, and run-cost values are reported in
Table~\ref{tab:experiment-implementation}. In the Stage-1 audit, the
shape-leak and audit-failure counts were recorded as zero. In the derangement
audit across the three weight-channel arms, no fixed point or donor mismatch
was observed, and the family distributions were recorded as byte-identical.
In the standing-diagnostics record, ema\_cos 0.9929 and erank 24.97 were
recorded for the JEPA embedding, while $\hat{\rho}=0.7771$ and
$\kappa_{\mathrm{eff}}=1.175$ were recorded for the allocator. The expected
and realized row counts matched, and all weight-channel arms were run on the
same evaluation stack. This single-stack arrangement removed the
Ollama/Transformers distinction present in earlier stages from the
weight-channel evaluation.

\begin{figure}[t]
\centering
\includegraphics[width=\linewidth]{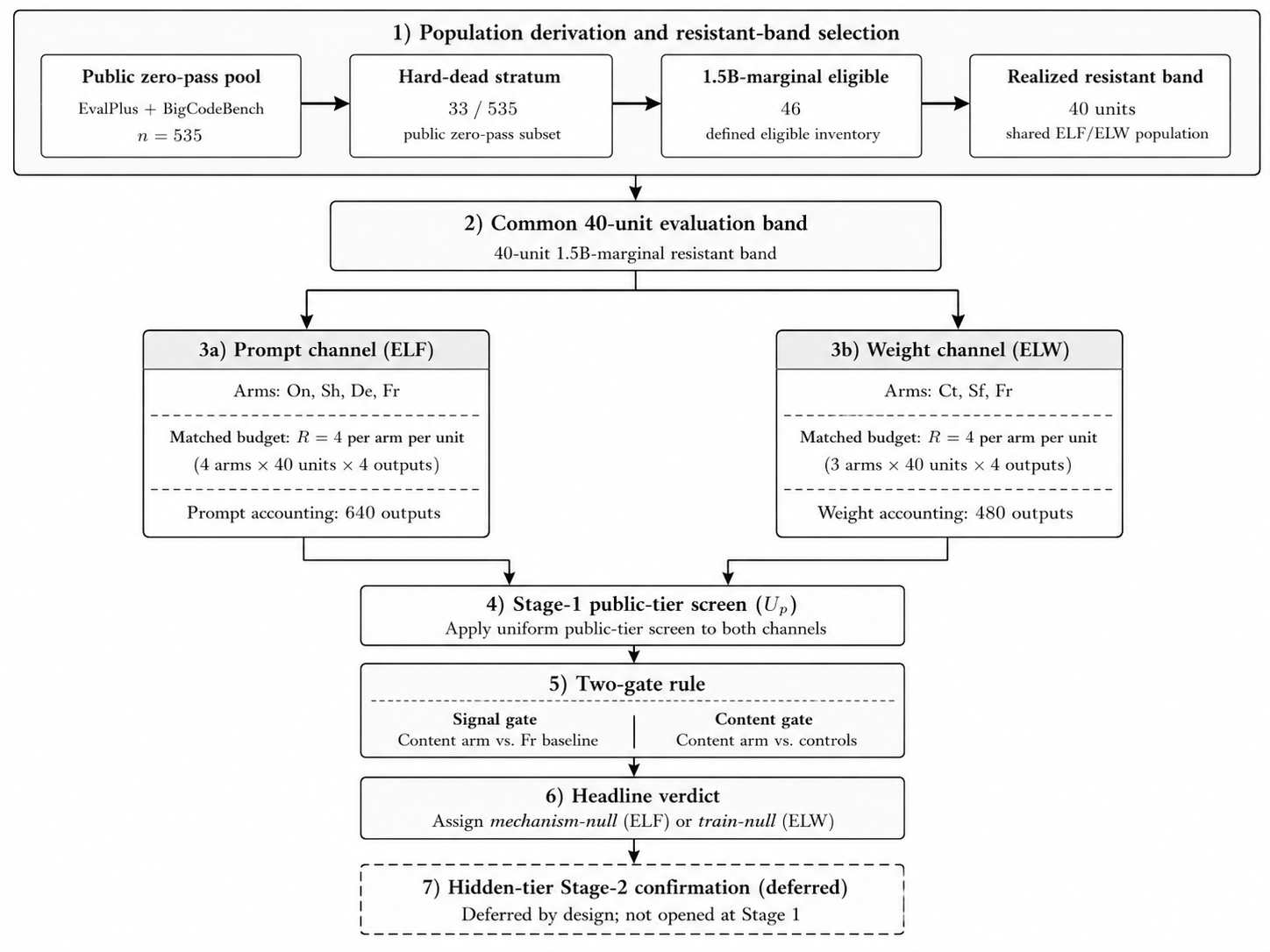}
\caption{Population derivation and channel accounting in the PoPE evaluation
flow are shown in two panels. On the 40-unit band derived from EvalPlus and
BigCodeBench, 640 generations were produced in the prompt channel and 480
generations were produced in the weight channel. The $U_p$ screen is linked
through the signal gate $G_s$ and content gate $G_c$ to the \vdMechNull{} and
\vdTrainNull{} verdicts, while the Stage-2 hidden tier is retained with
\vdDeferred{} status. The population boxes are not mutually exclusive
partitions of a single sample. They represent nested strata derived across
generations.}
\label{fig:pipeline}
\end{figure}

Population narrowing, matched channel budgets, the two-gate decision, and
hidden-tier deferral are shown in a single reporting flow in
Figure~\ref{fig:pipeline}.

\begin{table}[t]
\caption{Headline evaluation design. The population, arm, endpoint, and
matched output-generation-budget contracts of the two channels are shown.
The results belong to the public-tier $U_p$ screen, and hidden confirmation
has \vdDeferred{} status.}
\label{tab:experiment-config}
\centering\footnotesize\setlength{\tabcolsep}{4pt}
\fitwidth{\begin{tabular}{@{}p{3.2cm}
>{\raggedright\arraybackslash}p{5.6cm}
>{\raggedright\arraybackslash}p{5.6cm}@{}}
\toprule
Axis & Prompt channel (ELF, \genk{7}) & Weight channel (ELW, \genk{8}) \\
\midrule
Population &
40-unit 1.5B-marginal resistant band
(37 P0-marginal $+$ 9 S2-marginal $=$ 46 $\to$ realized 40) &
same band \\
Arms &
4 (\armOn{}/\armFr{}/\armSh{}/\armDe{}) &
3 (\armFr{}/\armCt{}/\armSf{}) \\
$R$ & 4 & 4 \\
Generations & 640 ($=4\times40\times4$) & 480 ($=3\times40\times4$) \\
Endpoint &
$U_p$: public-tier best-of-4 unlock \eqref{eq:pub4-unlock} &
$U_p$: public-tier best-of-4 unlock \eqref{eq:pub4-unlock} \\
Hidden confirmation & \vdDeferred{} & \vdDeferred{} \\
\bottomrule
\end{tabular}}
\end{table}

\begin{table}[t]
\caption{Channel-specific implementation and accounting records are shown in
blocks. The design and decoding fields compare the two channels along common
axes. Training, accounting, stack, and cost fields are retained within their
respective record classes. Dagger-marked values are contained in the program
cycle record and the record of İşcan (2026c).}
\label{tab:experiment-implementation}
\centering\footnotesize\setlength{\tabcolsep}{4pt}
\begin{tabular}{@{}
>{\raggedright\arraybackslash}p{0.21\linewidth}
>{\raggedright\arraybackslash}p{0.35\linewidth}
>{\raggedright\arraybackslash}p{0.35\linewidth}@{}}
\toprule
Implementation / accounting axis &
Prompt channel (ELF, \genk{7}) &
Weight channel (ELW, \genk{8}) \\
\midrule
\multicolumn{3}{@{}l}{\textit{Design}} \\
\addlinespace[2pt]
Hardware &
\multicolumn{2}{>{\raggedright\arraybackslash}p{0.72\linewidth}@{}}
{single NVIDIA RTX 3060 Laptop 6 GB GPU (WSL2, driver 591.74)} \\
Frozen model class &
\multicolumn{2}{>{\raggedright\arraybackslash}p{0.72\linewidth}@{}}
{0.5--1.5B code LLM
(Qwen2.5-Coder 0.5B/1.5B, DeepSeek-Coder 1.3B)} \\
Headline base model &
\multicolumn{2}{>{\raggedright\arraybackslash}p{0.72\linewidth}@{}}
{Qwen2.5-Coder-1.5B-Instruct} \\
Benchmark family &
\multicolumn{2}{>{\raggedright\arraybackslash}p{0.72\linewidth}@{}}
{HumanEval+/MBPP+ (EvalPlus), BigCodeBench (Python)} \\
\midrule
\multicolumn{3}{@{}l}{\textit{Decoding}} \\
\addlinespace[2pt]
Serving stack &
Ollama &
single 4-bit Transformers stack \\
Decoding temperature &
$0.8^{\dagger}$ &
$0.7$ \\
Other decoding settings &
generation length 1024 tokens, context 4096 tokens &
top\_p 0.9, generation length 512 tokens, sampling enabled \\
Round protocol &
single-pass best-of-4 (blind) &
rounds 1--2 blind, rounds 3--4 error-augmented \\
Base seed / derivation &
SHA-256 derivation, frozen namespace &
base 20260705, SHA-256 derivation \\
\midrule
\multicolumn{3}{@{}l}{\textit{Training}} \\
\addlinespace[2pt]
QLoRA adapter &
\multicolumn{2}{>{\raggedright\arraybackslash}p{0.72\linewidth}@{}}
{NF4, rank 16, $\alpha$ 32, and dropout 0.05 for the weight channel} \\
QLoRA optimization &
\multicolumn{2}{>{\raggedright\arraybackslash}p{0.72\linewidth}@{}}
{learning rate $2\times10^{-4}$, 2 epochs, sequence length 1024 tokens,
attention+MLP projection layers
(query/key/value/output, gate/up/down), and 491 optimizer steps} \\
Training corpus &
\multicolumn{2}{>{\raggedright\arraybackslash}p{0.72\linewidth}@{}}
{236 non-evaluation units ($\leq$30/unit).
1{,}964 content $+$ 1{,}964 shuffled pairs.
46 units excluded} \\
\midrule
\multicolumn{3}{@{}l}{\textit{Accounting}} \\
\addlinespace[2pt]
Row accounting &
\multicolumn{2}{>{\raggedright\arraybackslash}p{0.72\linewidth}@{}}
{For the weight channel, 480 $=$ 280 blind
(240 round-1/2 $+$ 40 early-unlock, 20 unit--arm pairs)
$+$ 200 augmented} \\
ERA smoke corpus &
\multicolumn{2}{>{\raggedright\arraybackslash}p{0.72\linewidth}@{}}
{1{,}920 candidates
(smoke basis only. Not the screen denominator)} \\
LG fresh-unit stratum &
\multicolumn{2}{>{\raggedright\arraybackslash}p{0.72\linewidth}@{}}
{$n = 80$ / $g = 8$} \\
\midrule
\multicolumn{3}{@{}l}{\textit{Stack and cost}} \\
\addlinespace[2pt]
torch (stack) &
$2.7.1{+}\mathrm{cu}118^{\dagger}$ &
$2.12.1{+}\mathrm{cu}130$ \\
transformers &
--- (Ollama serving) &
4.57.6 \\
Observed run cost &
no separate run-cost record for the prompt channel &
peak VRAM 3.516 GiB and wall time 5{,}561.5 s for the weight channel \\
\bottomrule
\end{tabular}
\end{table}

The endpoint labeling is binding throughout this section. The headline
verdicts for ELF and ELW were obtained using the $U_p$ screen
\eqref{eq:pub4-unlock}. Hidden-tier ($T_h$) true-unlock confirmation was
deferred by design, and no hidden-tier spend was incurred because the public
screen was not passed. No ELF or ELW count should therefore be read as a
``true unlock.'' The program-level true-unlock endpoint $U(u,a)$, defined as
public $\wedge$ prompt-hidden, was realized in the inherited m3 primary
\citep{iscan2026c}. LG P0 measures the unit-by-portfolio interaction in
public-tier dense progress, while hidden-partial calibration is retained as a
separate descriptive record. Each result below is explicitly labeled as
primary, secondary, or descriptive. Statistical and audit verdicts are
reported separately.

\subsection{Primary mechanism screens and arm profile}\label{sec:primary}

The two preregistered families were tested separately on the same 40-unit
resistant band. These were the prompt channel
(ELF, Table~\ref{tab:elf-family}) and the weight channel
(ELW, Table~\ref{tab:elw-family}). The headline verdict of each family is
based on the $U_p$ screen \eqref{eq:pub4-unlock} and has primary status.
Hidden-tier confirmation was deferred by design. Both families carry the
E-UNLOCK estimand. Net discordant unlock advantage over paired units was
evaluated using the exact one-sided McNemar test on pooled discordant pairs.
Holm correction was retained within each family
(ELF $m=3$, ELW $m=2$). The forgetting guard is a separate do-no-harm rule
and is not included in the McNemar hypothesis family. The frozen net threshold
was met for \armOn{}--\armFr{}, while the other primary net thresholds were
not met. None of the exact tests met the prespecified threshold at the raw
level, and the adjusted p-values therefore did not alter the inferential
reading. Discordant counts are reported under the m3 convention
($b_{01}$ = treatment-favored, namely \armOn{}/\armCt{}-favored,
$b_{10}$ = control-favored, and net $= b_{01} - b_{10}$). Because the frozen
record fields were labeled under the reverse convention, only the remapped
frozen values are reported here.

The promotion rule takes the two-gate form in both families
\eqref{eq:gate-signal}--\eqref{eq:gate-content}. A content arm is credited only
when both the signal gate
(content arm $>$ \armFr{}/blind) and the content gate
(content arm $>$ \armSh{}/deranged placebo) are passed. An arm that passes the
signal gate but remains below the content gate is classified as form under the
preregistered reading. This was the pattern recorded in the two family
tables. In the prompt channel, \armOn{} remained directionally positive
relative to \armFr{} but fell below \armSh{}
(Table~\ref{tab:elf-family}). In the weight channel, \armCt{} exceeded neither
\armFr{} nor its own placebo (Table~\ref{tab:elw-family}).

\begin{table}[t]
\caption{The \genk{7} ELF Stage-1 preregistered family was evaluated with 640
generations on the 40-unit 1.5B-marginal resistant band. The counts show the
$U_p$ screen and discordance under the m3 convention, with $b_{01}$
treatment-favored and $b_{10}$ control-favored. The frozen net threshold was
met for \armOn{}--\armFr{}, and no statistically detectable difference was
observed in the exact test. The net thresholds were not met for
\armOn{}--\armSh{} or \armOn{}--\armDe{}, and no contrast met the exact-test
threshold at the raw level. Holm correction did not alter the inferential
reading, and the family verdict was recorded as \vdMechNull{}. Hidden-tier
confirmation is not included. These results do not constitute evidence of
equivalence or non-inferiority. Equivalence was not tested separately.}
\label{tab:elf-family}
\centering\footnotesize\setlength{\tabcolsep}{4pt}
\fitwidth{%
\begin{tabular}{l r r r r c l}
\toprule
Contrast & $b_{01}$ & $b_{10}$ & Net & $p$ (raw) &
Gate (unit threshold) & Verdict \\
\midrule
\armOn{} $-$ \armSh{} &
2 & 4 & $-2$ & $1.0$ & 5 &
\emph{fail} --- content threshold not met \\
\armOn{} $-$ \armDe{} &
4 & 2 & $+2$ & $0.34375$ & 5 &
\emph{fail} --- net threshold not met \\
\armOn{} $-$ \armFr{} &
5 & 2 & $+3$ & $0.2266$ & 3 &
net threshold met, exact test \emph{n.s.} \\
\bottomrule
\end{tabular}
}
\end{table}

A net effect of $-2$ was observed in the \armOn{}--\armSh{} contrast, and the
content gate was not met. More units were unlocked under the content-ablated
\armSh{} placebo than under the learned controller. A net effect of $+2$ was
observed in the \armOn{}--\armDe{} contrast, but no statistically detectable
difference was obtained ($p = 0.34375$), and the $+5$-unit gate threshold was
not met. In the \armOn{}--\armFr{} contrast, the frozen unit-count threshold
was met with a net effect of $+3$, while no statistically detectable
difference was observed in the exact McNemar test ($p = 0.2266$). The frozen promotion rule was triggered under the
condition \armOn{} $\leq$ \armSh{} ($10 \leq 12$), and the family verdict was
recorded as \textbf{\vdMechNull{}}.

\begin{table}[t]
\caption{The \genk{8} ELW preregistered family was evaluated with 480
generations on the same 40-unit resistant band. The counts show the $U_p$
screen and discordant contrasts under the m3 convention. Holm correction was
retained within the family of two planned McNemar contrasts ($m=2$). The
forgetting guard is a separate do-no-harm rule and is not included in the
McNemar hypothesis family. The family verdict was recorded as
\vdTrainNull{}. Hidden-tier confirmation is not included. These results do not
constitute evidence of equivalence or non-inferiority. Equivalence was not
tested separately.}
\label{tab:elw-family}
\centering\footnotesize\setlength{\tabcolsep}{4pt}
\fitwidth{%
\begin{tabular}{l r r r r c l}
\toprule
Contrast & $b_{01}$ & $b_{10}$ & Net & $p$ (raw) &
Gate (unit threshold) & Verdict \\
\midrule
\armCt{} $-$ \armFr{} &
4 & 4 & $0$ & $1.0$ & 4 &
\emph{fail} --- signal threshold not met \\
\armCt{} $-$ \armSf{} &
1 & 3 & $-2$ & $1.0$ & 4 &
\emph{fail} --- content threshold not met \\
forgetting guard &
--- & --- &
content-blind 4 $\geq$ frozen-blind 3 (allow 2) &
--- & --- &
\emph{pass} --- within allowance 2 \\
\bottomrule
\end{tabular}
}
\end{table}

A net effect of exactly 0 was observed in the \armCt{}--\armFr{} contrast
($b_{01} = 4$, $b_{10} = 4$, $p = 1.0$). The predeclared $+4$-unit gate was
not met. In the \armCt{}--\armSf{} contrast, the SHA-deranged placebo LoRA was
numerically ahead (net $-2$, $p = 1.0$). The content gate was not met. The
preregistered forgetting guard was recorded as \vdPass{} and was not violated.
Broader forms of retention loss or downstream harm were not tested by this
guard. The
\armSf{}--\armFr{} contrast yielded a descriptive net effect of $+2$
($p = 0.376953125$). No statistically detectable difference was observed in
this contrast, and no claim that the placebo was beneficial is licensed. The
family verdict was recorded as \textbf{\vdTrainNull{}}.

The ties and negative net effects in the two channels were based on only
6--8 discordant pairs. They are therefore directional or consistency findings
under low power. The ELF design had been powered for preregistered effects of
+5/+5/+3 units, and the ELW design had been powered for effects of +4 units.
The observed effects were $-2$/+2/+3 and 0/$-2$, respectively. These results
do not constitute evidence of equivalence or non-inferiority. Equivalence was
not tested separately. No TOST margin is claimed. Each null should be read as
``no statistically detectable difference was observed.'' It should not be
interpreted as evidence of an absence of effect.

\phantomsection
\label{sec:perarm}

Under the same low-power qualification, the pooled per-arm counts showed the
placebo to be numerically equal to or ahead of content in both channels. In the
prompt channel, \armSh{} finished ahead of \armOn{}
(12 to 10, \vdMechNull{}), followed by \armDe{} (8) and \armFr{} (7). In the
weight channel, \armSf{} was numerically ahead
(10, with \armCt{} $=$ \armFr{} $= 8$), but the difference was not
discriminative. These rankings are descriptive. The primary conclusion is
derived only from the paired contrasts in Table~\ref{tab:elf-family} and
Table~\ref{tab:elw-family}. The total spread across the four arms in the prompt
channel was 5 units, from 7 to 12. The three arms in the weight channel were
contained within a 2-unit band, from 8 to 10.

The same 40-unit band was evaluated in both channels. At the per-cell level,
the unlock memberships of arms with the same pooled total were recorded as
non-identical. The units unlocked by one arm did not coincide exactly with
those unlocked by another arm. Agreement therefore occurred at the net-effect
level rather than at the unit level, and the pooled tie should not be read as
cell-by-cell equality. In the blind rounds of the weight-channel evaluation,
the per-arm unlock counts were recorded as 6/6/8. This count is also
descriptive.
The per-cell and per-arm unlock rates and the membership heat strip in
frozen-manifest order are provided in Figure~\ref{fig:a3} in
Appendix~\ref{app:c}.

The per-arm mean best-of-4 $q$ values were recorded as
\armOn{} .4507 / \armSh{} .4268 / \armFr{} .3859 / \armDe{} .3774
in the prompt channel. They were recorded as
\armSf{} .43850 / \armCt{} .40329 / \armFr{} .40329
in the weight channel. \armCt{} and \armFr{} are identical at
floating-point precision.

\begin{figure}[t]
\centering
\includegraphics[width=\linewidth]{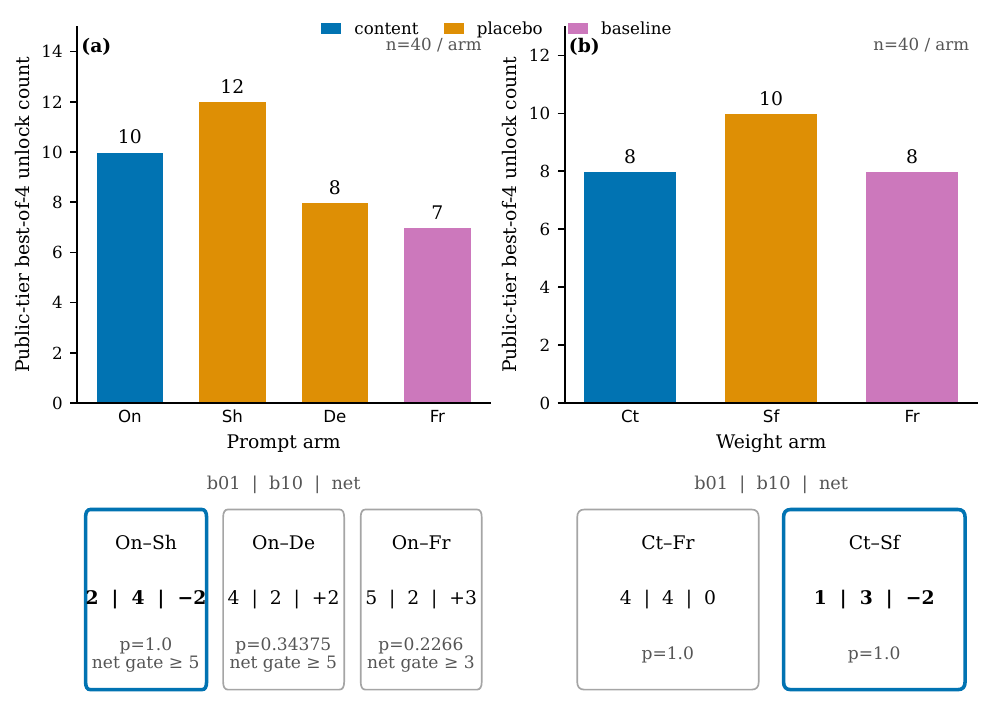}
\caption{Pooled per-arm $U_p$ screening counts \eqref{eq:pub4-unlock} are shown
for the two channels. In the prompt channel, 12 were recorded for \armSh{},
10 for \armOn{}, 8 for \armDe{}, and 7 for \armFr{}. Across the three arms in
the weight channel, 10 were recorded for \armSf{}, 8 for \armCt{}, and 8 for
\armFr{}. The same 40-unit band was evaluated in the two channels using 640
and 480 generations, respectively. In the lower contrast strip, $b_{01}$ is
treatment-favored, $b_{10}$ is control-favored, and
net $=b_{01}-b_{10}$ under the m3 convention. The primary content contrasts
are \armOn{}--\armSh{} and \armCt{}--\armSf{}. The paired evidence is limited
to 6--8 discordant pairs in each contrast. The unlock memberships of arms
with the same pooled total do not coincide exactly. The counts belong to the
$U_p$ screen and are not true unlocks. These results do not constitute
evidence of equivalence or non-inferiority. Equivalence was not tested
separately.}
\label{fig:headline}
\end{figure}

The pooled profile of the two channels is summarized in
Figure~\ref{fig:headline}. The positions of the two family verdicts in the
instrument ledger are reported in Table~\ref{tab:ledger}, while the per-unit
membership records are reported in Figure~\ref{fig:a3} in
Appendix~\ref{app:c}.

In Figure~\ref{fig:headline}, the placebo arms were numerically equal to or
ahead of the learned content arm in both channels. The primary decisions were
obtained from the paired contrasts and the two-gate rule rather than from the
pooled bar heights.

\subsection{Program ledger, standing diagnostics, and self-audit}
\label{sec:ledger}

The controlled headline-verdict set $\mathcal{V}$ is defined in
\eqref{eq:verdictset}. Supporting diagnostic, machinery, inherited, and audit
dispositions are reported in separate ledger fields and are not members of
the headline-verdict set.

\begin{equation}\label{eq:verdictset}
\begin{aligned}
\mathcal{V}
&= \mathcal{V}_{+}
\,\cup\, \mathcal{V}_{\mathrm{diag}}
\,\cup\, \mathcal{V}_{0}
\,\cup\, \mathcal{V}_{\mathrm{kill}},
\\[3pt]
\mathcal{V}_{+}
&= \bigl\{\text{\vdConfirmed{}}\bigr\},
\qquad
\mathcal{V}_{\mathrm{diag}}
= \bigl\{
  \text{\emph{standing}}^{+},
  \text{\vdDeferred{}(validity)}
\bigr\},\\[3pt]
\mathcal{V}_{0}
&= \bigl\{
  \text{\vdMechNull{}},
  \text{\vdTrainNull{}},
  \text{\vdReplayNull{}}
\bigr\},\\[3pt]
\mathcal{V}_{\mathrm{kill}}
&= \bigl\{
  \text{\vdGOneKill{} (gate-blocked)},
  \text{\vdPTwoNull{}},
  \text{\vdStratumNull{}},\\
&\qquad\ 
  \text{\vdDoa{}},
  \text{\vdUnrealizable{}},
  \text{\vdUnderpowered{}}
\bigr\}.
\end{aligned}
\end{equation}

Each headline evaluation stage is assigned a verdict in $\mathcal{V}$.
Supporting stages retain their frozen diagnostic, machinery, inherited, or
audit dispositions in separate fields. Here,
$\mathcal{V}_{+}$ contains the program's single confirmatory positive,
$\mathcal{V}_{\mathrm{diag}}$ contains descriptive standing and deferred
dispositions, $\mathcal{V}_{0}$ contains the placebo-controlled evaluation
nulls, and $\mathcal{V}_{\mathrm{kill}}$ contains the preregistered gates and
realizability kills that stopped live spend. The row for the legacy stage is
inherited and is quoted from m3 \S4.1, Table 15 \citep{iscan2026c}. It is not
reported again as a new result. The offline, live, allocation, architecture,
and public-screen tiers are explicitly separated within the same column.

\begin{table}[t]
\caption{Program ledger (Part~1: inherited, calibration, and diagnostic
stages). Stages \genk{0}--\genk{8} are grouped by stage class across this and
the continued table. For each row the instrument and channel, statistical
verdict, audit or validity status, key record, and scope are shown on separate
axes. The ELF and ELW records (in the continued table) belong to the $U_p$
screening definition \eqref{eq:pub4-unlock} on the 40-unit band. The remaining
rows have supporting or descriptive scope. \genk{0} is an inherited record, and
dagger-marked values are contained in the program cycle record and the record
of İşcan (2026c).}
\label{tab:ledger}
\centering\footnotesize
\setlength{\tabcolsep}{3pt}
\renewcommand{\arraystretch}{1.18}
\begin{tabular}{@{}
>{\raggedright\arraybackslash}p{0.040\linewidth}
>{\raggedright\arraybackslash}p{0.150\linewidth}
>{\raggedright\arraybackslash}p{0.150\linewidth}
>{\raggedright\arraybackslash}p{0.140\linewidth}
>{\raggedright\arraybackslash}p{0.285\linewidth}
>{\raggedright\arraybackslash}p{0.115\linewidth}@{}}
\toprule
$\mathcal{G}_k$ &
Instrument / channel &
Statistical verdict &
Audit or validity &
Key record &
Scope \\
\midrule
\multicolumn{6}{@{}l}{\textit{Inherited live baseline}} \\
0 &
legacy FJR/\allowbreak RIFT/\allowbreak DCH/\allowbreak ECK/\allowbreak
EOT/\allowbreak SR (m3 \S4.1) \newline prompt &
\emph{null} &
inherited &
FJR $-1$ vs \armSh{} $p = 0.773$ ($n = 46$).
SR +6 vs B $p = 0.073$, +1 vs \armSh{} $p = 0.50$
($n = 69$)$^{\dagger}$ &
live. None exceeded \armSh{} \\
\midrule
\multicolumn{6}{@{}l}{\textit{Calibration and diagnostic stages}} \\
1 &
HEF \newline offline &
\emph{standing}$^{+}$ &
\emph{saturation-null} &
content axis $p \approx 7\times10^{-14}$,
leave-one-cluster-out leakage control 0 &
offline synthetic \\
1.5 &
live-ollama \newline prompt &
\emph{live-null} &
--- &
form-not-content reproduced live &
live \\
2 &
FASTR \newline offline &
\emph{diagnostic-null} &
F4-caught &
\armDe{}$>$\armOn{} reversal. 33/535 at 0.5B &
live (thin) \\
3 &
DS \newline offline &
\vdPTwoNull{} &
offline gate &
scale 27.5\% $\to$ 60\%$^{\dagger}$.
Temp 0.8 scan-max$^{\dagger}$ &
supporting \\
\bottomrule
\end{tabular}
\end{table}

\begin{table}[t]
\ContinuedFloat
\caption{Program ledger (Part~2, continued: mechanism, allocation, and primary
channel stages). Columns are as in Part~1 (Table~\ref{tab:ledger}). The
\textbf{ELF} and \textbf{ELW} rows are the two primary $U_p$ screens on the
40-unit band.}
\label{tab:ledger-cont}
\centering\footnotesize
\setlength{\tabcolsep}{3pt}
\renewcommand{\arraystretch}{1.18}
\begin{tabular}{@{}
>{\raggedright\arraybackslash}p{0.040\linewidth}
>{\raggedright\arraybackslash}p{0.150\linewidth}
>{\raggedright\arraybackslash}p{0.150\linewidth}
>{\raggedright\arraybackslash}p{0.140\linewidth}
>{\raggedright\arraybackslash}p{0.285\linewidth}
>{\raggedright\arraybackslash}p{0.115\linewidth}@{}}
\toprule
$\mathcal{G}_k$ &
Instrument / channel &
Statistical verdict &
Audit or validity &
Key record &
Scope \\
\midrule
\multicolumn{6}{@{}l}{\textit{Mechanism and allocation stages}} \\
4 &
LG \newline offline &
\vdConfirmed{} (P0) &
\vdStratumNull{} &
$T_{\mathrm{int}}$ 0.016667,
$p_{\mathrm{perm}}$ 0.0001.
Hidden-partial calibration
($n=115$, $r \approx 0.65$--$0.70$) &
fresh units. Public dense unit-by-portfolio interaction and
separate descriptive record \\
4b &
AEG-BANDIT \newline allocation &
\vdReplayNull{} &
superseded $B=12$ $\to$ governing $B=8$ &
Superseded +6.0 pp and +0.001 control retained together.
Governing +0.02209 ($p=0.2022$),
$-0.00227$ ($p=0.6176$),
\vdReplayNull{}.
$\hat{\rho}$ 0.7771,
$\kappa_{\mathrm{eff}}$ 1.175 &
offline replay \\
5 &
FDP \newline offline &
\vdGOneKill{} &
both feature sets &
gate-blocked by design$^{\dagger}$ &
BigCodeBench \\
6 &
ERA \newline architecture &
built+smoke &
validity \vdDeferred{} &
$K = 93$, $\hat{M} \approx 0.03$ over 1,920 candidates.
Hedge $0.88 < 10.25$ &
descriptive \\
\midrule
\multicolumn{6}{@{}l}{\textit{Primary channel screens}} \\
\textbf{7} &
\textbf{ELF} \newline prompt &
\textbf{\vdMechNull{}} &
primary &
\armSh{} 12 $\geq$ \armOn{} 10 &
public screen \\
\textbf{8} &
\textbf{ELW} \newline weight &
\textbf{\vdTrainNull{}} &
primary &
\armCt{} 8 = \armFr{} 8, $p=1.0$ &
public screen \\
\bottomrule
\end{tabular}
\par\vspace{2pt}
\begin{minipage}{0.96\linewidth}\footnotesize
$\mathcal{G}_{k}$ denotes the $k$th stage of the program. Dagger-marked values
are based on the program cycle record and the record of İşcan (2026c), rather
than on the frozen results archive.
\end{minipage}
\end{table}

\begin{table}[t]
\caption{Supporting gate evidence. The rows are descriptive gate records.
The \genk{1.5} HEF-LIVE harm-guard record is underpowered and carries no harm
claim. The \genk{2} FASTR record has diagnostic status with $n\ll180$.}
\label{tab:gate-evidence}
\centering\footnotesize\setlength{\tabcolsep}{4pt}
\fitwidth{%
\begin{tabular}{p{2.5cm} p{10.4cm} p{3.0cm}}
\toprule
Gate & Frozen readout & Label \\
\midrule
\genk{1.5} HEF-LIVE &
B 15 $>$ \armSh{} 13 $>$ \armDe{} 12 $>$ \armOn{} 10 ($n=16$).
\armOn{}--B discordant $1$ vs $6$, net $-5$.
Harm guard $p=0.0625$
(descriptive, underpowered, no harm claim) &
descriptive \\
\addlinespace
\genk{2} FASTR &
\armDe{} 6 vs \armOn{} 1 unlock.
Discordant $0$ vs $5$.
Cluster-bootstrap RD $-0.333$ [$-0.600$, $-0.133$] &
diagnostic, $n\ll180$ \\
\addlinespace
\genk{3} DS &
$P_{\mathrm{conc}}=.107$,
$P_{\mathrm{mid}}=.157$,
$P_{\mathrm{spread}}=.132$.
Policy$-$marginal-best $-0.008$
(permutation $p=0.269$).
The eligible-33 live budget was not spent &
\vdPTwoNull{} \\
\addlinespace
\genk{5} FDP &
hand $A=0.2938$, $p=0.898$, \vdKill{}.
Embed $A=0.3812$, $p=0.091$,
$\operatorname{Val}(\pi)=0.6333$ vs
$\operatorname{Val}(\mathrm{BF})=0.566$.
Below the $A\geq0.50$ floor &
\vdKill{} \\
\bottomrule
\end{tabular}
}
\end{table}

The profile of the legacy ledger row, in its inherited form, was as follows
\citep{iscan2026c}$^{\dagger}$\footnote{This number is contained in the
program cycle record and the record of İşcan (2026c), rather than in the
frozen results archive.}. The FJR controller yielded a net effect of +3
relative to the blind-resampling baseline ($p = 0.274$), but this value fell
to $-1$ relative to the \armSh{} placebo
($p = 0.773$, $n = 46$). The self-repair pilot yielded a net effect of +6
relative to the baseline ($p = 0.073$) and remained at only +1 relative to
\armSh{} ($p = 0.50$, $n = 69$, leak 0/69). EOT yielded +3
($p = 0.27$) and +1 relative to \armSh{}
($p = 0.50$, $n = 69$). The code-free ECK variant was recorded as
\vdUnderpowered{} with +3 ($p = 0.29$, $n = 109$). RIFT was not triggered in
any of the 24 trials (0/24, \vdDoa{}). DCH remained \vdUnrealizable{} in the
absence of an offline corpus. No statistically detectable difference was
observed in any legacy contrast.

Within the instrument sequence in the ledger, the content axis of the HEF
detector was decisive only in the offline synthetic regime
($p \approx 7\times10^{-14}$, leave-one-cluster-out leakage control 0).
In the same offline record, the learned-signal permutation gate exceeded its
predeclared threshold
(AUC $= 0.628 \geq 0.60$, permutation $p = 0.012$). When the same detector
was transferred to live evaluation, the form-not-content null was observed
again. In FASTR, \armDe{} finished ahead of \armOn{}, and the diagnostic record
remained in the null direction. In the diversity-scheduling instrument,
z-conditioning was stopped by the preregistered offline gate before live spend
(\vdPTwoNull{}). No out-of-unit transfer was detected for the learned
portfolio phenotype from the same stage
(ICC 0.372, out-of-unit Spearman 0.152 $<$ permutation p95 0.255, AUC 0.546).
This record is descriptive. While LG yielded the single \vdConfirmed{}
positive, stratum depletion was recorded in the same evaluation
(\vdStratumNull{}, $46 < 60$). FDP was blocked by the preregistered first-draw
gate for both feature sets (\vdGOneKill{}), while ERA was retained with
built-and-smoke status and \vdDeferred{} validity. None of these rows is
confirmatory in isolation. Each is reported as the output of its own
preregistered rule.

The generator-scale observation is among the supporting values in the ledger
and is reported descriptively. In the same 40-unit pool, the any-portfolio
unlock rate was recorded as 27.5\% for the 0.5B generator and 60\% for the
1.5B generator. The same direction was retained in the per-portfolio rates.
In the 0.5B pool, $P_{\mathrm{conc}} = .107$,
$P_{\mathrm{mid}} = .157$, and $P_{\mathrm{spread}} = .132$ were measured.
In the 1.5B pool, $.350$, $.475$, and $.450$ were measured, respectively,
and the portfolio ordering remained the same at both scales. These rates are
descriptive. The temperature scan showed that 0.8 yielded the highest value among the scanned temperatures. This
temperature observation belongs to the Ollama stack of the
diversity-scheduling stage. The separate Transformers stack used for the
weight channel (ELW) employed its own preregistered decoding configuration
(Table~\ref{tab:experiment-implementation}). These two values are derived from
cycle logs and define only a direction across two points. No scaling-curve
extrapolation is performed in this section. The nulls for the conditioning and
diversity arms were observed again at both scales.

\begin{figure}[t]
\centering
\includegraphics[width=\linewidth]{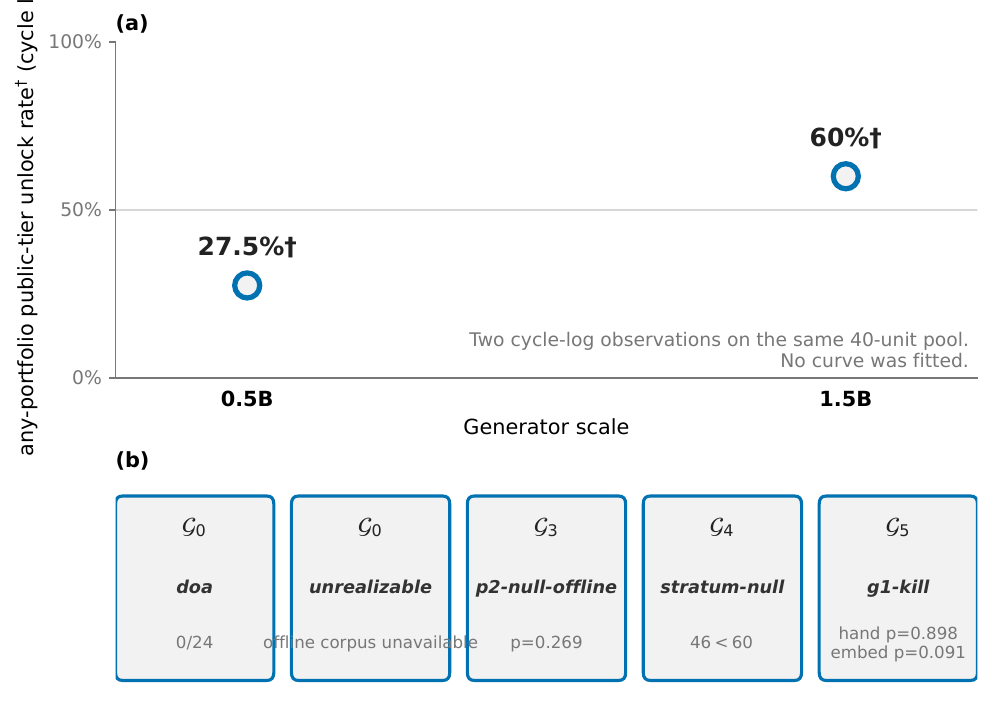}
\caption{Generator-scale observation and kill-gate strip. In the same
40-unit pool, the any-portfolio unlock rate from the cycle log was recorded as
$27.5\%^{\dagger}$ for the 0.5B generator and $60\%^{\dagger}$ for the 1.5B
generator. The two points are cycle-log observations obtained on the same 40-unit pool at different generator scales. No curve
was fitted, and no scaling-curve extrapolation is performed. In the lower
strip, the preregistered gates that stopped downstream spend are shown. These
gates are ordered as \vdDoa{} (0/24), \vdUnrealizable{}, \vdPTwoNull{},
\vdStratumNull{} ($46 < 60$), and \vdGOneKill{} for both feature sets.
The $^{\dagger}$ scale values are contained in the program cycle record rather
than in the frozen results archive.}
\label{fig:scalegates}
\end{figure}

The two generator-scale points are reported in the upper panel of
Figure~\ref{fig:scalegates}, while the preregistered gate by which each null
was produced is reported in the lower strip. A catalogue of gate outcomes,
rather than a mechanism attribution, is provided by the figure.

\phantomsection
\label{sec:standing}

Alongside the null-dominated profile of the ledger, the program's mechanism-
and measurement-level diagnostic records are reported in a separate table so
that the alternative explanation of instrument failure can be evaluated
(Table~\ref{tab:standing}). These rows carry the E-MECHANISM-DESCRIPTIVE label
and all have descriptive status. No row is presented as evidence of a
deployment gain or a ``validated joint-embedding predictive architecture
(JEPA)-RL reward.'' The single \vdConfirmed{} positive carrying the
E-INTERACTION estimand, LG P0, is reported in a separate paragraph after this
profile.

\begin{table}[t]
\caption{Standing diagnostics are shown in three epistemic blocks.
Instrument calibration, machinery liveness, and descriptive-structure
measurements are separated through their respective reference and scope
fields. The rows do not constitute evidence of deployment superiority or a
validated reward. The ERA rows have built+smoke status, and ERA validity is
\vdDeferred{}.}
\label{tab:standing}
\centering\footnotesize\setlength{\tabcolsep}{4pt}
\begin{tabular}{@{}
>{\raggedright\arraybackslash}p{0.20\linewidth}
>{\raggedright\arraybackslash}p{0.27\linewidth}
>{\raggedright\arraybackslash}p{0.27\linewidth}
>{\raggedright\arraybackslash}p{0.12\linewidth}@{}}
\toprule
Diagnostic & Key statistic & Reference or scope & Status \\
\midrule
\multicolumn{4}{@{}l}{\emph{A. Instrument calibration}} \\
Exact Hedge regret under vocabulary growth$^{a}$ &
worst 0.8836 $<$ bound 10.254 &
margin 9.37 &
descriptive bound check \\
\addlinespace
Outcome-calibrated severity$^{b}$ &
v(failure-atom) 0.5$\to$1.0,
v(pass-atom) 0.5$\to$0.0 &
corrected-sign projected SGD &
descriptive \\
\addlinespace
Decoy-separable transitions$^{c}$ &
real-state dist 0.3480 $<$ decoy-state dist 0.5176 &
gate \vdPass{} &
descriptive gate \vdPass{} \\
\addlinespace
Exact correlated-evidence allocator$^{d}$ &
$\hat{\rho}=0.7771$,
$\kappa_{\mathrm{eff}}=1.175$ &
--- &
descriptive \\
\midrule
\multicolumn{4}{@{}l}{\emph{B. Machinery liveness}} \\
JEPA embedding non-collapse$^{e}$ &
ema\_cos 0.9929, erank 24.97 &
$d_z=128$ &
descriptive liveness \\
\addlinespace
ELF Stage-1 controller liveness$^{f}$ &
LinUCB $6/6$ actions &
160 transitions, 40 unit updates &
descriptive liveness \\
&
\textsf{blind} 47,
\textsf{edge-first} 43,
\textsf{counterex-min} 25,
\textsf{type-contract} 22,
\textsf{high-diversity} 12,
\textsf{complexity} 11 &
22 atoms, 0 cap violations.
Hedge 1.1224 $<$ 45.52 &
\\
\addlinespace
Reward-permutation liveness$^{g}$ &
$P_{\mathrm{conc}}-P_{\mathrm{spread}} = +0.1293$,
permutation $p\approx1.0\times10^{-4}$ &
93-atom vocabulary.
1,280 scored rows &
mechanism liveness, not an endpoint result \\
\addlinespace
ERA reward-mechanics conformance$^{e}$ &
bounds hit rate 1.0,
cap saturation 1.0,
monotonicity 1.0/1.0 &
Kendall $\tau$ with reward--error 0.468 and reward--$q$ 0.327 &
descriptive mechanism check \\
\addlinespace
Ontology coverage$^{h}$ &
coverage $0.9989\geq0.98$, \vdPass{} &
494 generic F-value $\to$
388 F-numeric $+$ 106 F-container &
descriptive coverage gate \\
\midrule
\multicolumn{4}{@{}l}{\emph{C. Descriptive structure}} \\
Error-taxonomy saturation$^{i}$ &
$K = 93$, $\hat{M} \approx 0.03$ &
over 1,920 candidates &
descriptive. ERA validity \vdDeferred{} \\
\addlinespace
C-matrix co-occurrence structure$^{j}$ &
log loss 0.254$\to$0.185 &
27.3\% relative improvement &
descriptive offline gate \\
\addlinespace
Public--hidden dense calibration$^{k}$ &
Pearson $r \approx 0.65$ / Spearman $\approx 0.70$ &
$n=115$, monotone buckets &
descriptive, separate from P0 \\
\bottomrule
\end{tabular}
\par\vspace{2pt}
\begin{minipage}{0.96\linewidth}\footnotesize
The frozen fields correspond to
$^{a}$ERA regret record 2,
$^{b}$ELF/ERA,
$^{c}$ELF decoy-separation record 4,
$^{d}$AEG validation record,
$^{e}$ERA smoke record,
$^{f}$ELF Stage-1 record,
$^{g}$ELF smoke record for reward,
$^{h}$ELF smoke record for ontology,
$^{i}$ERA taxonomy record 1,
$^{j}$ELF C-matrix smoke record, and
$^{k}$LG calibration record.
\end{minipage}
\end{table}

The error taxonomy was observed to have reached saturation on the online
harvest of the JEPA-based ERA. Across 1,920 candidates, $K = 93$ categories
were recorded, and the Good--Turing missing mass was measured as
$\hat{M} \approx 0.03$. The Hedge regret bound held for every category born
under vocabulary growth, with worst 0.8836 $<$ bound 10.254 and a margin of
9.37. The planted-signal probe was passed under the corrected-sign projected
SGD update of learned severity. Failure-atom severity moved from 0.5 to 1.0,
while pass-atom severity moved from 0.5 to 0.0. The predeclared offline gate
for the co-occurrence structure of the C-matrix, whose threshold was 1\%, was
passed by a clear margin. Diagonal log loss decreased from 0.254 to an
interaction log loss of 0.185, a 27.3\% relative improvement was recorded,
and interaction matrix C was admitted for Stage-1.
The decoy-separation gate was recorded as \vdPass{}, constituting the first
positive JEPA-mechanism signal in the program
(real-state distance 0.3480 $<$ decoy-state distance 0.5176).

By contrast, in ERA's own smoke evaluation, the JEPA embedding remained
non-collapsed but without downstream signal. JEPA loss decreased from 48.2 to
24.0, and the embedding did not collapse
(ema\_cos 0.9929, erank 24.97).
However, out-of-fold ridge $R^{2} \approx 0.0096$ and decoder
$R^{2} \approx -133$ were recorded. ERA values are reported only as
descriptive machinery measurements, and ERA validity is retained with
\vdDeferred{} status.
In the weight-channel training record, \armCt{} loss was measured as
0.5406$\to$0.00176 (99.674\%), and \armSf{} loss was measured as
0.4942$\to$0.00135 (99.727\%). Both adapters completed 491 optimizer steps
with 0 NaNs and were 73.9 MB each.

The calibration profile was recorded descriptively as follows. A Pearson
correlation of $r \approx 0.65$ and a Spearman correlation of $\approx 0.70$
were observed between the public best-of-4 $q$ statistic and the hidden pass
rate ($n = 115$), and the bucket means followed a monotone staircase
(0.114 / 0.343 / 0.462 / 0.792). The context for the single
\vdConfirmed{} positive reported below is provided by this profile.

An independent conversion record was in the same direction. During the
diversity-portfolio stage of the 0.5B generator, conversion from public unlock
to hidden true unlock was measured as 8 of 13 (0.6154) in the
$P_{\mathrm{conc}}$ arm, 15 of 19 (0.7895) in the $P_{\mathrm{mid}}$ arm, and
12 of 16 (0.75) in the $P_{\mathrm{spread}}$ arm. In total, true unlock was
obtained for 35 of 48 units. These rates are descriptive and carry no claim
of portfolio superiority.

The program's single \vdConfirmed{} positive was LG P0, which carries the
E-INTERACTION estimand. It has primary status and is not an unlock-advantage
result. At the LG P0 gate, the primary statistic was obtained as
$T_{\mathrm{int}} = 0.016667$. The permutation test yielded
$p = 0.0001$ ($n_{\mathrm{perm}}$ 9999). The same verdict was retained in the
dedup-robust recomputation with $t_{\mathrm{int}} = 0.018255$. The per-arm
profile of the P0 record was descriptively monotone. The mean
dense-interaction fraction was measured as 0.4556 in the
$P_{\mathrm{conc}}$ arm, 0.3741 in the $P_{\mathrm{mid}}$ arm, and 0.3268 in
the $P_{\mathrm{spread}}$ arm. The design was well powered
(power $\geq 0.97$, $n = 80$ fresh units / $g = 8$), and this was the only
result in the program with this level of power. In the program-wide
sensitivity check, the family denominator was taken to be approximately 10,
and a threshold of $0.05/10 = 0.005$ was reported. Unless the exact family
universe is listed in the frozen ledger, this calculation is to be read as a
descriptive sensitivity check rather than as a confirmatory multiplicity
correction. For P0, $p = 0.0001 < 0.005$ was recorded. The LG P0 primary
statistic was recorded as $T_{\mathrm{int}}=0.016667$, with permutation
$p=0.0001$ and verdict \vdConfirmed{}. The statement that a unit-by-portfolio
interaction was observed in public-tier dense progress on fresh units is
licensed by this \vdPass{}. In the separate hidden-partial calibration record,
a descriptive association was observed between public and hidden partial
progress. No claim of controller superiority or P0-confirmed public--hidden
coupling is derived from this result.

\phantomsection
\label{sec:selfaudit}

Alongside this profile of positives, the built-in self-audit chain applied by
the program to its own positive findings is reported. The family of the
budgeted best-of-$B$ allocator (AEG-BANDIT) carries the E-ALLOCATION estimand.
Under an equal draw budget, adaptive--random net unlock was measured through
the exact zero-cost offline replay gate. The superseded replay record showed
+6.0 pp in favor of adaptive allocation and was retained together with the
+0.001 control from the same record
($\Delta=0.06025$, $p=0.01375$, budget 12). This initial verdict was later
explicitly invalidated in the frozen audit record by the self-audit conducted
under a matched draw budget of $B=8$. The fixed arms were limited to 8 draws
by the pool cap, while the adaptive arm received 12 draws. The comparison was
therefore made without matched draw budgets and was marked decision-invalid in
the persistent supersession record. The initial value was retained only as a
superseded audit step together with its retraction. The steps of the chain are
provided in Table~\ref{tab:aeg}.

\begin{table}[t]
\caption{AEG self-audit lineage. The $B=12$ rows are retained with
decision-invalid and superseded status. The +6.0 pp primary delta is shown
together with the +0.001 control from the same record and the supersession
record. The governing family verdict is \vdReplayNull{}, obtained from the
+0.02209 ($p=0.2022$) and $-0.00227$ ($p=0.6176$) contrasts in the matched
$B=8$ record.}
\label{tab:aeg}
\centering\footnotesize\setlength{\tabcolsep}{4pt}
\fitwidth{%
\begin{tabular}{l r p{4.4cm} r r p{3.8cm}}
\toprule
Record status & Budget & Comparison & $\Delta$ & $p$ & Decision \\
\midrule
superseded &
12 &
adaptive vs best fixed &
+0.06025 (+6.0 pp) &
0.01375 &
former \vdPass{}, superseded with +0.001 control, decision-invalid \\
\addlinespace
superseded &
12 &
adaptive vs random &
+0.00097 (+0.001) &
0.44664 &
superseded control \\
\addlinespace
governing &
8 &
adaptive vs best fixed &
+0.02209 &
0.2022 &
\vdNs{} \\
\addlinespace
governing &
8 &
adaptive vs random &
$-0.00227$ &
0.6176 &
\vdReplayNull{} \\
\addlinespace
governing sanity &
8 &
null world &
$-0.02667$ &
0.91354 &
sanity \vdPass{} \\
\bottomrule
\end{tabular}
}
\end{table}

\begin{figure}[t]
\centering
\includegraphics[width=\linewidth]{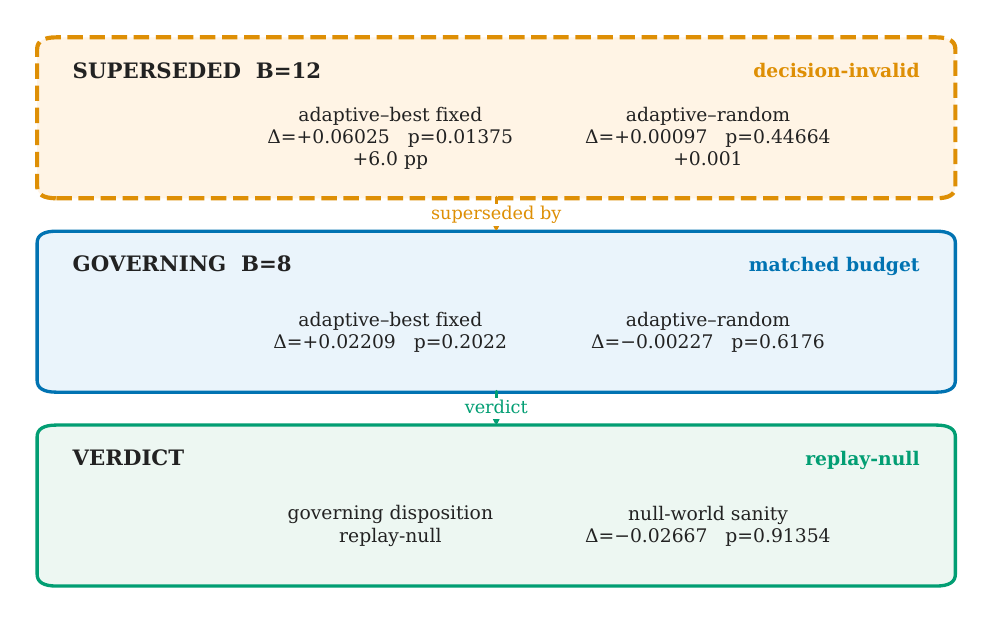}
\caption{AEG self-audit lineage and governing-record view. The +6.0 pp
primary delta in the superseded $B=12$ record
($\Delta=0.06025$, $p=0.01375$) and the +0.00097 (+0.001) adaptive--random
control from the same record ($p=0.44664$) are retained together. In the
governing matched $B=8$ record, adaptive--best fixed was recorded as +0.02209,
adaptive--random was recorded as $-0.00227$, and the family verdict was
\vdReplayNull{}. The null-world sanity record is shown with $-0.02667$ and
$p=0.91354$.}
\label{fig:aeg-lineage}
\end{figure}

In Figure~\ref{fig:aeg-lineage}, the +6.0 pp value and +0.001 control from the
superseded $B=12$ record are shown within the same supersession step, while the
+0.02209 and $-0.00227$ contrasts from the governing $B=8$ record are shown in
a separate governing-record step (source: Table~\ref{tab:aeg}).

In the governing matched draw-budget ($B = 8$) recomputation, no statistically
detectable difference was observed for the adaptive--best fixed contrast
($\Delta = +0.02209$, $p = 0.2022$). In the adaptive--random control of the
superseded ($B = 12$) record, +0.00097 and $p = 0.44664$ were recorded. In the
governing matched ($B = 8$) record, the adaptive--random contrast was recorded
as $-0.00227$ with $p = 0.6176$. The family verdict was recorded as
\textbf{\vdReplayNull{}}. The null-world sanity check yielded the expected
directional result ($\Delta = -0.02667$, $p = 0.91354$). The
correlated-evidence internals of the allocator are reported descriptively,
with $\hat{\rho} = 0.7771$ and $\kappa_{\mathrm{eff}} = 1.175$.
The value $\kappa_{\mathrm{eff}}=1.175$ was calculated using median $m=3.0$
and $\hat{\rho}=0.7771$.

The saturation pattern was separated descriptively by the budget scan from
the same frozen replay record. Identification probability increased with the
draw budget, from 0.5646 at $B=4$ to 0.6088 at $B=8$, with chance level 0.3333.
By contrast, the expected maximum of the adaptive arm did not separate from
that of random allocation, with 0.6169 versus 0.5967 at $B=4$ and 0.7040
versus 0.6891 at $B=8$. The model-based bootstrap extrapolation retained the
same pattern at larger budgets. At $B=48$, identification was 0.8858, while
the expected maximum remained 0.7754 versus 0.7773. These extrapolated values
are not exact replay values.

The +6.0 pp primary delta from the superseded ($B=12$) record and the +0.001
control from the same record are reported together with the supersession
marker. The valid $B=8$ analysis carries the +0.02209 and $-0.00227$ contrasts
and the \vdReplayNull{} verdict. A layer-by-layer account of the same
retraction discipline is reported among the audit outcomes in
Table~\ref{tab:audit}.

All AEG values are exact offline replay records on the frozen LG S1 cache
($n=80$). No live AEG candidate was generated.

\subsection{Audit outcomes}\label{sec:sensitivity}

None of the primary conclusions was changed by the sensitivity-level
observations. All of these observations have secondary or descriptive status.
The primary null in the weight channel (ELW) was also observed on the
continuous secondary endpoint. The continuous best-of-4 $q$ contrast for
\armCt{}--\armFr{} was zero at floating-point precision and yielded
permutation $p = 1.0$.
The accompanying continuous \armCt{}--\armSf{} contrast was recorded with
mean $-0.03521$ and permutation $p=0.43826$
($n_{\mathrm{perm}}=10000$).
The continuous best-of-4 $q$ contrasts for ELF were retained in the same
analysis record. The mean difference was +0.0239 for
\armOn{}--\armSh{} (permutation $p = 0.6863$),
+0.0733 for \armOn{}--\armDe{} ($p = 0.2140$), and
+0.0648 for \armOn{}--\armFr{} ($p = 0.2354$). In the prequential quartile
record, the last-minus-first unlock rate was recorded as $-0.10$ for
\armOn{} and $-0.10$ for \armFr{}, with a net quartile difference of 0.0.
These secondary records were not added to the confirmatory binary family.
The LG P0 verdict was retained in the dedup-robust recomputation
($t_{\mathrm{int}} = 0.018255$). The profile in which pooled ties were
supported by different unit sets is provided in the per-cell and per-arm heat
strip in the appendix (Figure~\ref{fig:a3}).

An earlier record of the same self-audit discipline is present in the
diagnostic layer. A learned-signal AUC value at the diagnostic stage was
reduced from 0.597 to 0.467 after a defect that counted ties as wins was
corrected (Mann--Whitney, tie $=0.5$). The corrected value is below chance,
and this record is not an effect claim.

The family distribution in the ELW corpus is provided in the compact table in
Appendix~\ref{app:c} (Table~\ref{tab:elw-family-distribution}).

Population accounting was recorded in the frozen audit as follows. The ELF
population manifest defined 37 P0-marginal and 9 S2-marginal units. Their union
yielded 46 units, and the realized Stage-1 count received \vdPass{} status at
40. The strict all-24 recount diagnostic in the same audit recorded a value of
28. Both counts were retained together in the frozen population audit, and the
denominator of the 40-unit screening tables was not changed. ELW used the same
frozen Stage-1 manifest.
The ELW protocol accounting matched the frozen record at
480 rows $=$ 280 blind
(240 round-1/2 $+$ 40 early-unlock rows $=$ 20 unit--arm pairs)
$+$ 200 error-augmented rows.
The ELF audit manifest recorded hidden access 0, prompt violation 0, and
shape leak 0.

This accounting is reported as program population, corpus, and candidate
accounting for the nested strata (Table~\ref{tab:consort}). The flow is not a
partition of a single pool. It is a sequence of strata derived by progressive
narrowing from the same dead-unit pool across generations. The 40-unit band of
the two headline families was formed through 1.5B-marginal selection after the
FASTR thin-stratum finding of 33/535 at 0.5B. The LG fresh-unit stratum yielded
46 units against the target of 60 and was reported as \vdStratumNull{} together
with the record that EvalPlus had been exhausted at 1.5B. No exclusion was
silently dropped. Each is shown as a recorded box.

In the ELF pilot, blind \armFr{} unlocked 4/10 of the FASTR-33 ``hard-dead''
units. The Stage-1 population screen was constructed as 1.5B-marginal after
this record.

\begin{table}[t]
\caption{Program population, corpus, and candidate accounting is shown in
layers. Unit, training-pair, and candidate denominators are separated by
per-layer dividers. The rows are not mutually exclusive partitions of a single
sample. They carry nested strata derived across generations. No exclusion was
dropped, and dagger-marked values are retained in the program cycle record and
the record of İşcan (2026c).}
\label{tab:consort}
\centering\footnotesize\setlength{\tabcolsep}{4pt}
\begin{tabular}{@{}
>{\raggedright\arraybackslash}p{0.08\linewidth}
>{\raggedright\arraybackslash}p{0.23\linewidth}
>{\raggedright\arraybackslash}p{0.17\linewidth}
>{\raggedright\arraybackslash}p{0.43\linewidth}@{}}
\toprule
Step & Stratum & Gen & Layer accounting \\
\midrule
\multicolumn{4}{@{}l}{\textit{Unit layer}} \\
1 &
Dead-unit pool (zero-pass-in-pool, public tier) &
--- &
initial pool \\
2 &
Search-bound stratum &
\genk{0} inherited &
$n = 46$ / 69 / 109$^{\dagger}$ \\
3 &
Hard-dead thin stratum (0.5B) &
\genk{2} FASTR &
535 defined $\to$ 33 realized. 33/535 \\
4 &
40-unit 1.5B-marginal resistant band &
\genk{7} ELF / \genk{8} ELW &
46 defined $\to$ 40 realized.
ELF 640 generations.
ELW 480 generations \\
5 &
Fresh-unit stratum &
\genk{4} LG &
60 defined $\to$ 46 realized.
$n=80$ / $g=8$.
381 fresh screened
$\to$ 359 unlocks with blind $k=4$
$\to$ 9 marginal after $k=8$
$+$ 37 frozen P0-reuse
$=46<60$.
\vdStratumNull{} \\
\midrule
\multicolumn{4}{@{}l}{\textit{Training-pair layer}} \\
6 &
ELW non-evaluation training split &
\genk{8} &
236 units
$\to$ 1,964 content $+$ 1,964 shuffled pairs
($\leq$30/unit, unit split, leakage 0, and 46 units excluded) \\
\midrule
\multicolumn{4}{@{}l}{\textit{Candidate layer}} \\
7 &
ERA online taxonomy corpus &
\genk{6} &
1,920 candidates \\
\midrule
\multicolumn{4}{@{}l}{\textit{Recorded exclusions}} \\
E1 &
LG fresh-unit stratum &
\genk{4} &
\vdStratumNull{} ($46 < 60$) \\
E2 &
DCH offline corpus &
\genk{0} &
\vdUnrealizable{} (offline corpus unavailable) \\
E3 &
RIFT trigger &
\genk{0} &
\vdDoa{} (dead on arrival, 0/24) \\
E4 &
FDP first-draw gate &
\genk{5} &
\vdGOneKill{} \\
\bottomrule
\end{tabular}
\end{table}

Two amendments were recorded in the audit layer of this flow and are reported
as frozen. In the ELF pilot, it was detected that the \armSh{} control
corrupted the task statement, producing the destroyed-task defect. The
scaffold was corrected, and Stage-1 was run with the corrected \armSh{}.
Zero shape leak was recorded in the Stage-1 audit. In ERA, two $\kappa$
defects were detected and corrected at the G6 falsification step. These
amendments were not post-hoc interventions that changed the results. They were
trace-preserving corrections closed before the primary runs. Both corrections
are recorded with dates in the audit trail in Appendix~\ref{app:a}.

The statistical verdict and audit verdict are reported separately for each
layer (Table~\ref{tab:audit}). The known open defect is also included in this
table without suppression and with the status
\emph{as-frozen failing (tracked)}. The HEF decoy-synthetic test has failed
deterministically since the Torch 2.12 migration. HEF is a supporting offline
layer and is not used as the source of any endpoint-confirmatory claim.

\begin{table}[t]
\caption{The statistical verdict and audit status are shown by layer.
Headline, supporting, amendment, and open-defect records are retained in
separate blocks. Audit-evidence and note fields are separate, and no audit
status is used in place of a statistical verdict. The HEF decoy-synthetic test
is retained with \emph{as-frozen failing (tracked)} status.}
\label{tab:audit}
\centering\footnotesize
\setlength{\tabcolsep}{2pt}
\begin{tabular}{@{}
>{\raggedright\arraybackslash}p{0.16\linewidth}
>{\raggedright\arraybackslash}p{0.15\linewidth}
>{\raggedright\arraybackslash}p{0.14\linewidth}
>{\raggedright\arraybackslash}p{0.27\linewidth}
>{\raggedright\arraybackslash}p{0.16\linewidth}@{}}
\toprule
Layer &
Statistical verdict &
Audit status &
Evidence &
Note \\
\midrule
\multicolumn{5}{@{}l}{\textit{Headline layers}} \\
\genk{7} ELF Stage-1 &
\vdMechNull{} &
\emph{clean} &
zero shape leak, no audit failure, recorded count 0 &
destroyed-task \armSh{} defect detected and corrected in the pilot \\
\genk{8} ELW evaluation &
\vdTrainNull{} &
\emph{clean} &
adapter-effect \vdPass{},
12/12 prompt-hash rebuilds,
derangement audited,
leakage 0 &
forgetting guard \vdPass{} \\
\midrule
\multicolumn{5}{@{}l}{\textit{Supporting layers}} \\
\genk{4} LG P0 &
\vdConfirmed{} &
\emph{clean} &
dedup-robust, pool audit &
single well-powered positive \\
\genk{4\mathrm{b}} AEG &
\vdReplayNull{} &
\emph{self-audited} &
superseded $B{=}12$ +6.0 pp with +0.001 control retained together, governing $B{=}8$ +0.02209 and $-0.00227$, persistent supersession pointer &
matched draw-budget correction \\
\midrule
\multicolumn{5}{@{}l}{\textit{Amendment and deferred layers}} \\
\genk{6} ERA &
\vdDeferred{} (validity) &
amended &
2 $\kappa$ defects falsified and fixed at G6 &
JEPA shuffle gap $=0.018<0.05$ \\
\genk{5} FDP &
\vdGOneKill{} &
gate-block &
the gate block is a result &
BigCodeBench \\
\genk{2} FASTR &
diagnostic null &
F4-catch &
\armDe{}$>$\armOn{} reversal recorded &
live thin stratum \\
\midrule
\multicolumn{5}{@{}l}{\textit{Open defect}} \\
\genk{1} HEF &
\emph{standing}$^{+}$ / saturation null &
\emph{as-frozen failing (tracked)} &
decoy-synthetic test, deterministic since Torch 2.12 &
HEF is not used as a primary numerical claim \\
\bottomrule
\end{tabular}
\end{table}

All results in this section are restricted to frozen 0.5--1.5B models,
HumanEval+/MBPP+/BigCodeBench (Python)
\citep{chen2021,austin2021,liu2023}, the zero-pass-in-pool regime, a single
scaffold, the matched output-generation budget ($R = 4$), and the $U_p$
screening endpoint \eqref{eq:pub4-unlock}. The mechanistic and epistemic
readings of the pattern observed within these boundaries are provided only in
the Discussion (\S\ref{sec:discussion}).

\section{Discussion}\label{sec:discussion}

This study constitutes a PoPE evaluation of whether frozen small code LLMs can
operationally use execution evidence that falsifies their own outputs. Learned
error content was tested against form-matched placebos on the same units in
both the prompt and weight channels. The only \vdConfirmed{} result in the
program was the unit-by-portfolio interaction observed in public-tier dense
progress. Public--hidden partial-progress calibration is retained as separate
descriptive evidence. In addition, the allocation gain that appeared strongest
was withdrawn through the built-in self-audit.

The results obtained within this framework address a narrower and more
fundamental question than whether learned error conditioning is generally
useful. Is the limited set of draws directed by learned error content, or by
the form of the retry and the scale of the generator? Three principal patterns
emerged together in relation to this question. First, on the 40-unit
1.5B-marginal resistant band, error content did not outperform the
form-matched placebo in either the prompt or the weight channel. Both verdicts
were obtained on the preregistered $U_p$ screen
(Definition~\ref{def:pub4}), and hidden-tier true-unlock confirmation remains
deferred by design (\S\ref{sec:primary},
Tables~\ref{tab:elf-family} and~\ref{tab:elw-family}). Second, a descriptive
increase in the 1.5B direction was observed between the two within-program
scale points (27.5\% $\to$ 60\%$^{\dagger}$ when moving from 0.5B to
1.5B\footnote{This number is contained in the program cycle record and the
record of İşcan (2026c), rather than in the frozen results archive.},
\S\ref{sec:ledger}, Table~\ref{tab:ledger}). This observation is not presented
as a lever causally isolated from the other interventions or as a scaling
curve. Third, the +6.0 pp primary delta from the superseded ($B=12$) record was
retained together with the +0.001 control from the same record and its
supersession marker. In the governing matched ($B=8$) record, the
\vdReplayNull{} verdict was recorded with contrasts of +0.02209 and
$-0.00227$ (\S\ref{sec:selfaudit}). When read together, these three patterns
yield the following four regularities.

\phantomsection\label{sec:d-laws}%
The program findings are compressed into four regularities, presented below
as \emph{Regularities 1--4}. These regularities are presented not as positive
laws, but as failure boundaries and design observations observed within the
program, each restricted by its own scope guard. None is advanced as an
impossibility proof or as a general law \citep{popper1959,mayo2006}. This
choice is motivated by the asymmetry of falsification. A failure boundary
observed on a limited band is not a universal negative law, but a local
refutation that records the region in which a claim fails, and any
generalization must be established through new tests. The four regularities
cover four axes of the same question: unconfirmed attribution is considered
along the axis of content-attributable superiority (Regularity~1), the factor
that moved is considered along the scale axis (Regularity~2), the time of
transfer is considered along the pre-first-draw position axis
(Regularity~3), and the channel in which transfer was not obtained is
considered along the small-data weight-channel axis (Regularity~4).
Immediately after each regularity, the scope guard that narrows the claim is
stated explicitly. Content-attributable superiority was not confirmed. The
subsequent readings concerning form, scale, and timing are not themselves this
negative result, but an explicitly bounded layer of interpretation.

The first is \emph{Regularity 1} (the two-channel attribution failure
boundary). Superiority attributable to error content was not confirmed in
either deployment channel. This failure boundary is termed the ``form wall''
within the program. On the resistant band, the arm carrying error content did
not outperform its channel-specific placebo twin in either channel. In the
prompt channel, the pooled $U_p$ screening counts were
\armSh{} 12 $\geq$ \armOn{} 10. No statistically detectable difference was
observed in the \armOn{} $-$ \armSh{} contrast
(net $-2$, $b_{01} = 2$, $b_{10} = 4$, McNemar $p = 1.0$,
\vdMechNull{}). In the weight channel, \armCt{} 8 = \armFr{} 8 was observed
(net 0, $b_{01} = 4$, $b_{10} = 4$, $p = 1.0$, \vdTrainNull{}), while the
SHA-deranged placebo LoRA
(LoRA, low-rank adaptation, \armSf{} 10) was numerically highest. These two
results are not interpreted as an absence of effect or as equivalence. Both
primary paired findings rest on 6--8 discordant pairs, and equivalence was not
tested separately. As a scope guard, this regularity is restricted to the 40-unit
1.5B-marginal resistant band, the $U_p$ screening endpoint, the tested single
scaffold, and the matched output-generation budget $R=4$. It is not generalized
to the broader population.

A descriptive increase in the 1.5B direction was observed between the two
within-program scale points. The second is \emph{Regularity 2} (the two-point
scale direction). The direction
$s_{1.5\mathrm{B}} > s_{0.5\mathrm{B}}$ was observed. This observation is not
presented as a lever causally isolated from the other interventions or as a
scaling curve. While the same conditioning and diversity nulls were reproduced
at both scales, the solution rate increased from 27.5\% to 60\% when moving
from 0.5B to 1.5B. Two points define a direction. They do not define a scaling
curve. The existence of scale dependence is well established in the
literature \citep{kaplan2020,hoffmann2022}. However, those anchors were fitted
to pretraining cross-entropy and carry no quantitative prediction for the
cells considered here. As a guard, no scaling-law extrapolation is performed.
Only a testable conjecture for 7B is formulated below.

Although support was added by scale, the manner in which the same budget was
spent within a unit remained a separate question. The third is
\emph{Regularity 3} (order-statistic saturation). Under a matched draw budget,
$\Delta_{\mathrm{adaptive-random}} \approx 0$ was observed, which is consistent
with a saturation-compatible pattern in the order statistic under best-of-B
sampling. Under the matched
draw-budget ($B = 8$) self-audit, within-unit adaptation left approximately
zero net gain over uniform mixing. The +6.0 pp allocation result in the
program's superseded ($B=12$) record
($\Delta = 0.06025$, $p = 0.01375$, budget 12) was retained together with the
+0.00097 adaptive--random control from the same record
(+0.001, $p = 0.44664$) and the supersession marker. In the governing matched
($B=8$) record, adaptive--best fixed was recorded as
$\Delta = +0.02209$ ($p = 0.2022$), adaptive--random was recorded as
$\Delta = -0.00227$ ($p = 0.6176$), and the verdict was \vdReplayNull{}
(\S\ref{sec:selfaudit}). Best-of-$n$ selection has been characterized in terms of its output
distribution and reward--KL trade-off \citep{beirami2025}. That analysis
provides broader context and does not by itself establish the within-unit
saturation mechanism inferred here. It also adds a limiting condition to the
inference-scaling account under which a small model combined with sampling can
be Pareto optimal \citep{wu2025}. Under the correlated-draw regime studied here, no statistically detectable
adaptive-allocation advantage over the preregistered controls was observed. As a
guard, this observation is restricted to a single \vdReplayNull{}.
Positioning a learned prior before the first draw is retained as a design
hypothesis alternative to adaptation between draws and is not presented as a
theorem or a confirmed regularity.

The remaining candidate channel for content that was not transferred between
draws was the weights. The fourth is \emph{Regularity 4} (the small-data
weight boundary). No transfer superiority was observed while
$\mathrm{loss}_{\mathrm{train}} \downarrow$. The quantized LoRA (QLoRA)
adapter trained on 1,964 content and 1,964 shuffled pairs derived from
236 non-evaluation units (NF4, rank 16/$\alpha$ 32, 491 steps) reduced the
training loss to 0.002. Nevertheless, content-attributable transfer superiority was not confirmed
on the evaluation band (\armCt{} 8 = \armFr{} 8). It was recorded that the adapter met the offline
training and machinery criteria but did not produce content-attributable
operational superiority. When read together with analyses of knowledge
capacity in large language models, this distinction indicates that
acquisition, exposure, task diversity, adapter capacity, and training-target
structure were not disentangled for the 236-unit corpus
\citep{allenzhu2024}. As a guard, this null is restricted to a single rank, a
single set of target layers, a single 236-unit corpus, and a single scale.
Limited task diversity was not separated from adapter capacity or
training-target structure and is retained as one possible explanation. No
general claim is made about the transfer capacity of the weight channel.

\phantomsection\label{sec:d-why}%
The four regularities state what was not transferred. Why it was not
transferred can be read only at the level of an account. The reading from this
point onward is an interpretation. It is not a mechanism directly identified
by the design. An explanation consistent with the findings but not directly
identified by the design can be stated as follows. At the 1.5B scale, the
error distribution around a resistant unit is not reshaped merely by naming
or recoding error content. The movement observed in the sampling
distribution is consistent with sensitivity to the form of the retry and the scale of the generator. This
reading is consistent with two observations. First, the band is not inert: at
the descriptive level, \armSh{} produced 12 unlocks, whereas \armFr{}
produced 7. Movement was observed at the descriptive level. Because form superiority was
not established through a separate positive mechanism test, this pattern is
treated as consistent with form sensitivity rather than as proof of a form
effect. Second, although the same content could be measured and
separated offline
(C-matrix log loss 0.254 $\to$ 0.185, 27.3\% admitted,
decoy-separation gate), no online refuting force was produced.

This pattern is consistent with the anchoring account of the preceding stage.
When the failed solution or its abstractions are shown to the model again in
the prompt, the sampling distribution is drawn toward the local neighborhood
of the failure \citep{tversky1974,jones2022,dinh2023}. Findings of
self-preference in large language model (LLM) evaluators point in the same
direction \citep{panickssery2024}. The same distributional narrowing is also
visible in the offline supervised fine-tuning (SFT) regime in which feedback
is transferred into the weight channel. SFT over self-generated correction
traces has been found insufficient because of distribution mismatch and mode
collapse, and online RL has been required \citep{kumar2025}. This pattern was
not changed by the learned controller. Only the representation from which the
pull was exerted was replaced by a learned lattice or adapter.

The same account can also be combined with the mechanistic negatives from the
second stage. The coverage wall showed that deeper sampling did not rescue
systematic failures. The capability scissors showed that a capable generator
left no discriminable error margin among visibly passing outputs
\citep{iscan2026b}. The wall observed here can be read as the learned-side
view of the same wall. The tested object was not merely the addition of error
text to a prompt. It was tested whether compiling criticism derived from
execution outcomes into a lattice, adapter weights, or allocator state
produced a discriminating force over new generations that exceeded the
placebo. When no discriminable signal is available to a leakage-free
selector, no conditionable signal may be available to a learned controller
either. Conditioning can reweight support in the model distribution, but the
resistant band was defined as a region in which the required support had not
been observed in the pool. Scale adds support, form changes the search
neighborhood, and, under this account, redescribing content does not add
support. By retaining the learned object and its form-twin placebo within the
same training and evaluation channel, it is made testable whether the observed
movement is attributable merely to optimization or to the existence of a
representation. The same form wall was also observed in the legacy stage of
the program. There, too, the two strongest content channels failed to
outperform a content-free shape placebo and only tied it
\citep{iscan2026c}.

The rivals to this account must be named explicitly. The first rival is an
encoding bottleneck. The lattice or adapter encoding may compress the
refuting detail of the counterexample in a lossy manner. The second is corpus
narrowness. The confound stated for
\emph{Regularity 4 (the small-data weight boundary)} also applies here. The
third is the screening endpoint. Because hidden-tier confirmation was
deferred, a content effect that was not visible in the public-tier screen has
not been ruled out. A negative public-tier screen licenses only the
termination of the preregistered promotion. The fourth is band-selection
heterogeneity. Because the resistant band was constructed only from a history
of public zero-pass outcomes, unit types that might respond differently to
content may have been collected within the same stratum. These explanations
cannot be fully distinguished by the present design. No mechanism-level claim
is therefore made.

\phantomsection\label{sec:d-memorization}%
Among the rival explanations, corpus narrowness becomes more concrete when it
is read together with the weight channel's own finding. Two observations were
recorded together in the weight-channel adapter experiment. Both adapter
variants symmetrically reduced the training loss to 0.002 within 491 steps.
Actual adapter activation was confirmed by the adapter-effect probe: when the
adapter was disabled, the base output was restored byte-for-byte, and 12 of
12 sampled rows were reproduced exactly in the prompt-hash rebuild probes.
Nevertheless, no statistically detectable difference was observed in the
\armCt{} $-$ \armFr{} contrast (net 0, $p = 1.0$). The combination of
near-zero training loss and unconfirmed evaluation-band superiority is
consistent with a memorize-don't-generalize interpretation. However, target
memorization, scaffold adaptation, corpus diversity, adapter capacity, and
semantic error-relation learning were not separated from one another by this
design. This is the precise expression of the learned throughline in the
weight channel. What was tested was not whether the corpus had been memorized,
but whether what had been learned opened a solution on the resistant band.
The combination of near-zero training loss and unconfirmed content-attributable transfer records that the
latter was not observed. Pair multiplicity must also be noted. Although the
1,964 pairs increase the visible sample size, independent task support is not
expanded in the same proportion. The 1,964-pair count should therefore not
obscure the limit of 236 independent units, and the zero-leakage audit should
not be read as a diversity audit.

This reading is consistent with recent findings on parameter-efficient
fine-tuning (PEFT). The closest of these findings is the study documenting
failure modes of small-data LoRA injection, including regression toward
overrepresented answers and degradation on external benchmarks when new facts
are packed into the adapter. This boundary is extended by ELW to an
unconfirmed-transfer null at 1.5B through its SHA-deranged placebo LoRA and paired
McNemar test \citep{pletenev2025}. Low-rank adaptation has been reported to
learn less and forget less than full fine-tuning \citep{biderman2024}. Here,
too, the forgetting guard was passed
(content-blind 4 $\geq$ frozen-blind 3). The preregistered forgetting guard was
therefore not violated, and broader harm was not tested. It has also been shown that narrowly
targeted fine-tuning can instill the target behavior while eroding general
correctness \citep{ibrahim2026}. The weight-channel experiment therefore
provides a symmetric negative result relative to this literature. Narrow
error-content training in the small-data regime did not violate the forgetting
guard, but it also did not produce content-attributable transfer superiority.
Memorization is retained as an explanation consistent with this result and is
not presented as a confirmed mechanism. This memorization boundary is
consistent with the finding that LoRA instruction tuning primarily learns
response-start and style tokens. The form interpretation thereby also receives
an external comparator in the weight channel without requiring a positive
\armCt{} result \citep{ghosh2024}. Positive findings in the self-training
literature \citep{zelikman2022,gulcehre2023,singh2024}, by contrast, arise in
large-scale regimes with correct-answer filtering. No comparison in that
literature tests error-structure conditioning with small data in a frozen
small model. The result is therefore stated under the following boundaries:
one rank (16), one QLoRA configuration (NF4, $\alpha$ 32), one 236-unit
corpus, one model scale (1.5B), and the $U_p$ screening endpoint.

\phantomsection\label{sec:d-firstdraw}%
Given that training fit did not yield confirmed content-attributable transfer
superiority in the weight channel, and that prompt-channel content did not
outperform its placebo between draws, a timing question remains. When can a learned prior become useful? The self-audit chain of the
budgeted best-of-B allocator (AEG-BANDIT)
(\S\ref{sec:selfaudit}, Table~\ref{tab:aeg}) records that the initial positive
appearance depended on budget inequality. The +6.0 pp value from the
superseded comparison is retained together with the +0.001 control from the
same record and carries a budget confound caused by the adaptive arm receiving
12 draws while the fixed arms were pool-capped at 8. In the matched
draw-budget ($B = 8$) replay, the adaptive--best fixed contrast was recorded
as +0.02209. The governing adaptive--random contrast was $-0.00227$. The
+0.00097 (+0.001) control from the superseded ($B=12$) record is retained only
as the preceding step in the audit lineage, together with the +6.0 pp primary
delta and the supersession record. The null-world calibration
($\Delta = -0.02667$, $p = 0.91354$) was consistent with the reading that the
test bench did not generate a false positive. The interpretation of this
chain is stated as follows and is explicitly labeled as an interpretation.
What constrains within-unit adaptation is not a failure of identification, but
saturation of the order statistic. Per-test passes within a candidate were
strongly correlated ($\hat{\rho} = 0.7771$), and the effective evidential
value of an observation containing median $m = 3.0$ public tests was
compressed to $\kappa_{\mathrm{eff}} = 1.175$. Additional draws therefore
purchase nearly the same evidence again. Under this reading, the design does not identify the unique binding
constraint on within-unit adaptation, though an information limit is more
consistent with the observations than a budget limit.

This reading points to a boundary of classical allocation theory. In the
fixed-budget best-arm-identification framework
\citep{audibert2010,karnin2013,lattimore2020}, the bottleneck is modeled as
identification. What was observed here, however, was a max-of-B statistic that
saturated before identification became operative. Recent theoretical
negatives within the same framework are also consistent with this reading.
It has been shown that, when a relevant complexity exists in fixed-budget
identification, it is attained by the best non-adaptive procedure
\citep{degenne2023}, and that no stable-consistent algorithm uniformly
outperforms uniform allocation in the two-arm fixed-budget setting
\citep{wang2024b}. These are the closest theoretical neighbors of the
observation that the allocator left zero net gain over mixing. The distinction
between existence and allocation also becomes clear at this point. Best-of-N
existence results state that a correct candidate may be present in the pool
\citep{brown2024}. The question here was how a fixed budget should be spent
within a unit, and the observed answer was that spending order became
practically irrelevant on a saturated order statistic. The independent
finding that the effectiveness of iterative debugging declines with the
number of rounds \citep{adnan2025} is consistent with the same saturation
interpretation. Pre-first-draw placement is retained as a design hypothesis
consistent with \vdReplayNull{}. This hypothesis is not presented as a
theorem or a confirmed regularity. Indeed, the attempt to place a prior before
the first draw (FDP) received \vdKill{} for both feature sets at its own
preregistered G1 gate$^{\dagger}$.

\phantomsection\label{sec:d-epistemic-core}%
The aggregate of these three channel findings cannot be reduced to a single
number. The program's principal epistemic inference arises not from any single
number, but from the placebo hierarchy as a whole. The third stage
distinguished model-internal verbal self-critique from a counterexample derived
from execution. The former is a textual artifact drawn from the same
distribution, whereas the latter is an external comparison performed by a
computational procedure outside the model \citep{iscan2026c}. That distinction
is extended inward in this study. The criticism considered here is genuinely
external in origin. An error taxonomy harvested from 1,920 real execution
outcomes and saturated at $K = 93$ categories, an outcome-calibrated severity
measure, and a C-matrix that passed a predeclared gate were all available.
Nevertheless, once this criticism was compiled into controller state, whether
as a prompt lattice
(the learned error-lattice prompt controller, ELF), a weight adapter
(the error-lattice QLoRA weight adapter, ELW), or an architectural state
(the error-set architecture, ERA), its behavioral signature could not be
distinguished from that of the placebo. What was actually tested must
therefore be stated explicitly. Once criticism is compiled into a learned
object, such as a lattice, adapter, or architecture, the question being tested
becomes whether the content of that criticism can open a solution through that
object. The form-matched placebo twins make this test meaningful. Because the
twin preserves predeclared scaffold components such as the task prompt,
entrypoint, block layout, and injection position while removing the refuting
content, exact token length and lexical distribution were not matched
one-to-one, and residual surface aliasing was not ruled out. Under these
constraints, a difference by which the learned object exceeded its twin was
treated as the required measurement for content attribution, but no such
difference was observed in either channel.

The observed pattern is consistent with prior evidence on form sensitivity. An older and established
epistemic distinction between form and content is operationalized in both
deployment channels of this program. In in-context learning, random
replacement of ground-truth labels was found to reduce performance only
slightly, indicating that the result was driven not by content but by label
space, input distribution, and format \citep{min2022}. The form sensitivity
documented in that literature provides an attributional rationale for the
\armSh{} control. The result reported here is not presented as a direct
replication of the content-irrelevance phenomenon reported for classification
demonstrations. A prompt-semantics antecedent of this content irrelevance has
also been documented. Models have been reported to learn from misleading or
pathologically incorrect instructions at nearly the same rate as from
well-formed instructions \citep{webson2022}. The \armSf{} and SHA-deranged
placebo-LoRA controls constitute the weight-channel counterpart of this
observation. Under this reading, the observed movement is associated with the
form of conditioning rather than with content. The \armSh{} arm, which
preserved the predeclared scaffold components while removing content, finished
ahead of \armOn{} (12 $\geq$ 10). The placebo LoRA, in which the
correspondence between error blocks was disrupted through a SHA-seeded
derangement, was numerically highest in the weight channel (\armSf{} 10).
The \armDe{} arm, in which transitions were deliberately mismatched, remained
close to \armFr{} (8 to 7). This observation is operationalized by the
two-gate promotion rule: any arm that passes the signal gate but remains below
the content gate is assigned the form label by definition. The prompt channel
ended in this pattern, whereas the weight channel failed the signal gate as
well. The ability of a placebo to produce movement without carrying content
makes the external origin of compiled evidence insufficient on its own for
operational credit.

The epistemological reading of this pattern is bounded as an interpretation.
The externality of criticism is a property not of its provenance, but of its
operational position. The execution oracle remains external at test time. By
contrast, a prior learned from the oracle and compiled into the generation
state is no longer the oracle. A feature used within the same generative
process can be a representation of criticism, but it does not occupy the external epistemic position of criticism.
The same execution trace becomes a counterexample relative to the oracle, a
target in the training corpus, a token sequence in the prompt lattice, and a
gradient contribution in the adapter. These transformations may provide
useful compression, but they do not automatically preserve the
counterexample's function of testing a new output against an independent
procedure. The philosophical importance of the placebo twins arises here.
When no operational difference is found between a representation of criticism
and a twin that carries the same learned and formal load without carrying the
falsifying content, credit cannot be assigned to the content of criticism.
Externality is thereby read not as an immutable property of the data source,
but as a live relation between evidence and conjecture. Compiled criticism is
preserved as error-derived information. However, unless the new conjecture is
reconnected to the external oracle, it carries the role of conditioning
rather than that of independent criticism.

In Popperian terms, criticism performs its work at the point of refutation,
namely at execution. Compiling it at the point of generation turns criticism
into a prior belief. This transformation removes the division of labor
between conjecture and refutation. In the weight channel this conversion is
made explicit, since the training objective \eqref{eq:qlora-ce} treats each
error atom harvested from the oracle as a likelihood target, so that the
content of criticism is compiled by the gradient into a prior encoded in the
adapter weights. The epistemic force of criticism derives
not from the truth of its content, but from its status as an independent
comparison procedure standing outside the conjecture \citep{popper1963}.
Compiled criticism is thereby caught one level higher by the exposure
diagnosis that was applied to bare failing code in the third stage. Once a
representation learned from the oracle is shown again through the prompt or
the weights, it is no longer a test but a condition \citep{iscan2026c}. The
gap between offline measurability and online refuting force is also consistent
with this reading. The fact that content can be separated
(decoy separation, C-matrix) did not imply that it would perform
discriminating work at generation time. The epistemic validity of criticism
and its generative efficacy must therefore be kept separate. An executable
oracle can continue to show that a conjecture is false. A frozen generator,
by contrast, may lack the representational freedom required to produce an
alternative correct candidate from the same counterexample. This distinction
between ``measurable'' and ``operative'' is consistent with the
standing-diagnostics records of the program
(\S\ref{sec:standing}, Table~\ref{tab:standing}) coexisting with the transfer
nulls without contradiction. In the tested regime, criticism did not perform
its work as a vocabulary, as model-internal verbal critique, or as a compiled
error representation. The function that was confirmed was the renewed testing
of every new generation against the execution oracle. This reading is
directly connected to the program's contributions. The channel-agnostic
placebo hierarchy that separates content from form establishes the measurement
standard through which this distinction is made visible, while the self-audit
that withdraws the program's own positive within the experiment applies the
same standard to the researcher's own claims. The output of the study is not
a repair algorithm, but a retestable measurement standard. The role of
criticism is recovered when each new conjecture is reconnected to the external
oracle.

For this result to become visible, falsification had to be applied at two
levels. At the object level, model outputs were refuted through execution. At
the meta level, the researcher's own claims were made refutable through
preregistration, form-matched placebos, a matched output-generation budget
($R = 4$), and executable audit invariants \citep{iscan2026c}. The finding
that compiled criticism did not produce content-attributable superiority beyond
the placebo could not have been produced without the meta level. Without the \armSh{}, \armDe{}, and SHA-deranged
placebo-LoRA controls, the +3 direction of \armOn{} relative to \armFr{} or
the reduction of the \armCt{} adapter's training loss to 0.002 could have
been reported as a working content mechanism. The placebo hierarchy is
therefore not decorative. It is the measurement apparatus that carries the
claim itself. The contribution is consequently formulated not as a repair
algorithm, but as a reflexive measurement instrument.

The same epistemic discipline must be applied to the borrowed architecture.
A borrowed architecture is not a validated result. The joint-embedding
predictive architecture (JEPA) head of ERA is a design borrowed from the vision
literature \citep{lecun2022,bardes2022,assran2023}, and it operated as intended
at the smoke-test level. The embedding did not collapse, and training
progressed. Nevertheless, no downstream signal was produced
(ridge $R^{2}_{\mathrm{oof}} \approx 0.0096$,
decoder $R^{2}_{\mathrm{oof}} \approx -133$). The closest realization of JEPA
on the language and code side adds a training-time embedding-prediction
objective to next-token loss and reports gains on code-adjacent datasets
\citep{huang2025}. The null reported here instead concerns error content
injected at inference time through the prompt lattice or through small-data
QLoRA in the weight adapter. The distinction is therefore between
training-time and inference-time channels of the same architectural family.
The two observations do not conflict. The guarantees of VICReg concern
embedding non-collapse, not downstream utility. A non-collapsed but unhelpful
embedding is consistent with the architecture operating as designed while
failing to provide utility in this regime. The statement that ``JEPA should
be positioned before the first draw'' is also an analogy within this
framework. It refers to a world-model-before-acting vision and is not an
empirical result established by that literature in this regime. Each of these
distinctions explains why ERA validity was left \vdDeferred{}.

When this reading is combined with
\emph{Regularity 3 (order-statistic saturation)}, a consistent design lesson
emerges. If compiled criticism is ineffective between draws because of
saturation, and if it did not produce content-attributable superiority beyond
the form-matched placebo once compiled, one remaining testable position within
the program at which a learned error prior can be tested is before
the first draw. Even there, it must outperform a form-matched placebo. This
account is consistent with the view that falsification is not a vocabulary or
stored taxonomy, but a relation through which every new conjecture is reopened
to executable comparison. The account itself is left falsifiable. A single
demonstration in which a pre-first-draw prior outperforms its form control
would weaken the reading developed in this section. The present evidence
contains no such demonstration. The \vdGOneKill{} result of FDP is the first
attempt on this front to have been stopped at the gate.

\phantomsection\label{sec:d-distinctions}%
Four distinctions systematically preserved by the program's reporting
discipline prevent these readings from being expanded beyond their support.
First, the statistical verdict and audit verdict are separate axes
(\S\ref{sec:sensitivity}, Table~\ref{tab:audit}). The p-value of a contrast
and the integrity of the pipeline that produced that contrast are reported
separately. As-frozen audit statuses are not overwritten retrospectively. The
HEF decoy-synthetic test, which has deterministically \emph{failed} since
Torch 2.12, was therefore not concealed and was retained as a tracked open
defect in Table~\ref{tab:audit}. HEF is not used to support any
endpoint-confirmatory claim.

Second, a gate block, stratum exhaustion, or public-screen null is a
first-class result. The \vdGOneKill{} result of FDP for both feature sets, the
\vdStratumNull{} result of LG
($46 < 60$, with EvalPlus exhausted at 1.5B), and the ELF/ELW $U_p$ screens
are not failed attempts, but preregistered measurements that stopped
expenditure at the correct point. A gate is not an incomplete experiment. It
is a design object that prevents progression to the wrong epistemic stage.
Because a preregistered gate gives a claim a genuine chance to die, a
\vdGOneKill{} result is, in this sense, a record imposed on the program itself
by severe-testing discipline, not an incomplete experiment
\citep{mayo2006,mayo2025}. The kill reasons for the legacy controllers at the
beginning of the program carry the same status. RIFT was recorded as
\vdDoa{} (0/24), DCH as \vdUnrealizable{}, and ECK as
\vdUnderpowered{}. Each is a reported box rather than a silent drop
(\S\ref{sec:sensitivity}, Table~\ref{tab:consort}).

The complement to this distinction is the manner in which the program's single
\vdConfirmed{} result is protected. In LG P0, permutation $p = 0.0001$ was
recorded for the unit-by-portfolio interaction in public-tier dense progress.
In the program-wide sensitivity check, the family denominator was taken to be
approximately 10, and a threshold of $0.05/10 = 0.005$ was reported. Because
the exact family universe is not listed in the frozen ledger, this calculation
is retained as a descriptive sensitivity check rather than as a confirmatory
multiplicity correction (\S\ref{sec:standing}). This reporting rests on the
standard basis by which confirmatory and exploratory analyses are separated
and hypotheses are frozen before the data are observed \citep{nosek2018}.
P0 is an E-INTERACTION result, not an unlock-advantage result. Public--hidden
partial-progress calibration is retained as separate descriptive evidence
($r \approx 0.65$--$0.70$). No reading of a coupling substrate that could be
exploited by a controller, or of a ``working controller,'' is licensed by
this result.

Third, a tie is not sameness, and offline decisiveness is not live transfer.
The results \armSh{} 12 $\geq$ \armOn{} 10 and
\armCt{} 8 = \armFr{} 8 ($p = 1.0$) rest on 6--8 discordant pairs. These
observations do not constitute evidence of equivalence or non-inferiority.
Equivalence was not tested separately, and no TOST margin was defined. A tie
over 6--8 discordant pairs is not a demonstration of sameness, but an absence
of discrimination. Because corroboration is asymmetric with respect to
refutation, ``failed to confirm'' and ``refuted'' are distinct expressions
\citep{popper1959,mayo2006}. In this framework, corroboration is not a degree
of verification, but a record indexed to the severity of the test survived.
Because survival is cheap under a low-powered test, no high-severity survival
status is assigned to either paired finding \citep{mayo2025}. Similarly, the
content-axis result of HEF in the offline synthetic environment
($p \approx 7\times10^{-14}$, leave-one-cluster-out leakage control 0) was
decisive offline. However, the Ollama probe of the same detector reproduced
the form-not-content null in live evaluation. Offline certainty is nowhere
used as evidence of live transfer.

Fourth, the principle from the first stage that
``naming severity is not having it'' \citep{iscan2026a} is deepened in this
program. In m1, naming severity was insufficient. Here, severity was fitted
from real outcomes and calibrated by ERA, yet it still did not exceed form
once the order statistic had saturated. Having a measured severity is
therefore not the same as having an operational severity. A three-step ladder
thereby emerges, with each step carrying its own falsifier: named severity,
which was refuted as a discriminating signal in m1; measured and calibrated
severity, which was obtained in ERA; and operational severity, defined as
exceeding form after the order statistic has saturated, which was not obtained
in this program. This observation is consistent with the literature on the
limits of self-correction \citep{huang2024}. The presence of an apparently
external signal does not guarantee that it is usable. This limit on usability
is narrowed further by the finding that small models can correct themselves
only in the presence of a strong external verifier \citep{zhang2024c}.

\phantomsection\label{sec:d-positioning}%
These distinctions also determine the position of the study relative to prior
work. The study is distinguished from prior work not by method names, but by
the experimental controls retained, continuing the positioning discipline of
the third stage \citep{iscan2026c}. Positive results have been reported within
their respective regimes by the self-repair and self-refinement line
\citep{madaan2023,shinn2023,chen2024,olausson2024}, the execution-feedback RL
line \citep{le2022,shojaee2023,gehring2024}, the inference-aware best-of-N
line \citep{brown2024,snell2025,chow2025}, the PEFT and self-training line
\citep{hu2021,dettmers2023,zelikman2022,singh2024,biderman2024}, and the
progressive-refinement line \citep{du2025}. The point of distinction is that
none of these lines jointly includes a placebo separating content from form,
a matched output-generation budget ($R = 4$), tests of both channels on the
same units against their channel-specific placebo twins, and a self-audit that
withdraws its own positive within the experiment. A controls-coverage map for
selected prior work is provided as Table~\ref{tab:positioning} in
Appendix~\ref{app:g}. The marks are coded from the published method
descriptions of the cited studies. The map is descriptive rather than an
exhaustive certification and carries no claim of outcome superiority.

The simultaneous presence of placebo control, a matched output budget, two
channels on shared units, and self-audit in the PoPE row of
Table~\ref{tab:positioning} explains why the nulls reported here define a
narrower failure boundary without directly contradicting prior positive
findings. Controls-based positioning also keeps visible the regime mismatch
between the positive literature and the nulls reported here.
Execution-feedback RL and self-training systems operate on larger models, with
online updates, or on broader corpora \citep{gehring2024,singh2024}. The
inference-aware best-of-N line links budget allocation to model weights, but
does not pose the content-versus-form attribution question on the same
endpoint \citep{chow2025,damani2025}. The same regime difference is visible in
positive reports concerning error-content feedback itself. Returning
error-type feedback to the model improved code quality at ChatGPT scale
\citep{liu2024}, whereas fine-tuning on outputs from stronger models copied
style without closing the content gap across the 1.5B--13B range
\citep{gudibande2024}. These two ends jointly bound the sub-2B error-corpus
corner considered here, a corner that remains uncharacterized with respect to
scale, endpoint, and placebo controls.

\phantomsection\label{sec:d-implications}%
Three bounded practical implications follow from the findings, and all three
are formulated only for the tested regime. The scope comprises frozen
0.5--1.5B code models, the zero-pass-in-pool regime,
HumanEval+/MBPP+/BigCodeBench in Python, a single prompt scaffold, a matched
output-generation budget ($R = 4$), and the $U_p$ screening endpoint.

First, a learned error prior is not a reliable default against form, scale, or
mixing in this regime. On the evidence observed for a frozen coder of
$\leq$1.5B, larger-generator (\emph{Regularity 2, the two-point scale direction}) and
simple-mixing (\emph{Regularity 3, order-statistic saturation}) baselines
should be evaluated before a learned controller is credited. Because repeated draws from
smaller models have independently been shown to exceed a single draw from a
larger model under a matched budget only when unit-test selection is available
\citep{hassid2024}, this ordering is specific to a deployment regime in which
the execution oracle is available. At the level of the observed evidence, no
measurable contribution beyond these two alternatives was added by the learned
conditioning layer. For the practitioner, the default comparison should
therefore not be a weak arm without a controller, but the strongest simple
alternative using the same matched output-generation budget ($R = 4$).
Second, before a content arm is advanced, it should be tested on a resistant
band against a form-matched placebo within its own deployment channel. In the
data from this program, gains measured only against \armFr{} or blind sampling
could not be distinguished from form effects under the two-gate rule. Third,
small-data QLoRA training over error content did not produce
content-attributable transfer superiority in this regime. Task diversity
should be isolated before additional optimization steps are prioritized.
Because of the confound stated for
\emph{Regularity 4 (the small-data weight boundary)}, this recommendation
retains the status of a hypothesis and is linked to the follow-up experiment
specified below.

\phantomsection\label{sec:d-followups}%
The natural extension of these bounded implications is a set of open research
directions and preregistration-ready follow-up studies. The four regularities and their
guards directly determine four follow-up lines left preregistration-ready by
the program. First, task-diverse corpora can be used to test task diversity
directly, the named prime suspect under
\emph{Regularity 4 (the small-data weight boundary)}. The same QLoRA
configuration should be retrained on a corpus with greater source diversity
while the number of units is held fixed. Findings on instruction diversity
\citep{zhang2024a,zhang2024b} show in larger regimes that diversity can be
more consequential than volume. It has also been reported that weaker but
cheaper generators can produce compute-optimal data through high coverage and
diversity \citep{bansal2025}. The side effects of narrow fine-tuning have been
demonstrated independently \citep{ibrahim2026}. These anchors motivate the
hypothesis but do not validate it at this scale or under this data budget.

Second, a 7B generator should be evaluated.
\emph{Regularity 2 (the two-point scale direction)} records a descriptive
cycle-log increase. Repetition of the same resistant-band protocol at 7B on a
fresh band constructed separately from the original 40 units would test
whether the attribution boundary shifts with scale or with the regime.

Third, the deferred ERA validity test should be conducted. ERA currently
constitutes only built and smoke-proven machinery
($K = 93$, $\hat{M} \approx 0.03$, Hedge bound, and no signal in the JEPA
head, \S\ref{sec:d-epistemic-core}). No ERA component should be regarded as
validated before its preregistered reward-validity test has been run.

Fourth, the allocator should be repositioned as a pre-first-draw prior. The
design rule obtained from
\emph{Regularity 3 (order-statistic saturation)} suggests moving the exact
allocator machinery of AEG-BANDIT from between draws to before the first draw.
The finding of declining debugging effectiveness independently motivates this
move \citep{adnan2025}, and the \vdGOneKill{} result of FDP has already
established the gate discipline for this front. This ``before the first draw''
position is shared with a recent realization of the world-model-before-acting
vision, in which a latent world model performs zero-shot planning before
actions are selected \citep{assran2025,hafner2025}. Its use here is analogical
and is not an empirical result established by that literature in this regime.
Positive findings from the progressive-refinement line \citep{du2025} provide
a natural external comparator in this setting. Unless the same claim is tested
in this regime under a form-matched placebo and a matched output-generation
budget ($R = 4$), it is not directly comparable with the nulls reported here.
These four lines are designed to sharpen the boundary of the primary result.
None changes the evidential status of the current verdicts unless and until it
is run and reported.

\phantomsection\label{sec:d-validity}%
To determine which claim could be upgraded by these follow-up lines, the
validity boundaries of the findings are considered next. The boundaries are
reported across four facets: construct, conclusion, internal, and external
validity. With respect to construct validity, the program endpoint is true
unlock, defined as public $\wedge$ prompt-hidden. However, dead and resistant
status is defined only through the public tier, and the two decisive verdicts,
ELF and ELW, are based on the $U_p$ screen. Hidden-tier confirmation was
deferred by design because of the negative public-tier screen. ERA validity is
deferred. ``Severity'' is a fitted rather than directly measured quantity and
should be read under this construct boundary.

At the level of conclusion validity, all primary contrasts were reported
against the preregistered gate thresholds using exact one-sided McNemar tests
with within-family Holm correction. The ties on the 40-unit band rest on
6--8 discordant pairs. The low-powered families are labeled
``failed to confirm,'' and none is read as ``refuted.'' Equivalence and
non-inferiority were not tested for any tie. By contrast, LG P0 is the
program's only well-powered result
(power $\geq 0.97$, $n = 80$/$g = 8$, permutation $p = 0.0001$) and licenses
only the claim of a unit-by-portfolio interaction in public-tier dense
progress.

With respect to internal validity, the mirror and derangement controls preserve
predeclared scaffold components such as the task prompt, entrypoint, block
layout, and injection position, but do not hold exact token length or lexical
distribution fixed. A residual degree of aliasing therefore remains between
content and prompt statistics. The budget is defined as a matched
output-generation budget ($R = 4$). Equality in FLOPs or input tokens was not
achieved. The weight-channel experiment does not separate adapter capacity
from the task diversity of the corpus. This boundary is retained in the guard
for \emph{Regularity 4 (the small-data weight boundary)}. The +6.0 pp and
+0.001 control from the superseded $B=12$ AEG-BANDIT record were identified
together with their supersession as arising under a pool-cap confound, and the
record was corrected within the paper. The correction itself records that the
internal-validity mechanism operated.

With respect to external validity, the findings are restricted along the
following axes. Model scale is 0.5--1.5B
(Qwen2.5-Coder 0.5B/1.5B, DeepSeek-Coder 1.3B, Ollama). The regime is
zero-pass-in-pool. The benchmarks are HumanEval+/MBPP+ through EvalPlus and
BigCodeBench, and only Python is covered. A single prompt scaffold and a
single QLoRA configuration (NF4, r16/$\alpha$32) were used. The same direction
or magnitude is not claimed for larger models, different languages, different
scaffolds, or different adaptation configurations.

The four facets are not independent checklist items. A boundary in one facet
directly changes the claim language in another. The public-only construct
requires the results to be read at the conclusion level as superiority
screens. The diversity confound requires capacity language to be withdrawn at
the internal-validity level. Resistant-band selection and small model scale
jointly limit external transport. A clean audit does not close this transport
gap. This reading is consistent with an account under which boundary reporting
is not a list of limitations, but a design map specifying which future
contrast could upgrade which claim. The four concrete lines of this design
map were stated above in preregistration-ready form.

Under these boundaries, the central result is restricted as follows. On the
40-unit 1.5B-marginal resistant band and at the $U_p$ screening endpoint,
error content did not outperform its form-matched placebo in either the prompt
or the weight channel. In the two-point contrast derived from the cycle log, a
descriptive scale increase in the 1.5B direction was observed, and the
governing matched draw-budget ($B = 8$) self-audit was recorded as
\vdReplayNull{}. None of these statements carries a claim of equivalence,
non-inferiority, or general impossibility. Hidden-tier true-unlock confirmation
is deferred by design.

\phantomsection\label{sec:d-selfaudit}%
The discipline of falsification was also applied to the program's own positive
results \citep{iscan2026b}. AEG-BANDIT provides the example. The superseded
($B=12$) +6.0 pp allocation result that could have been reported with
$p = 0.01375$ was retained together with the +0.001 control from the same
record and the persistent supersession marker. In the governing matched
($B=8$) record, \vdReplayNull{} was recorded with contrasts of +0.02209 and
$-0.00227$. The initial decision was preserved as invalid in the persistent
supersession record and was reported together with the valid $B=8$ analysis.
This recording discipline prevents negative evidence from being neutralized
post hoc. The same discipline was applied throughout this paper. No working
JEPA-RL controller is claimed. A program of negative-result and
measurement-methodology character is presented. The nulls are not no-go
theorems, but design rules that direct the expenditure of the next design
toward the appropriate target. The standing diagnostics record that the
instrument was operational and reduce the plausibility of a gross
instrument-failure explanation; they do not exclude measurement error. This
reading also renders testable the question of whether a sequence of nulls can
count as progressive within a research programme \citep{lakatos1968}. The
programme can be read as a progressive problem shift only for as long as each
null narrows the question posed by the next test. Otherwise, continued
development must be recorded as degeneration.

\section{Conclusion}\label{sec:conclusion}

PoPE (Popperian Placebo-controlled Evaluation), the original evaluation methodology introduced here, was used to assess whether frozen small code large language models (LLMs, 0.5--1.5B) could operationally use evidence that falsified their own generated code. The operational value of learned, error-conditioned self-repair was tested across two separate deployment channels and was evaluated through a learned error-lattice controller in the prompt channel and a quantized low-rank adaptation (QLoRA) adapter in the weight channel. Every measurement was rendered falsifiable by two structural components of the methodology: a low-cost preregistered gate applied at each stage before live expenditure was opened, and a channel-agnostic control hierarchy in which each content arm was paired with a scaffold-matched placebo twin whose content was either ablated or deranged within the same deployment channel. All comparisons were conducted under a matched output-generation budget ($R = 4$) on the 40-unit 1.5B-marginal resistant band.

Under this evaluation, error content did not outperform its channel-specific placebo twin in either channel. In the prompt channel, \armSh{} finished ahead of \armOn{} (12 to 10, \vdMechNull{}). In the weight channel, the content arm and the intervention-free baseline were numerically tied, while the SHA-deranged placebo remained ahead (\armCt{} $=$ \armFr{} $= 8$, \armSf{} $= 10$, \vdTrainNull{}). These results do not constitute evidence of equivalence or non-inferiority. Equivalence was not tested separately. The transferable contribution retained from the study is therefore not a working controller, but a reflexive evaluation apparatus through which each negative result can be retested. The findings are interpreted not as showing that compiled criticism disappears as information, but as showing that its external epistemic role in independently testing a new conjecture is lost. Within the same evaluation sequence, only the unit-by-portfolio interaction in public-tier dense progress was recorded as \vdConfirmed{}. Public--hidden partial-progress calibration was retained as separate descriptive evidence. This result was not interpreted as evidence of controller superiority. In the two-point scale contrast obtained from the cycle log, a descriptive increase in the 1.5B direction was observed, with the any-portfolio unlock rate rising from 27.5\% to 60\%$^{\dagger}$\footnote{This value is contained in the cycle log and the record of İşcan (2026c), rather than in the frozen result records.}. This observation was not interpreted as evidence of a scaling law or of a single causal lever. The principal contribution to the reliability of the methodology was provided by its ability to withdraw its own apparently strongest positive result. The +6.0 pp primary delta in the superseded ($B=12$) record was retained together with the +0.001 control from the same record, and the \vdReplayNull{} verdict was made persistent in the governing matched ($B=8$) record through the +0.02209 and $-0.00227$ contrasts.

The open research directions indicated by this evidential standard are task diversity and generator scale. Although training fit was achieved in the weight channel, the failure to confirm content-attributable transfer superiority leaves limited task diversity as one possible explanation. This explanation was not separated from adapter capacity or training-target structure. In the first follow-up study, task-diverse corpora should therefore be tested against the same deranged control. The descriptive two-point scale observation suggests that a larger generator (7B) could be tested as a separate preregistered intervention under the same attribution controls. Task diversity and scale should not be bundled again within the same run. The hidden-tier Stage-2 confirmation deferred by design and the validity test of the joint-embedding predictive architecture (JEPA)-based reward remain available to be conducted at low cost under the same preregistered discipline. The reported nulls are not no-go theorems, but design rules. A future learned error prior can be tested for whether, positioned before the first draw, it outperforms its placebo twin within its own deployment channel when task diversity and generator scale are varied as separate interventions. Within this framework, the idea of JEPA-based reinforcement learning (RL) is neither confirmed nor refuted. Only the boundaries within which that idea might become operationally effective are delineated.


\section*{Reproducibility and Data Availability}

The research data, result manifests, generated outputs, prompt renders, seed payloads,
audit records, pre-registration and amendment records, generation and evaluation code,
falsification suites, and figure/table generation scripts produced and analyzed in this
study have not been released through a public repository, DOI, or external deposit. These
materials are available from the corresponding author upon reasonable request for academic
verification and replication. Sharing of the requested materials may be limited by file size,
technical transfer constraints, third-party benchmark/model license conditions, and
institutional policy constraints. The open-weight models and EvalPlus benchmarks used in the
study are already available from their respective public sources.

\section*{Ethics and Generative-AI Use Disclosure}

The study evaluates publicly released open-weight code models on public benchmarks. No human
subjects or personal data are involved. Large language models were used in two declared roles:
as the \emph{objects} of the study, namely the three tested frozen models, and as
manuscript-preparation assistants for drafting and editing under the author's direction and
review. All experimental design decisions, pre-registrations, and verdicts belong to the
author. All reported numbers were produced either by the executable pipeline or by the explicitly marked program-cycle records.

\section*{Funding}

This work was supported by the Scientific and Technological Research Council of Türkiye
(TÜBİTAK) under the 1001 programme, project no. 225M316, “A Tilt-Trirotor Vertical Take-Off
and Landing Controller Enabling Task-Oriented Transfer and Rapid Adaptation.
Hardware-in-the-Loop and Real Validation of a Meta-Learning--Based Reinforcement Learning
Architecture.” The meta-learning code developed in that project was integrated into the test
harness together with the frozen language models studied here and was used during the
agent-design phase to optimize the algorithms designed by the author.

\section*{Conflict of Interest}

The author declares that there is no conflict of interest.

\printbibliography

\appendix
\part*{APPENDIX}
\phantomsection
\addcontentsline{toc}{part}{APPENDIX}

These appendices have been organized so that the records supporting the claims
in the main text can be traced by the reader. In Appendix~\ref{app:a}, the
audit and amendment trail for all stages of the program is presented in
chronological order. Appendix~\ref{app:b} summarizes the scaffold roles and placebo-construction
rules for the prompt and weight channels. In
Appendix~\ref{app:c}, the per-cell and per-arm unlock rates on the resistant
band are reported at full resolution. Appendix~\ref{app:d} records
provenance notes associated with the main-text program ledger.
Appendix~\ref{app:e} summarizes the calibration and supersession chain of
AEG-BANDIT. Appendix~\ref{app:f} summarizes the derivational objects used by
ERA together with the deferred-validity record. In Appendix~\ref{app:g}, the study is
positioned relative to prior work through the experimental controls retained.

\section{Audit and amendment trail}\label{app:a}

In this appendix, the audit events associated with all stages of the program
(\genk{0}--\genk{8}) are recorded in chronological order together with their
dispositions and evidential statuses. The purpose is not merely to report the
final amended verdict, but also to make the failure modes of the evaluation
pipeline reproducible and open to criticism. Accordingly, the original audit
failures were not removed, dated amendments were reported separately, and
post-amendment verdicts were not substituted for the original as-frozen
verdicts. Each correction is presented in the sequence
original rule $\rightarrow$ failure $\rightarrow$ amendment
$\rightarrow$ amended rule.

\begin{enumerate}[label=(\roman*),leftmargin=2em]

\item \emph{Pre-run falsification suites and gate firings.}
Before each program generation was initiated, its implementation was tested
using its own executable falsification suite. Preregistered gate firings,
including the \vdGOneKill{} result triggered for both feature sets at the FDP
stage, were retained in the run record as dated dispositions rather than as
silent exclusions.

\item \emph{FASTR sandbox amendment (\genk{2}).}
At the diagnostic stage, a sandbox defect caused by restricted built-ins had
rendered 589 of 1,912 candidates all-exception casualties. After the defect
was corrected and rescoring was performed, the dead-unit count decreased from
189 to 154, and the eligible search-bound stratum decreased from 38 to 33
(33/535). This amendment was closed before the primary runs and was retained
as a trace-preserving data-quality correction.

\item \emph{FASTR learned-signal amendment (\genk{2}).}
At the same stage, the learned-signal AUC was reduced from 0.597 to 0.467 after
the rule under which ties had been counted as wins was corrected. The corrected
value remained below chance.

\item \emph{ERA $\kappa$ defects (\genk{6}).}
Two $\kappa$ (kappa) defects were detected and corrected at the G6
falsification step. The corrections were closed before the primary runs and
were retained in the trail.

\item \emph{ELF severity-sign correction (\genk{7}, design stage).}
The sign of the outcome-calibrated severity update used by the \armOn{}
controller was corrected from the reversed sign in the design document, and
the correction was verified using SymPy. The correction was applied before
the primary run and was recorded in this trail.

\item \emph{ELF destroyed-task \armSh{} correction (\genk{7}, pilot).}
During the Cycle-5 liveness pilot, it was detected that the original pilot
version of \armSh{} also corrupted the task statement, thereby producing the
destroyed-task defect. The scaffold was corrected through the registered
second shape-control amendment, and Stage-1 was run using the corrected
\armSh{}. In the Stage-1 audit, the shape-leak count was recorded as zero, and
no audit failure was recorded.

\item \emph{AEG supersession (\genk{4b}).}
The initial \vdPass{} decision from the naive replay reading
(+6.0 pp) was linked to its matched draw-budget successor through the
+0.001 control from the same record and a persistent supersession record.
The initial \vdPass{} decision was marked decision-invalid and was not
interpreted as the valid result. A summary of the chain is
provided in Appendix~\ref{app:e}.

\item \emph{HEF known failure (tracked).}
The decoy-synthetic test in the HEF learning test suite has failed
deterministically since the migration to Torch 2.12. The defect was not
suppressed. It was retained with tracked status and is preserved as
\emph{as-frozen failing (tracked)} in the audit-status-by-layer table
(Table~\ref{tab:audit}).

\end{enumerate}

A layer-level summary of this trail is collected in the audit-status-by-layer
table in the main text (Table~\ref{tab:audit}). The trail is retained to
document, at the row level, that statistical verdicts and audit verdicts carry
different epistemic statuses. As-frozen audit states are not overwritten by
any correction, and the conjectures of the evaluation system itself remain
open to executable counterexamples.

\section{Prompt/LoRA scaffold and placebo summaries}\label{app:b}

In this appendix, the frozen scaffold records for all arms in the two channels
are summarized. The roles of the \armOn{}/\armSh{}/\armDe{}/\armFr{} templates are summarized
for the prompt channel, and the training and evaluation configurations for
\armCt{}/\armSf{}/\armFr{} are summarized for the weight channel. A single
scaffold family is shared by the arms, and differentiation is introduced only
through the conditioning packet or training data. The definitions were derived
from the archived prompt-generation rules and the frozen training and
evaluation configurations.

\paragraph{\armOn{}.}
The live learned error-lattice controller is instantiated as the \armOn{} arm.
An error-conditioned repair prompt is generated and updated within the run.
The error content harvested from the unit itself is carried in the conditioning
block.

\paragraph{\armSh{}.}
A structure-matched, content-free placebo is instantiated as \armSh{}. The
genuine task prompt, entrypoint, layout, and scaffold are held fixed, while
only the public-facts block is replaced by a content-free block with the same
layout. The destroyed-task defect in the original pilot
version of \armSh{} and the registered shape-control amendment are recorded in
the trail in Appendix~\ref{app:a}.

\paragraph{\armDe{}.}
The donor-error-facts construction is instantiated as \armDe{}. Genuine-looking
but mismatched error content taken from another evaluation unit is used. The
task$\to$error mapping is disrupted while representational form is preserved.

\paragraph{\armFr{}.}
In the prompt channel, \armFr{} is instantiated as the pretrained state of the
same controller, and no online update is applied. In the weight channel,
\armFr{} is instantiated as the base model without an adapter
(Qwen2.5-Coder-1.5B-Instruct). Byte-identical restoration is verified in the
adapter-effect probe through the call by which the adapter is disabled.

\paragraph{\armCt{}.}
The \armCt{} arm is instantiated as a QLoRA adapter trained on 1,964 genuine
task--failing-code--error-atom pairs. The corpus was constructed from
236 non-evaluation units, with 46 evaluation/reserve units excluded and a cap
of $\leq$30 examples per unit.

\paragraph{\armSf{}.}
The \armSf{} arm is instantiated as a SHA-deranged placebo LoRA. Training is
performed on a SHA-seeded complete derangement of the same error blocks. The
number of examples, surface form, family marginals, and training
hyperparameters are identical to those of \armCt{}. The task$\to$content
mapping is removed.

The definitions of the executable audits used to preserve placebo integrity
are frozen together with these scaffolds. Byte-level SHA verification and the
forbidden-token audit are defined for the prompt channel. The derangement audit
with no fixed points and byte-identical family distributions, the unit-level
split, and the adapter-disable probe are defined for the weight channel. The
Stage-1 and evaluation outputs of these audits are reported in the main text
(Table~\ref{tab:audit}).

\section{Per-cell and per-arm unlock rates}\label{app:c}

\begin{figure}[t]
\centering
\includegraphics[width=\linewidth]{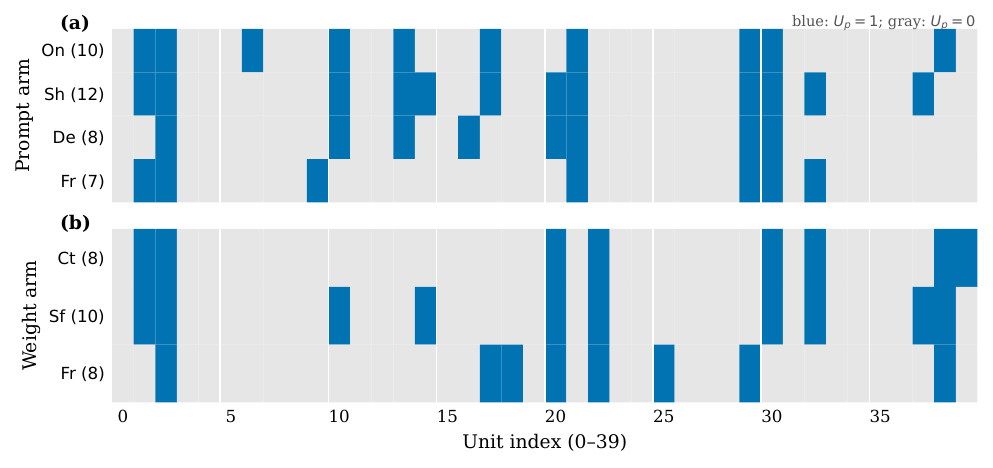}
\caption{Per-unit $U_p$ outcomes are shown in frozen-manifest unit order. The
columns have not been reordered by effect size. The prompt-channel row totals
are \armOn{} 10, \armSh{} 12, \armDe{} 8, and \armFr{} 7 units. The
weight-channel row totals are \armCt{} 8, \armSf{} 10, and \armFr{} 8 units.
Blue denotes $U_p=1$, and gray denotes $U_p=0$. The unlock memberships of
\armCt{} and \armFr{}, despite their identical pooled totals, do not coincide
exactly. The results belong to the public-tier screen on the 40-unit band and
do not include hidden-tier confirmation.}
\label{fig:a3}
\end{figure}

The preregistered per-cell and per-arm $U_p$ screening rates
(Definition~\ref{def:pub4}) on the 40-unit 1.5B-marginal resistant band are
provided here at full resolution together with paired discordant membership.
The purpose is to allow verification of whether pooled ties rest on different
sets of units. A zero or negative pooled difference is not interpreted as
evidence of equivalence or non-inferiority. All numbers belong to the $U_p$
screening endpoint \eqref{eq:pub4-unlock}. Hidden-tier true-unlock confirmation
was deferred by design, and unspent cells are marked ``not spent.'' The
unit-level membership records of arms with the same pooled total are shown in
the heat strip in Figure~\ref{fig:a3}.

\begin{center}
\captionof{table}{Distribution of error families in the ELW training corpus
(descriptive). The \armCt{} and \armSf{} marginals are identical. The number
of derangement fixed points is 0, and the 46 excluded units were never read
during training, as required by policy.}
\label{tab:elw-family-distribution}
\footnotesize
\fitwidth{%
\begin{tabular}{l r@{\qquad}l r@{\qquad}l r}
\toprule
Family & Rows & Family & Rows & Family & Rows \\
\midrule
F-runtime & 831 & F-value & 561 & F-container & 288 \\
F-type & 197 & F-numeric & 95 & F-syntax & 39 \\
F-interface & 31 & F-performance & 31 & F-string & 26 \\
\bottomrule
\end{tabular}}
\end{center}

This table and heat strip are retained to show the per-cell structure of the
mechanism screen defined by Definition~\ref{def:pub4} in the main text. They
do not replace the pooled results.

\section{Program-ledger provenance notes}\label{app:d}

The provenance fields associated with the program ledger are summarized here.
The complete ledger remains in Table~\ref{tab:ledger}. The controlled verdict set is
defined in \eqref{eq:verdictset}. The LEGACY rows are linked to
\citet{iscan2026c} as inherited records. Numbers that exist only in the cycle
log are marked with $^{\dagger}$, and it is stated that these numbers are not
contained in the frozen results archive. The \vdDoa{}, \vdUnrealizable{},
\vdUnderpowered{}, and \vdStratumNull{} dispositions are shown as separate
rows rather than as silent exclusions. The supporting view of the ledger,
comprising the generator-scale observation and offline-gate kills, is
collected in the strip shown in Figure~\ref{fig:scalegates} in the Results.

\section{AEG calibration and supersession summary}\label{app:e}

In this appendix, the instrument-calibration and supersession records of
AEG-BANDIT are summarized together. The closed-form acquisition function
itself is defined in Eq.~\eqref{eq:zib-ei}. The allocator's own falsification suite was
run at the level of instrument calibration before the supersession chain was
opened. The planted-signal positive control was triggered
($\Delta = 0.31611$, $p = 0.00067$), the difference remained null in the null
world ($\Delta = -0.01167$, $p = 0.6942$), and the closed-form validation
recorded a maximum absolute error of $1.02\times10^{-10}$ together with zero
submodularity violations across 60 trials. The planted-signal value is not an
allocation gain. It records that the instrument was capable of detecting the
planted signal.

The closed-form zero-inflated Beta expected-improvement derivation and the
correlated-evidence correction
($\hat{\rho} = 0.7771$, $\kappa_{\mathrm{eff}} = 1.175$) of the budgeted
best-of-B allocator are summarized here. The naive +6.0 pp
\vdPass{} in the superseded ($B=12$) record
($\Delta = 0.06025$, $p = 0.01375$, budget 12) is restated from the frozen record together with the
+0.00097 (+0.001) adaptive--random control from the same record
($p = 0.44664$) and the supersession marker. In the governing matched
draw-budget ($B = 8$) record, adaptive--best fixed is retained as
$\Delta = +0.02209$ ($p = 0.2022$), adaptive--random as
$\Delta = -0.00227$ ($p = 0.6176$), and the null-world sanity check as
$\Delta = -0.02667$ ($p = 0.91354$). The persistent supersession chain is
presented as the traceable record of the self-audit. The initial \vdPass{}
decision is not interpreted as the valid result.

This derivation and persistent supersession chain constitute the
machine-traceable record supporting the \vdReplayNull{} verdict reported in
the main text.

\section{ERA derivation summary and deferred validity}\label{app:f}

The derivational objects used by the error-set architecture (ERA) are
summarized here. They
comprise the arrival-order prior, the $1/(K+1)$ reservoir mass, the exact Hedge
regret bound under growing vocabulary
(worst 0.8836 $<$ bound 10.254), and the incidence-form Good--Turing missing
mass. The taxonomy-saturation values
($K = 93$, $\hat{M} \approx 0.03$) were calculated over 1,920 harvested
candidates. Two independently produced derivation records are preserved for comparison. The ERA validity test, defined
as the three-label out-of-fold first-draw ARM-REGRET statistic, was
preregistered. It was not run in this generation because the deferral rule was
triggered (\vdDeferred{}, \S\ref{sec:era}). All ERA outputs therefore have
descriptive-machinery status and are nowhere presented as a validated JEPA-RL
reward. For completeness, the JEPA no-signal finding in ERA's own smoke record
is repeated here
(ridge $R^{2}_{\mathrm{oof}} \approx 0.0096$,
decoder $R^{2}_{\mathrm{oof}} \approx -133$). These derivations and the smoke
record support the deferred-validity decision reported under
\S\ref{sec:era} in the main text.

\section{Positioning by controls}\label{app:g}

The positioning-by-controls table cited in the main text
(Table~\ref{tab:positioning}) is provided here. The study is positioned
relative to prior work not through terminology, but through the experimental
controls retained. The columns indicate the presence of a content-versus-form
placebo, a matched output-generation budget, a hidden-tier endpoint, paired
prompt and weight channels, and a SHA-deranged placebo LoRA. Self-audit coverage
of the study's own positive result is recorded in the scope rows rather than as
a separate axis. The table is a controls-coverage comparison. It does not carry a claim of outcome superiority. The marks were
coded from the published method descriptions of the cited studies. The map is
descriptive and does not constitute exhaustive certification.

\begin{sidewaystable}[p]
\caption{A controls-coverage map is shown for selected prior work. The matrix
separates the content-versus-form placebo, matched output budget, hidden
endpoint, two-channel, and weight-space placebo axes. The marks were coded
from the published method descriptions of the cited studies. The map is
descriptive and does not constitute exhaustive certification, evidence of
outcome superiority, or evidence of equivalence.}
\label{tab:positioning}
\centering\footnotesize\setlength{\tabcolsep}{6pt}
\begin{tabular}{@{}>{\raggedright\arraybackslash}p{0.34\linewidth}
*{5}{>{\centering\arraybackslash}p{0.105\linewidth}}@{}}
\toprule
Research cluster &
Content-versus-form placebo &
Matched output budget &
Hidden endpoint &
Two channels &
Weight-space placebo \\
\midrule
Self-repair / self-refinement
\citep{madaan2023,shinn2023,chen2024,olausson2024}
& --- & $\sim$ & $\sim$ & --- & --- \\

Execution-feedback RL
\citep{le2022,shojaee2023,gehring2024}
& --- & --- & $\sim$ & --- & --- \\

Inference-aware BoN / test-time compute
\citep{brown2024,snell2025,chow2025}
& --- & $\sim$ & $\sim$ & --- & --- \\

PEFT / self-training
\citep{hu2021,dettmers2023,zelikman2022,singh2024,biderman2024}
& --- & --- & $\sim$ & --- & --- \\

Progressive refinement
\citep{du2025}
& --- & $\sim$ & --- & --- & --- \\

\textbf{This study (gen 0--8)}
& \checkmark & \checkmark & $\sim^{*}$ & \checkmark & \checkmark \\
\bottomrule
\end{tabular}

\vspace{8pt}
\begin{tabular}{@{}
>{\raggedright\arraybackslash}p{0.24\linewidth}
>{\raggedright\arraybackslash}p{0.24\linewidth}
>{\raggedright\arraybackslash}p{0.45\linewidth}@{}}
\toprule
Research cluster & Regime & Scope and limitation \\
\midrule
Self-repair / self-refinement
\citep{madaan2023,shinn2023,chen2024,olausson2024}
&
Positive within its own regime
&
Prompt channel only. No content-versus-form placebo or self-audit is included
\\

Execution-feedback RL
\citep{le2022,shojaee2023,gehring2024}
&
Positive with larger models and online updates
&
Weight channel only. No matched budget or placebo is included
\\

Inference-aware BoN / test-time compute
\citep{brown2024,snell2025,chow2025}
&
Positive existence and budget-allocation results
&
The content-versus-form attribution question is not posed on the same endpoint
\\

PEFT / self-training
\citep{hu2021,dettmers2023,zelikman2022,singh2024,biderman2024}
&
Large-scale positive regimes with correct-answer filtering
&
Weight channel only. No weight-space placebo is included
\\

Progressive refinement
\citep{du2025}
&
Progressive-refinement regime
&
The positive results have not been tested in this regime under a form-matched
placebo and a matched output-generation budget
\\

\textbf{This study (gen 0--8)}
&
ELF and ELW were evaluated on the same 40-unit band against the same placebo
hierarchy
&
\armSh{}/\armDe{}/\armFr{} and a SHA-deranged placebo LoRA were used with
$R = 4$/arm/unit. In the self-audit of the study's own positive, the
superseded +6.0 pp and +0.001 records were retained together under a persistent
supersession pointer. The scope is limited to a small scale, one language, and
one scaffold
\\
\bottomrule
\end{tabular}

\par\vspace{5pt}
\begin{minipage}{0.94\linewidth}\footnotesize
A \checkmark indicates that the control was retained, $\sim$ indicates that it
was retained partially or conditionally, and --- indicates that it was not
present in the reported method description. A matched output budget denotes
the allocation of the same number of output generations to the feedback and
sampling conditions. A placebo denotes a channel-specific control that preserves the predeclared
scaffold components while task-relevant error content is ablated or the
task--error assignment is deranged. $^{*}$ True unlock was realized
in the inherited m3 primary. LG P0 measures the public-tier dense
unit-by-portfolio interaction, and hidden-partial calibration is retained as a
separate descriptive record. The ELF and ELW headline verdicts belong to the
$U_p$ screen, and hidden-tier confirmation was left \vdDeferred{}.
\end{minipage}
\end{sidewaystable}

\end{document}